\documentclass[fleqn,usenatbib]{mnras}

\usepackage{newtxtext,newtxmath}
\usepackage[T1]{fontenc}
\usepackage{graphicx}   % Including figure files
\usepackage{amsmath}    % Advanced maths commands
\usepackage{comment}
\usepackage{epsfig}
\usepackage{xcolor}
\usepackage{tabularx}
\usepackage{threeparttable}
\usepackage{hyperref}
\hypersetup{colorlinks=true, citecolor=blue, linkcolor=blue}
\usepackage{xspace}
\let\oldAA\AA
\renewcommand{\AA}{\text{\oldAA}\xspace}

\newcommand{\redtxt}[1]{\textcolor{black}{#1}}
\newcommand{\rrtxt}[1]{\textcolor{black}{#1}}
\newcommand{\purpletxt}[1]{\textcolor{purple}{#1}}
\newcommand{\hii}{H\,{\sc ii}}
\newcommand{\heii}{He\,{\sc ii}}
\newcommand{\neiii}{[Ne\,{\sc iii}]}
\newcommand{\oiii}{[O\,{\sc iii}]}
\newcommand{\oiiip}{O\,{\sc iii}}
\newcommand{\oii}{[O\,{\sc ii}]}
\newcommand{\ciii}{C\,{\sc iii}]}
\newcommand{\civ}{C\,{\sc iv}}
\newcommand{\feii}{Fe\,{\sc ii}}
\newcommand{\niv}{N\,{\sc iv}]}
\newcommand{\niii}{N\,{\sc iii}]}

\newcommand{\ergscm}{$\rm erg~s^{-1}~cm^{-2}$}

\newcommand{\cloudy}{\textsc{Cloudy}}
\newcommand{\pyneb}{\textsc{PyNeb}}

\title[The small blue bump in GN-z11]{\centering JADES - The small blue bump in GN-z11: \\
insights into the nuclear region of a galaxy at z=10.6}

\author[Ji et al.]{Xihan Ji,$^{1,2}$\thanks{E-mail: \href{mailto:xj274@cam.ac.uk}{xj274@cam.ac.uk}}
Roberto Maiolino,$^{1,2,3}$
Gary Ferland,$^{4}$
Francesco D'Eugenio,$^{1,2}$
Rachana Bhatawdekar,$^{5}$
\newauthor
Stéphane Charlot,$^{6}$
Jacopo Chevallard,$^{7}$
Mirko Curti,$^{8}$
Emma Curtis-Lake,$^{9}$
Kevin Hainline,$^{10}$
Zhiyuan Ji,$^{10}$
\newauthor
Brant Robertson,$^{11}$
Bruno Rodr\'iguez Del Pino,$^{12}$
Jan Scholtz,$^{1,2}$
Sandro Tacchella,$^{1,2}$
Christina C. Williams,$^{13}$
\newauthor
Joris Witstok$^{14,15,1,2}$
\\
$^{1}$Kavli Institute for Cosmology, University of Cambridge, Madingley Road, Cambridge, CB3 0HA, UK\\
$^{2}$Cavendish Laboratory, University of Cambridge, 19 JJ Thomson Avenue, Cambridge, CB3 0HE, UK\\
$^{3}$Department of Physics and Astronomy, University College London, Gower Street, London WC1E 6BT, UK\\
$^{4}$Department of Physics and Astronomy, University of Kentucky, 505 Rose Street, Lexington, KY 40506, USA\\
$^{5}$European Space Agency (ESA), European Space Astronomy Centre (ESAC), Camino Bajo del Castillo s/n, 28692 Villanueva de la Cañada, Madrid, Spain\\
$^{6}$Sorbonne Universit\'e, CNRS, UMR 7095, Institut d'Astrophysique de Paris, 98 bis bd Arago, 75014 Paris, France\\
$^{7}$Department of Physics, University of Oxford, Denys Wilkinson Building, Keble Road, Oxford OX1 3RH, UK\\
$^{8}$European Southern Observatory, Karl-Schwarzschild-Strasse 2, 85748 Garching, Germany\\
$^{9}$Centre for Astrophysics Research, Department of Physics, Astronomy and Mathematics, University of Hertfordshire, Hatfield AL10 9AB, UK\\
$^{10}$Steward Observatory, University of Arizona, 933 N. Cherry Avenue, Tucson, AZ 85721, USA\\
$^{11}$Department of Astronomy and Astrophysics University of California, Santa Cruz, 1156 High Street, Santa Cruz CA 96054, USA\\
$^{12}$Centro de Astrobiolog\'ia (CAB), CSIC–INTA, Cra. de Ajalvir Km.~4, 28850- Torrej\'on de Ardoz, Madrid, Spain\\
$^{13}$NSF’s National Optical-Infrared Astronomy Research Laboratory, 950 North Cherry Avenue, Tucson, AZ 85719, USA\\
\redtxt{$^{14}$Cosmic Dawn Center (DAWN), Copenhagen, Denmark}\\
\redtxt{$^{15}$Niels Bohr Institute, University of Copenhagen, Jagtvej 128, DK-2200, Copenhagen, Denmark}
}

\begin{document} 
\label{firstpage}
\pagerange{\pageref{firstpage}--\pageref{lastpage}}
\maketitle
 
\begin{abstract}
  We report the detection of continuum excess in the rest-frame UV between 3000 \AA\ and 3550 \AA\ in the JWST/NIRSpec spectrum of GN-z11, \redtxt{a luminous galaxy $z=10.603$.}
  The shape of the continuum excess resembles a Balmer continuum but has a break around 3546 \AA.
  %in the rest frame.
  \rrtxt{The fitting result of this excess depends on the assumed origin of the continuum.}
  \redtxt{If the continuum of GN-z11 is dominated by a stellar population with a small Balmer break, the apparent blueshift of the Balmer continuum is not significant and the best-fit Balmer continuum model indicates a temperature of $T_e = 1.78^{+0.25}_{-0.21}\times 10^4$ K.}
  \redtxt{In contrast, if the continuum is dominated by AGN emission,}
  a \redtxt{nebular} continuum model cannot fit the spectrum \redtxt{properly}.
  %, implying a different origin for the continuum excess.
  The absence of the Balmer jump indicates an electron temperature of $\sim 3\times 10^4$ K, significantly higher than the temperature of \redtxt{$T_{e}({\rm O^{2+}}) = 1.36\pm 0.13\times 10^{4}$ K inferred from \oiii$\lambda 4363$ and \oiii$\lambda 5007$}.
  The temperature difference \redtxt{can} result from mixing of different ionized regions: the Balmer emission mainly arises from dense and hot clouds in the Broad Line Region, whereas the forbidden lines originate from less dense and colder gas. 
  \redtxt{An alternative explanation for the observed continuum excess is the} \feii\ emission, which shows a characteristic jump blueward of the Balmer limit as previously seen in the spectra of many lower-redshift quasars. 
  Through comparisons with \textsc{Cloudy} models, we show an Fe abundance 
  above {$\sim 1/3$} solar is likely needed, \rrtxt{which could} be achieved via enrichment from Type-Ia supernovae, hypernovae, or pair-instability supernovae.
\end{abstract}

%
%-------------------------------------------------------------------

\section{Introduction}

The UV spectra of Type-1 active galactic nuclei (AGNs) often show prominent continuum excess extending from 2000 \AA\ to 4000 \AA, which is also known as the ``3000 \AA\ excess'' or the ``small blue bump'' \citep{neugebauer1979,richstone1980,grandi1982}.
Early observations show the small blue bumps in quasars exhibit a variety of shapes and are found to show variability weakly correlated with that of the UV continuum \citep[e.g.,][]{ulrich1980,ulrich1997}.
In addition, polarization analyses of the spectrum of the Seyfert-1 galaxy NGC 4151 show the small blue bump has a similar polarization as that of broad emission lines, implying a photoionization origin in the broad-line region \citep[BLR;][]{schmidt1980}.

The commonly adopted explanation for the UV continuum excess is a combination of \feii\ emission and Balmer continuum \citep[e.g.,][]{grandi1981,grandi1982,malkan1982,wills1985,dietrich2003,kovacevic2014}.
In BLRs, collisional excitation of Fe$^+$, fluorescence by \feii\ lines as well other emission lines with similar wavelengths, and radiative excitation by the AGN continuum are able to produce strong \feii\ emission \citep{netzer1983,wills1985,penston1987}.
Another important factor that enables detectable \feii\ emission in BLRs is the sublimation of dust grains, as Fe is usually significantly depleted onto dust in the interstellar medium \citep[ISM;][]{jenkins2009}.
However, it is also suggested that the varying strength of \feii\ in observations might indicate the occasional overlap of the \feii\ emitting region with regions outside the dust sublimation radius \citep{shields2010}.
Due to the dense energy levels of Fe$^+$ and fast motions of clouds within BLRs, \feii\ emission lines are usually highly blended in observations and form a pseudo-continuum spanning from the rest-frame UV to optical.
Still, in both observations and theoretical models of the \feii\ pseudo-continuum, the \feii\ emission can be divided by groups of broad emission centering at different wavelengths, whose relative strengths are determined by a series parameters including the optical depth, the velocity of microturbulence, the metallicity, and so forth \citep[e.g.,][]{baldwin2004,ferland2009,sarkar2021}.
The edges of the \feii\ groups, especially between the UV groups and optical groups, are prominent in some quasar spectra and resemble sharp jumps in the pseudo-continuum excess \citep[e.g.,][]{tsuzuki2006}.

Besides the \feii\ emission, a Balmer continuum is often needed for a complete fit of the UV spectra of Type-1 AGNs. The Balmer continuum contributes to a smoothly rising component of the continuum until the Balmer limit at 3646 \AA.
The Balmer continuum is produced by the recombination of free electrons to the $n = 2$ level of hydrogen.
The presence of the Balmer continuum provides an independent probe of the gas temperature.
This is because the ratio between the intensity of the Balmer continuum at the Balmer limit and that of any given Balmer line decreases with increasing temperature roughly following a power-law relation \citep{peimbert1967}.
%and has the following approximate form under the Case B approximation
%\begin{equation}
%    \rm I(Bac~3646~\AA)/I(Hn) \propto T_e^{-0.7}.
%\end{equation}
As temperature increases, while the whole nebular continuum becomes stronger, the relative strength of the Balmer jump becomes weaker with respect to the intensity of Balmer lines.
%more difficult to observe at higher temperatures at fixed signal-to-noise ratios (S/N) of Balmer lines.
%It is also noteworthy that when the Balmer continuum becomes partially optically thick, the functional form of the Balmer continuum changes and the interpretation of the electron temperature would hinge on the optical depth \citep{grandi1982}.
%\redtxt{Regardless, it remains debatable whether the Balmer continuum can reach a significant optical depth, which depends on its emitting region in the AGN.}
According to the widely adopted BLR model with a stratified ionization structure, H\,{\sc i} emission including the Balmer continuum comes from clouds closer to the accretion disk compared to \feii\ emission, leading to the stronger variability of H\,{\sc i} emission in observations \citep{ulrich1997}.
\redtxt{The Balmer continuum is also visible in nearby \hii\ regions and planetary nebulae \citep{peimbert2017} and has been used as an independent temperature probe \citep[e.g.,][]{liu_balte_2001}.
At high redshift, as revealed recently by JWST observations, strong Balmer continua were reported in several systems from either AGN or young stellar populations \citep[e.g.,][]{cameron_9422_2024,ji2024,katz2024,tacchella2024}.
}

\redtxt{On one hand, for AGNs, understanding} the origin of the small blue bump and decomposing the contributions from different nebular emissions are important for constraining the Fe abundance as well as the electron temperature of the BLR.
While the flux ratio of Mg\,{\sc ii}$\lambda \lambda 2796,2803$/\feii\ measured in the rest-frame UV has been used as an empirical proxy of the abundance ratio of $\alpha$/Fe in Type-1 AGNs, the accurate determination of the Fe abundance requires detailed modeling of the physical conditions within BLRs \citep{maiolino2019}.
Previous theoretical modeling attempts have shown
%observations of AGNs throughout cosmic time have shown a variety of morphologies and strengths for the continuum excess, which are difficult to reproduce by models and simulations.
it is particularly challenging to reproduce the various strengths of the observed \feii\ complex \citep{gaskell2022}.
Although significant progress has been made by implementing complex energy levels of Fe$^{+}$ as well as physical processes such as self fluorescence and continuum pumping 
%and self-shielding 
in photoionization simulations, the current application of the theoretical approach is limited to a few well observed sources and certain parameter regimes \citep[e.g., I Zwicky 1;][]{sarkar2021}.
As the UV spectrum becomes increasingly important for spectroscopic studies of AGNs in the early Universe, it is vital to understand the UV signatures of AGNs over a broader parameter space.
On the other hand, \redtxt{for both AGNs and star-forming (SF) galaxies,} independent temperature determinations from the Balmer jump can provide important clues on the temperature structures and chemical abundance patterns.
The main limiting factor for constraining the Balmer temperature in BLRs, however, is the degeneracy between the strength of the Balmer continuum and that of the \feii\ complex.

With the advent of the James Webb Space Telescope \citep[JWST;][]{jwst0,jwst1}, a diverse population of AGNs has been revealed at $z\sim 2 - 11$ \citep{furtak2023,greene2023,kocevski2023,maiolino2023,maiolino2023b,matthee2023,onoue2023,scholtz2023,ubler2023b,ubler2023a}.
Meanwhile, emission-line studies based on JWST/NIRSpec \citep{jakobsen2022} have shown the prevalence of highly ionized and metal-poor gas in the early Universe \citep[e.g.,][]{cameron_jwstism_2023,
curti2023b,curti2023a,laseter2023,sanders2023a,sanders2023b,tacchella2023_ismstar}.
%a surprisingly high fraction of auroral-line detection is confirmed by JWST/NIRSpec \citep{jakobsen2022}, indicating the prevalence of highly ionized and metal-poor gas in the early Universe \citep{laseter2023}.
At this parameter regime, the behavior of the small blue bump remains unexplored. A naive expectation might be that the small blue bump becomes less important and even vanishes in the BLRs of very early AGNs because of low Fe abundances and high electron temperatures.
However, in this manuscript we report the discovery of a significantly detected UV continuum excess in the JWST/NIRSpec spectrum of GN-z11, a luminous galaxy at
%\redtxt{potentially} hosting an AGN 
$z = 10.603$ \citep{oesch2016,bunker2023,tacchella2023,maiolino2023}, which partly resembles the shape of the small blue bumps previously observed in lower-redshift AGNs.

\redtxt{Notably, the JWST observations of the UV nebular emission of GN-z11 can be well explained by ionization of an accreting black hole according to the recent works by \citet{Maiolino2023_HeII,maiolino2023,scholtz2023_gnz11}.
A careful analysis on the multiplet ratios of \niii$\lambda 1746-1754$ and the doublet ratio of \niv$\lambda \lambda 1483,1486$ measured from the medium-resolution NIRSpec spectrum of GN-z11 indicates a high electron density of $n_e \gtrsim 10^{9}~{\rm cm^{-3}}$, typical of the physical conditions in BLRs while unseen in the ISM \citep{maiolino2023}.
The AGN identification is further supported by the detection of transitions typical of AGN, such as the C\,{\sc ii}$^*$1335 fluorescent emission and [Ne\,{\sc iv}]$\lambda 2424$,
high-velocity ionized winds seen in \civ$\lambda \lambda 1548,1550$ \citep{maiolino2023},
ionization cones in \ciii$\lambda \lambda 1906,1908$ \citep{Maiolino2023_HeII}, and an extended Ly$\alpha$ halo, whose brightness and extent is consistent with what observed in quasars at $z\sim 6$ and inconsistent with typical star forming galaxies at similar redshifts \citep{scholtz2023_gnz11}. 
The measured widths of permitted lines in GN-z11 from NIRSpec median-resolution spectra are narrower than those typically found in lower-$z$ Type-1 AGNs \citep{maiolino2023}, but in the range of those seen in the so-called Narrow Line Seyfert 1s \citep[NLSy1s;][]{osterbrock1985,goodrich1989,leighly1999}, and compatible with a low-mass black hole with $\rm M_{BH} \sim 1.6\times 10^6~M_{\odot}$.
Meanwhile, \cite{maiolino2023} estimated an Eddington ratio of 5 (with an uncertainty of a factor of 2), further showing that GN-z11 is similar to NLSy1s, which are usually highly accreting systems. It is worth noticing, in the context of this paper, that NLSy1s tend to show prominent \feii\ emissions \citep[e.g.,][]{veron-cetty2001,vestergaard2001,komossa2008}.
Interestingly, the existence of strong nitrogen lines suggests a super-solar N/O $\sim 430$ Myr after the Big Bang, indicative of either a exotic chemical enrichment history in this galaxy or a past chemical enrichment within the small BLR \citep{bunker2023,cameron2023,maiolino2023,senchyna2023}.
Since the first Type-Ia supernovae (SNe) that are capable of enriching Fe can occur as early as $\sim 30$ Myr after the initial star formation \citep[e.g.,][]{maiolino2019}, if there has been past chemical enrichment in the BLR, one might expect Fe has also been enriched by a few SNe given the likely active star formation in GN-z11 over the past 20--50 Myr \citep{tacchella2023}.}

\redtxt{We also note that some other works argue young stellar populations from a recent starburst might instead dominate the light of GN-z11 \citep[e.g.,][]{bunker2023,cameron2023,charbonnel2023,senchyna2023,alvarez-marquez_gnz11_2024}.
For example, \citet{bunker2023} show that the UV line ratios of GN-z11 lie in the transition zone between AGN and SF models of \citet{feltre2016} and \citet{gutkin2016} in the UV diagnostic diagram proposed by \citet{nakajima2018}, although we caution that the possibility of a super-Eddington accreting black hole in GN-z11 might lead to softer ionizing radiation compared to typical AGNs \citep[e.g.,][]{lambrides2024}. Also, it remains unclear whether the UV lines contain a broad component.
More recently, \citet{alvarez-marquez_gnz11_2024} reported non-detection of any broad component in H$\alpha$ using the JWST/Mid-Infrared Instrument (MIRI) medium-resolution spectroscopy (MRS) observations of GN-z11.
Still, \citet{alvarez-marquez_gnz11_2024} claimed an underlying broad component in H$\alpha$ is not ruled out due to the limited signal-to-noise (S/N).
In fact, according to \citet{alvarez-marquez_gnz11_2024}, the broad H$\alpha$ associated with 
AGN identified in \citet{maiolino2023} would not be detectable in their MIRI spectrum and, therefore, it is plausible that the AGN light is more dominant in the rest-frame UV.}
%in an AGN observed by JWST/NIRSpec at $z = 10.603$ (also known as GN-z11), we found a significantly detected continuum excess in the UV, resembling part of a small blue bump in lower-redshift AGNs.

\redtxt{Regardless of the dominant ionizing source in GN-z11, the existence of a} ``small blue bump'' \redtxt{at $z=10.603$} raises questions about whether the chemical enrichment processes and temperature structures in early \redtxt{galaxies and/or} AGNs differ from those at later cosmic epochs.
We show that such a bump is \redtxt{potentially} inconsistent with being associated with a Balmer jump; this implies that the temperature associated with the Balmer lines must be much higher than that inferred from the auroral (forbidden) transition, \oiii$\lambda 4363$, further confirming that the Balmer lines predominantly come from the BLR. We then interpret the small bump similarly as in other Type-1 AGNs (i.e., associated with the Fe complex, although likely with different properties relative to local AGN).

%Among the spectroscopically identified AGNs, we found one broad-line AGN at $z = 10.603$ in the Goods North (also known as GN-z11) and one narrow-line AGN at $z = 9.433$ in the Goods South (hereafter GS-z9) showing continuum excess in the UV.
%Interestingly, the UV excess in these targets seems limited in a narrower wavelength window compared to the typical small blue bump.
%In this manuscript, we present our analyses on the ``tiny blue bumps'' in the two galaxies.
%We propose a atypical \feii\ complex being the main contributor to the tiny blue bumps and constrain the strengths of the underlying Balmer continua.
%These analyses provide the first peek into the Fe abundance pattern and nebular conditions in AGNs at $z > 9$.

The layout of the manuscript is as follows. 
In Sections~\ref{sec:data_models} and \ref{sec:measure}, we describe the observational data we use \redtxt{as well as the spectral measurements}.
In Section~\ref{sec:method}, we introduce our method to analyze the observed continuum.
In Sections~\ref{sec:balmer_con} to \ref{sec:ems_continuumexcess}, we compare the observed UV continuum excess with nebular emission models as well as previously observed broad-line AGN spectra to identify potential contributions from the Balmer continuum, \feii\ emission, and other nebular emission. 
We discuss the physical interpretations for the shape and variation of the UV continuum excess as well as the temperature structures in the ionized clouds in Section~\ref{sec:discussion}.
We also comment on potential emission from other atoms and ions in this wavelength regime.
In Section~\ref{sec:conclude}, we present our conclusions and outlooks.

%\section{Observational data and main spectral features}
\section{Observations and data reduction}
\label{sec:data_models}

\begin{table}
        \centering
        \setlength{\tabcolsep}{4pt}
        \caption{Summary of GN-z11 observations used in this work.}
        \label{tab:obser}
        \begin{tabular}{l c c c c c}
            \hline
                \hline
                NIRSpec ID & Date & P.A.$^{\rm a}$ & Exp. Time & $\delta x^{\rm b}$   & $\delta y$ \\
                ---        & ---  & (deg)    & (hr)       &  (arcsec)      & (arcsec)   \\
                \hline
                3991       & 07-Feb-2023 &  19.57 &  6.9 & 0.025 & 0.021      \\
                5591       & 05-May-2023 & 289.05 &  2.6 & 0.002 & 0.041      \\
                \hline
        \end{tabular}
        
    \raggedright{$^{\rm a}$ NIRSpec position angle. $^{\rm b}$ In the MSA reference, with
    $x$ increasing from Q1 to Q3 and $y$ increasing from Q1 to Q2 \citep{ferruit2022}. The uncertainties
    on these values is due to a combination of uncertainties in the source position, astrometry, and
    the precision of the target acquisition (in both observations, the maximum allowed of eight target
    acquisition sources was used).}
\end{table}

\begin{table}
        \centering
        \caption{Fluxes measured from the Prism spectrum of GN-z11.}
        \label{tab:fluxes}
        \begin{tabular}{l c c}
                \hline
                Line & Wavelength [\AA] & Flux [$10^{-20}~{\rm erg~s^{-1}~cm^{-2}}$] \\
                \hline
                N\,{\sc iv}] & 1483.98, 1487.15 & $100\pm 11$  \\
                He\,{\sc ii}+O\,{\sc iii}] & 1641.03-1666.76 & $71\pm 12$  \\
                N\,{\sc iii}] & 1747.44-1752.77 & $87.0\pm 8.2$  \\
                C\,{\sc iii}]+Si\,{\sc iii}] & 1883.34-1909.37 & $113\pm 9$  \\
                $[$O\,{\sc iii}$]$ (tentative) & 2321.66 & $18.2\pm 5.1$  \\
                Mg\,{\sc ii} & 2796.35, 2803.53 & $54.3\pm 4.0$  \\
                O\,{\sc iii} (tentative) & 3133.70 & $8.5\pm 2.7$  \\
                $[$O\,{\sc ii}$]$ & 3727.09, 3729.88 & $92.5\pm 3.9$  \\
                H9 & 3836.48 & $11.8\pm 2.9$ \\
                $[$Ne\,{\sc iii}$]$ & 3869.86 & $98.4\pm 3.8$ \\
                He\,{\sc i}+H8 & 3889.74, 3890.16 & $32.9\pm 3.5$  \\
                $[$Ne\,{\sc iii}$]$+H$\epsilon$ & 3968.59, 3971.20 & $61.5\pm 3.8$ \\
                H$\delta$ & 4102.90 & $49.4\pm 3.5$ \\
                H$\gamma$ & 4341.69 & $93.6\pm 4.8$ \\
                $[$O\,{\sc iii}$]$ & 4364.44 & $21.4\pm 4.4$ \\
                \hline
        \end{tabular}
\end{table}

\begin{figure*}
    \centering\includegraphics[width=0.95\textwidth]{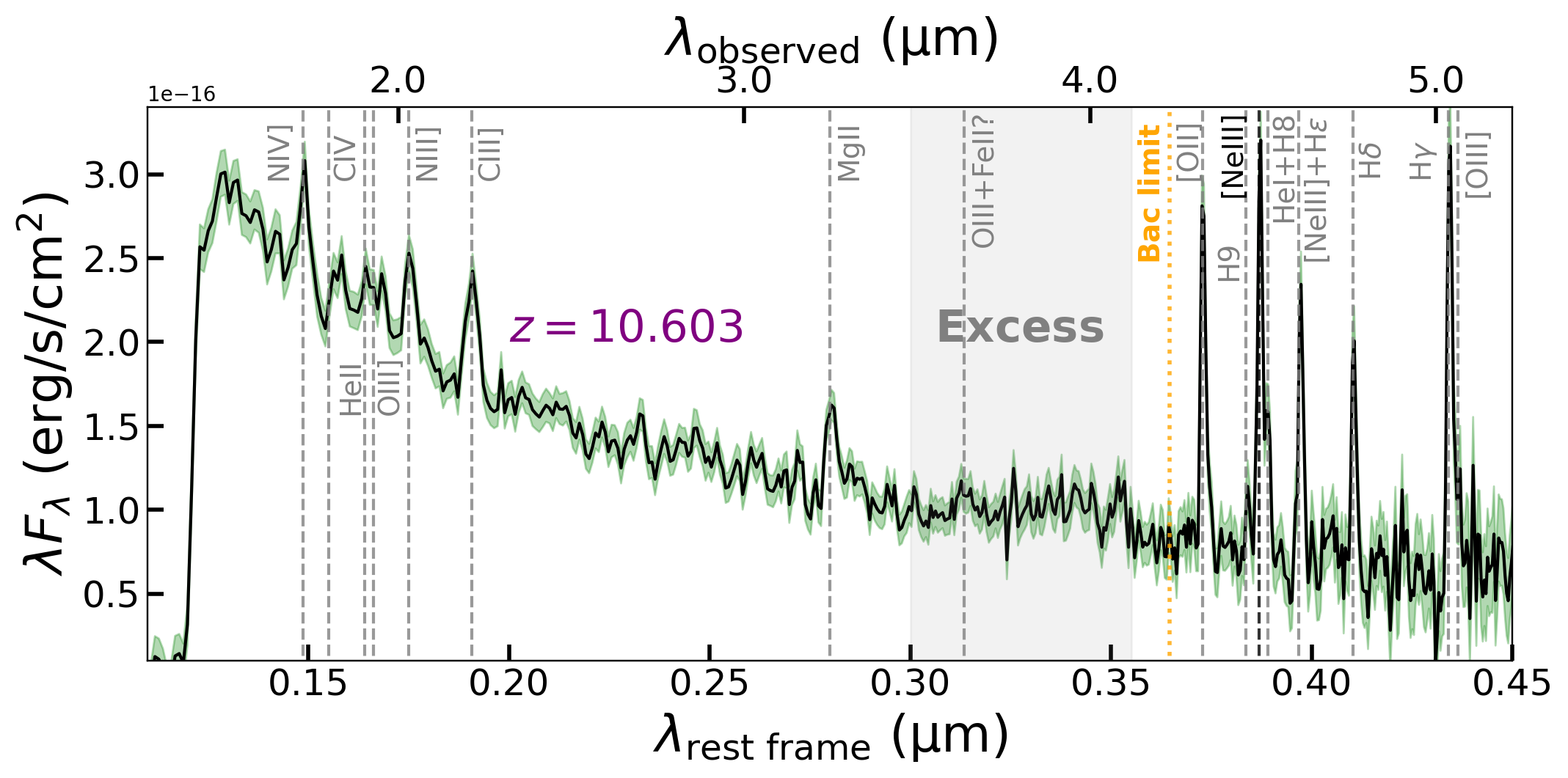}
    \centering\includegraphics[width=0.9\textwidth]{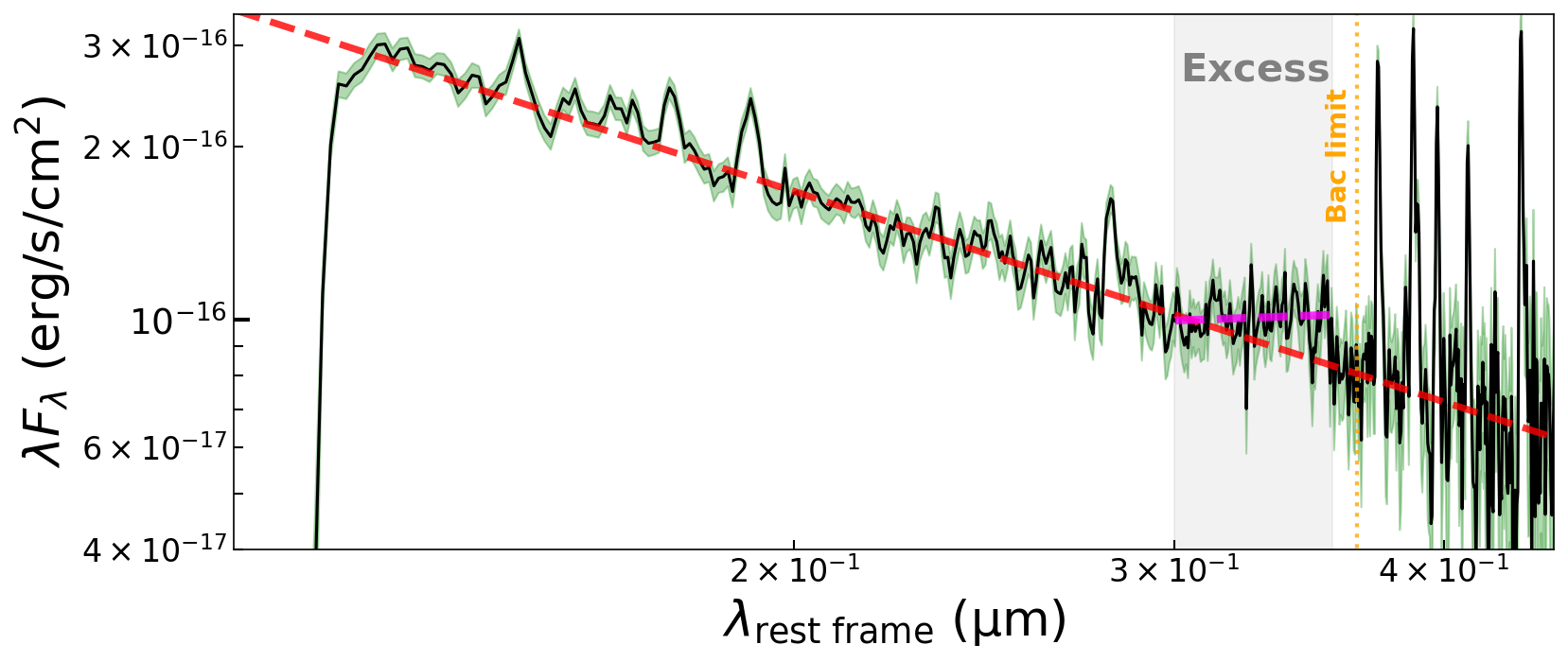}
    
    \caption{\textit{Top}: 
    NIRSpec/Prism spectrum of GN-z11.
    %from the Cycle 2 Program 1181.
    Locations of detected UV and optical emission lines are marked by vertical dashed grey lines.
    Near the location of \civ, there are both a blueshifted absorption trough and a redshifted emission feature, indicating the presence of outflow \redtxt{if the UV continuum of GN-z11 is AGN dominated} \citep{maiolino2023}.
    The grey shaded band indicates the visually identified spectral region showing continuum excess.
    The orange dotted line indicates the location of the Balmer continuum (Bac) limit at 3646 \AA.
    \textit{Bottom}: the same spectrum plotted on a log scale.
    We also plotted two power-law functions that fit the spectrum excluding major emission lines and the continuum excess (red dashed line) and the continuum excess only (magenta dashed line), respectively.
    %NIRSpec/Prism spectrum of GS-z9 from the Cycle 1 Program 1210.
    %The continuum excess, as indicated by the shaded grey region, is less prominent compared to that in GN-z11 but still visible.
    %\redtxt{TBD: 2D spectra?}
    }
    \label{fig:gnz11_prism}
\end{figure*}

%\begin{figure}
%    \includegraphics[width=0.49\textwidth]{Figs/prism_com3.png}
%    \caption{Comparison between the UV humps in the Prism spectra of GN-z11 and GS-z9.
%    The spectra are shown in log scale.
%    The hump region is indicated by the grey shaded band.
%    The orange dotted line marks the location of the Balmer limit at 3646 \AA.
%    For GS-z9, spectra from two Cycle 2 programs at two different epochs are shown, where the spectrum from Program 3215 is obtained at a later epoch ($\sim 1.1$ months later in the rest frame).
%    }
%    \label{fig:hump_com}
%\end{figure}

\begin{figure*}
    \centering\includegraphics[width=1.04\columnwidth]{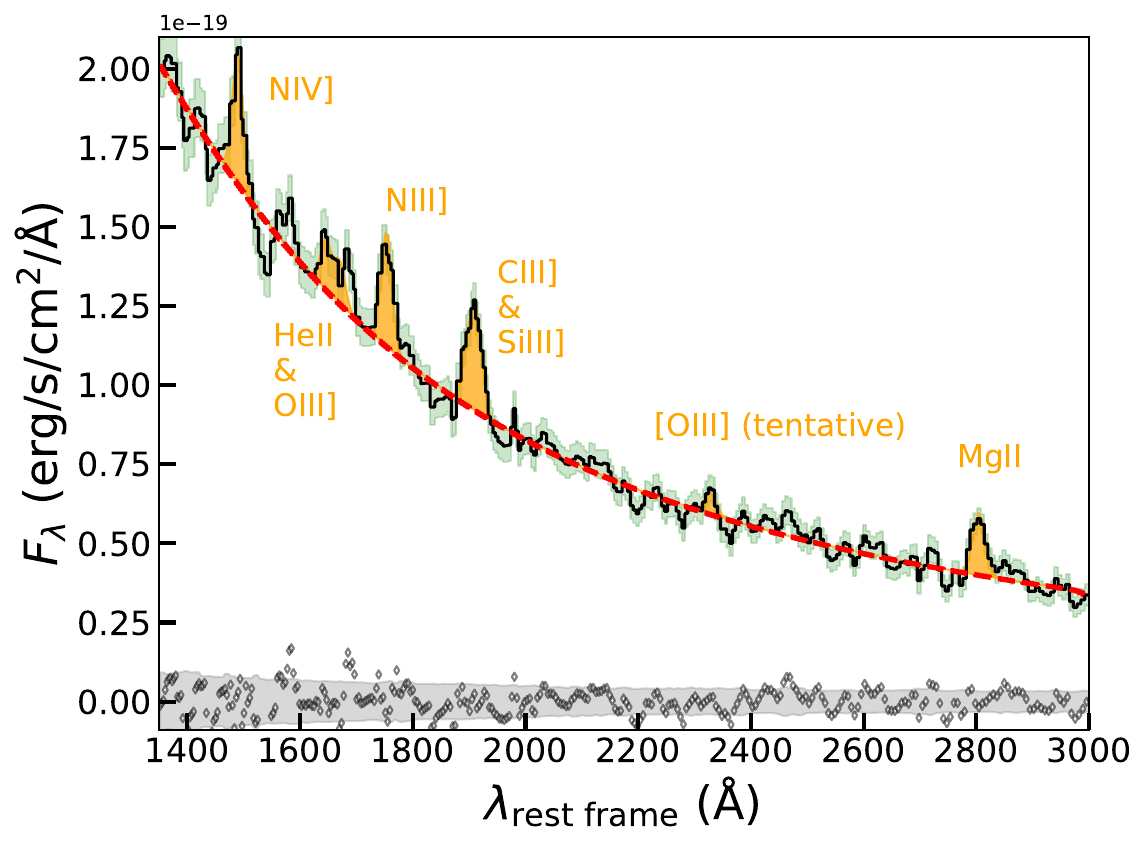}
    \includegraphics[width=0.96\columnwidth]{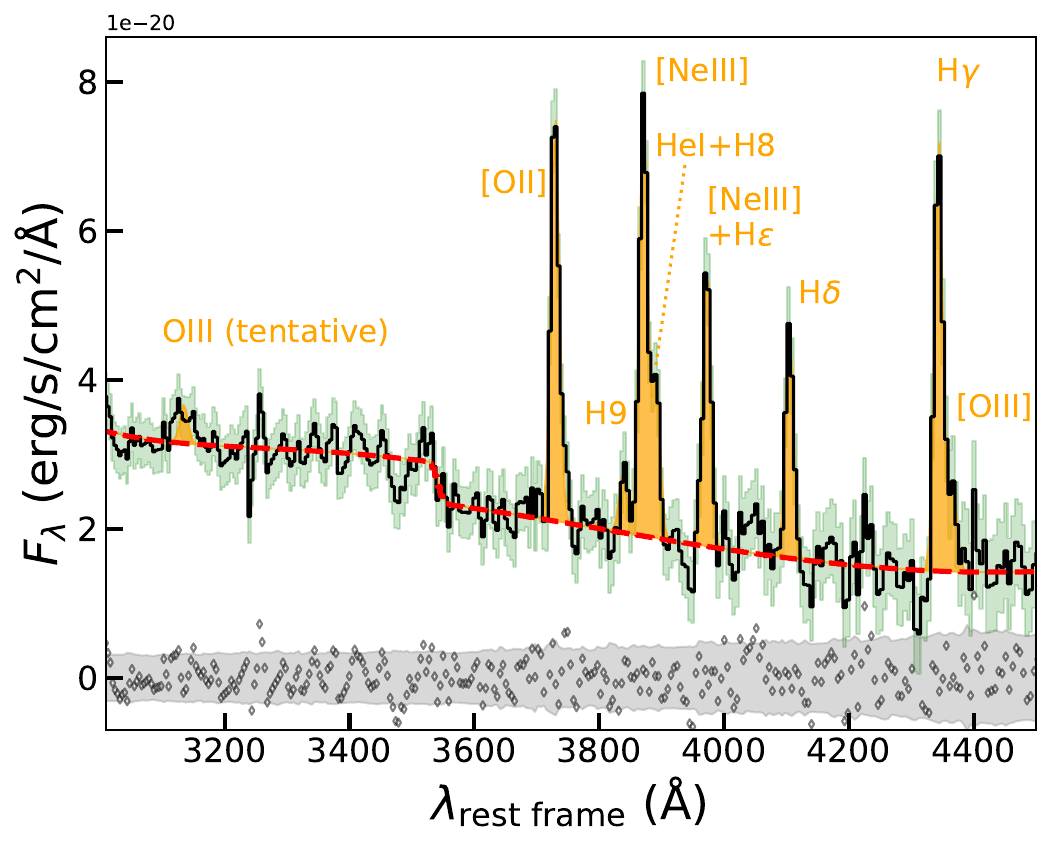}
    \includegraphics[width=0.95\columnwidth]{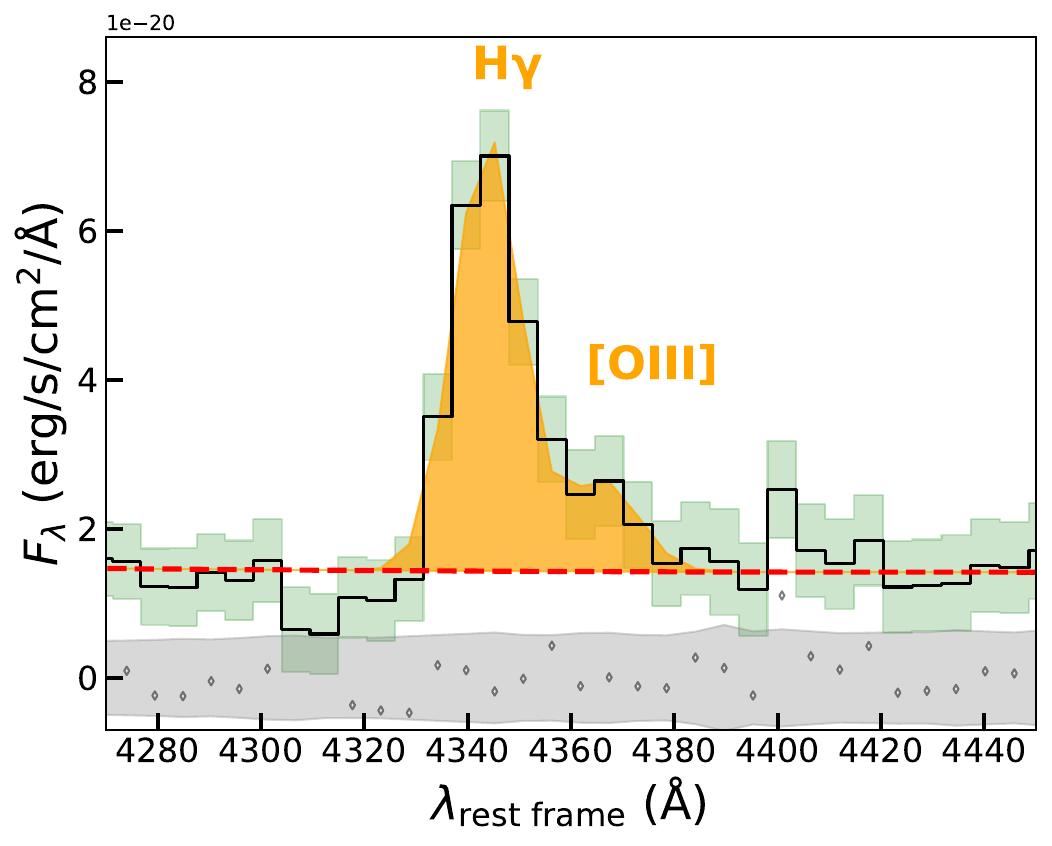}
    \includegraphics[width=0.95\columnwidth]{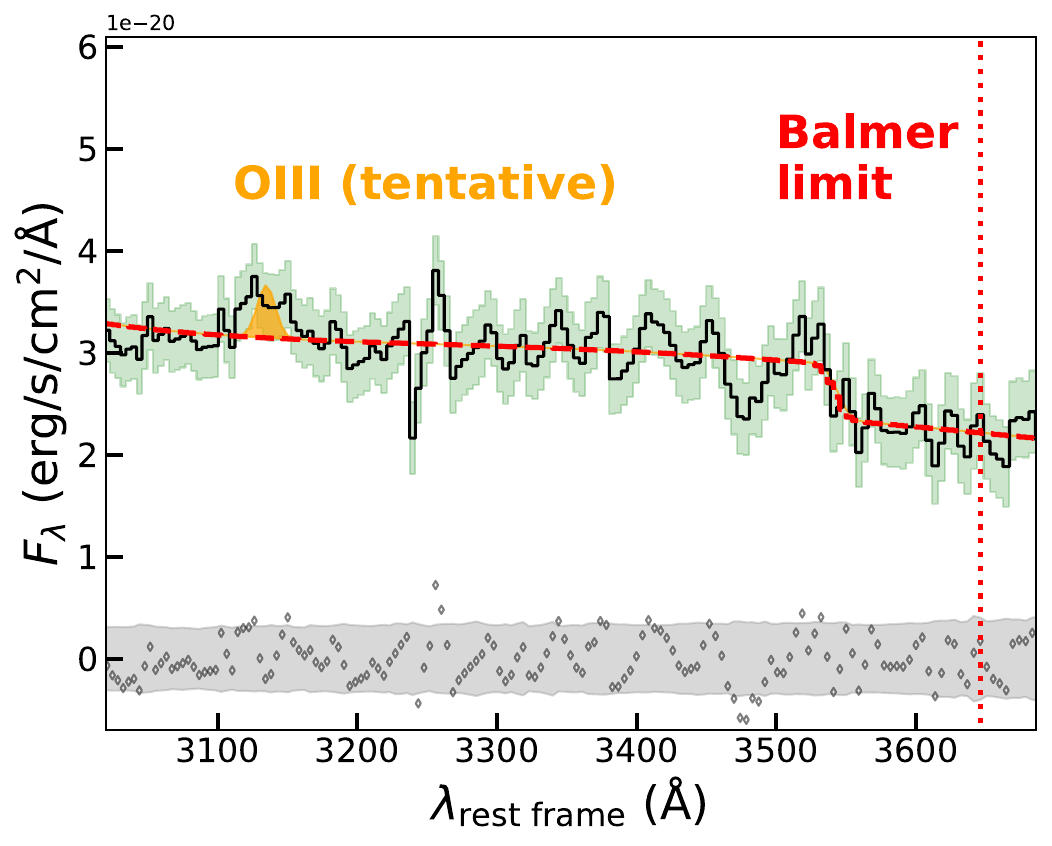}
    
    \caption{Spectral fitting results from \textsc{pPXF}. The solid black line and the green shaded region represent the observed Prism spectrum and spectral uncertainties of GN-z11 logarithmically rebinned by \textsc{pPXF}.
    The dashed line is the best fit continuum model with a functional form of a $4^{\rm th}$-order polynomial. The orange shaded areas correspond to emission line models.
    The gray band represents the $1\sigma$ rebinned uncertainties and the open diamonds are the fitting residuals.
    \textit{Top left:} Fitting results for the spectral range of 1350-3000 \AA in the rest frame of GN-z11.
    \textit{Top right:} Fitting results for the spectral range of 3000-4550 \AA in the rest frame of GN-z11.
    \textit{Top right:} Fitting results around H$\gamma$.
    \textit{Bottom:} Fitting results in the rest-frame near UV. The vertical dotted line indicates the Balmer limit.
    }
    \label{fig:gnz11_ppxffit}
\end{figure*}

In this work we use observational data obtained by JWST/NIRSpec.
\redtxt{We focus on the low-resolution Prism/CLEAR observations of GN-z11 that cover the rest-frame UV-to-optical regime.}
Our sample galaxy, GN-z11, was originally detected by the Hubble Space Telescope (HST) in the CANDELS GOODS North field \citep{grogin2011,koekemoer2011,oesch2016} and was later observed in the JWST Advanced Deep Extragalactic Survey (JADES) under the Cycle 2 Program ID 1181 (P.I.: D. Eisenstein; \citealp{eisenstein2023}).
The spectrum was obtained by combining observations from two different epochs and the detailed descriptions of the observations and data reduction are given by \cite{bunker2023} and \cite{maiolino2023}. In brief, there are two sets of observations obtained in February 2023 and May 2023, respectively.
The February observation consists of four NIRSpec micro-shutter assembly \citep[MSA;][]{jakobsen2022,ferruit2022,boker2023} configurations, with 14 groups per integration, one integration per exposure, and three nods each. Each configuration was repeated two times, for a total of $6.9$ hr exposure time. The May observation consists of three MSA configurations, again with 14 groups per integration, one integration per exposure and three nods, but each configuration was used only one time, for a total of $2.6$ hr exposure time.
\redtxt{The basic information of the two observations is summarized in Table~\ref{tab:obser}.}
%\purpletxt{In Appendix~\ref{appendix:a}, we show the spectra from individual observations.}

\redtxt{The reduction of these observations were carried out using the JADES data reduction pipeline \citep[DRP,][]{deugenio_jadesdr3_2024}, which is the same as the one described in \citet{curtis-lake2023}, \citet{curti2023b}, and \citet{bunker_dr1_2024}. The initial processing of the raw data frames used the STScI pipeline {(version 1.16.1, reference file \texttt{jwst\_1303.pmap})}. At this stage, we remove the detector bias and dark current, identify snowballs, and convert the multiple reads to count rates. As part of this latter step, saturation and jumps due to cosmic rays are identified and removed. At this stage, we continue the reduction using a custom
pipeline developed by the ESA NIRSpec Science Operations Team, whose essential steps are described in \citet{alvesdeoliveira2018} and \citet{ferruit2022}. A full description will be presented in S.~Carniani et al. (in~prep.).
The background is removed in the 2-d detector images, by subtracting from each shutter the average of the other two shutters, exploiting the 3-nod observing pattern.
This procedure takes into account the presence of neighbours/interlopers and of disobedient shutters.
After this step, we possess a set of 2-d spectra, which are
individually flat fielded and corrected for wavelength-dependent losses due to the telescope and instrument optics.
The 2-d spectra are then flux calibrated using the static calibration files, which take into account optical distortions.
The individual 2-d spectra are also corrected for position-dependent slit losses, assuming a point-source geometry.
This latter step is not entirely appropriate for GN-z11, given the presence of an extended component (half-light radius 0.05~arcsec) in addition to the point source \citep{tacchella2023}. Yet the point-source contribution is dominant (0.6~mag brighter), and does display an flux excess at 4~$\mu$m \citep[unlike the extended component;][their Fig.~6]{tacchella2023}.
To take into account the uncertain position of the source inside the micro-shutter (due to a combination of uncertainties in the source centre, in the relative astrometry, and in
the target acquisition), we tested the effect of random offsets in the nominal source position, drawing from a uniform distribution with width 0.05~arcsec; these changes have
negligible effect on localized spectral features (W.~Baker et al. in~prep.).
After the path-loss corrections are applied, the 2-d spectra are rectified.
To avoid excessive correlation, we Nyquist sample the instrument line spread function \citep{jakobsen2022}, which results in a non-monotonic wavelength grid, with the widest spectral pixels around $\lambda = 1.5~\mu$m, where the spectral resolution reaches its minimum. The procedure takes into account the nominal intra-shutter position (Table~\ref{tab:obser}) of each observation for the effect of slit under-filling \citep{ferruit2022} on the wavelength calibration. We note that our correction still leaves wavelength offsets at the level of 0.1 pixel \citep{deugenio_jadesdr3_2024}, but these can only be identified statistically using a large sample, and their sub-pixel magnitude means that they are unlikely to change our results.
%A summary of the observations is provided in Table~\ref{tab:obser}. 
Uncertainties on the surface brightness
(and data-quality flags) are propagated at each step; we use the variance-conserving resampling \citep{dorner2016} to take into account correlated noise between adjacent spectral pixels due to the
finite spectral resolution\footnote{We note that this effect of the correlated noise is not taken into account by other pipelines, which therefore produce NIRSpec spectra which typically have, only apparently, lower noise.}. By comparing the pipeline noise to the local standard deviation of the data (after taking into account spectral variations such as the spectral slope and emission lines), we find that the pipeline noise is typically 
%>>>XIHAN CHECK, GOING BY MEMMORY<<<
15 percent larger than the local standard deviation; this is expected, given that the latter does not account for correlated noise, which reduces the standard deviation of any local sample. 
%>>> XIHAN COULD CHECK BY COMBING THE WINDOW<<<. 
For a direct estimate of the covariance matrix between spectral pixels see \citet{witstok_nature_2024} and P. Jakobsen et al. (in~prep.).
The 1-d spectrum is obtained by extracting from each of the 2-d spectra using a 5-pixel box-car extraction, and then combining the resulting 1-d spectra using inverse variance weighting and 5-$\sigma$ clipping. Note that inverse-variance weighting does not conserve the flux for features dominated by shot noise. However, GN-z11 like most high-redshift sources is dominated by detector noise, so this issue is not a concern.
}

\section{Spectral measurements}
\label{sec:measure}

Figure~\ref{fig:gnz11_prism} shows the 1D NIRSpec/Prism spectrum of GN-z11, where we marked locations of prominent emission lines.
We measured the fluxes of important emission lines including H$\gamma$, H$\delta$, \oiii$\lambda 4363$, \neiii$\lambda 3869$, and \oii$\lambda \lambda 3726,3729$ in the Prism spectrum for the derivation of the electron temperature and oxygen abundance.
To measure the emission lines, we used the the penalized pixel-fitting code \citep[\textsc{pPXF};][]{cappellari2004,cappellari2017} and fitting emission lines as Gaussians \redtxt{convolved with the line-spread function (LSFs) for point sources observed by MSA Prism given by \citet{degraaff2023}.
We limited our fit to the spectral regime of 1350-4550 \AA in the rest frame (1.57-5.68 $\mu$m in the observed frame) to avoid the region near the Ly$\alpha$ damping wing.
}
During the fit, we tied the kinematics of Balmer lines and forbidden lines separately to account for their potentially different origins in the galaxy.
\redtxt{For the continuum, we approximate it as a piecewise power-law with break locations at $\lambda = 3000$ \AA\ and 3550 \AA\ multiplied by a $4^{\rm th}$-order polynomial.
The first break location is to describe the change in the slope shown in Figure~\ref{fig:gnz11_prism}, and the second break location is to describe the jump in the spectrum also shown in Figure~\ref{fig:gnz11_prism}.
We also tried setting the second break location at the Balmer limit of $\lambda = 3646$ \AA, which does not change the line fluxes significantly but results in a larger reduced $\chi ^2$.
The choice of the flexible modeling for the continuum is to measure the line fluxes without assuming any physical origin for the continuum.
The physically motivated fits for the continuum are discussed in Sections~\ref{sec:balmer_con} and \ref{sec:ems_continuumexcess}.
}
%To estimate the flux uncertainties, we manually added noise to the original spectrum according to the values reported by the data reduction pipeline and then refitted emission lines.
\redtxt{To estimate the flux uncertainties, we manually added noise to the original spectrum. For each pixel, we drew the corresponding noise from a Gaussian distribution with a mean of zero and a standard deviation that equals to the uncertainty of the flux density reported by the data reduction pipeline.
We then fitted the spectrum with \textsc{pPXF} to obtain line fluxes.}
The above procedure was repeated 500 times and for each emission line, we took the 68\% confidence interval of the flux distribution and divided it by $2\sqrt{2}$ to obtain the final uncertainty estimate\footnote{The scaling factor of $2\sqrt{2}$ is motivated by the fact that the total noise has been enlarged by a factor of $\sqrt{2}$ after adding noise to the observed spectrum.}.
\redtxt{While \textsc{pPXF} provides formal errors for line fluxes propagated from the input flux density uncertainties, these uncertainties are computed through the logarithmically rebinned spectrum in the wavelength space \citep{cappellari2017} and the bootstrapping method we adopted is generally more robust for estimating uncertainties.
We summarize the fluxes of major emission lines in Table~\ref{tab:fluxes} and spectral fitting results in Figure~\ref{fig:gnz11_ppxffit}.
We emphasize that the continuum model we adopted here has no physical meaning and is only used as a smooth baseline for extracting emission-line fluxes.
}
We compared our flux measurements with those by \cite{bunker2023} and \cite{maiolino2023} and confirmed good consistency.
%We summarized the measurement of major emission lines in Table \redtxt{[TBA]}.
\redtxt{Later in our analyses, we also combine the recent JWST/MIRI MRS measurement of the optical \oiii$\lambda 5007$ flux of GN-z11 reported in \citet{alvarez-marquez_gnz11_2024}, which is $1.36\pm 0.14\times 10^{-17}$ \ergscm.}

\redtxt{Besides the prominent emission lines mentioned above,}
the spectrum of GN-z11 shows a clear excess above the simple power law continuum, roughly between 3000 \AA\ and 3550 \AA\ in the rest frame. We marked the region showing the continuum excess in
Figure~\ref{fig:gnz11_prism}, where the spectrum is plotted both in linear (top panel) and logarithmic (bottom panel) scales.
To demonstrate the significance of the excess, we fit two power-law functions to the continuum excluding major emission lines (with at least $\sim 30$ \AA\ space around each emission line) and the excess region and the excess region only.

The best-fit power-law functions follow $F_{\lambda} \propto \lambda ^{-2.21 \pm0.04}$ for the majority of the continuum, and $F_{\lambda} \propto \lambda ^{-0.86 \pm 0.22}$ for the excess region.
Although the pixel-wise differences between the continuum excess and the power law followed by the majority of the continuum are mostly on the level of $1\sigma$, the cumulative difference over the 550 \AA\ spectral window of the continuum excess is $7.8\sigma$.
\redtxt{In comparison, if we evaluate the continuum excess at $2000-2500$ \AA\ with the same method, the significance at this regime is only $0.2\sigma-0.8\sigma$ depending on whether the excess is calculated over the same power law or over a different power law fitted by excluding the $2000-2500$ \AA\ spectral region.}
%Furthermore, the difference between the two power-law indices is \redtxt{$n\sigma$} based on an MCMC calculation.
%It is clear that the excess region exhibits a flatter continuum.
{To check whether the bump is driven by any artefact in observations, we Jackknifed the Prism spectra using individual measurements and still found significant continuum excess.
We show the Jackknifed spectra in Appendix~\ref{appendix:a}.}
%(TBD: simply show both observations if not enough visits to Jackknife?)}
\begin{comment}
Such flattening of the continuum is expected if a Balmer continuum exists and dominates the part of the continuum close to the Balmer limit.
However, as shown by Figure~\ref{fig:gnz11_prism}, the Balmer limit is $\sim 100$ \AA\ (or 19 pixels) redwards to the observed edge of the continuum excess, hence excluding that the excess is associated with the Balmer continuum.
\feii\ emission, on the other hand, could contribute to the flattening of the observed continuum as well.
However, typical UV \feii\ emission in quasars show larger intensities blueward of 3000 \AA\ rather than redwards to 3000 \AA\ \citep[e.g.,][]{tsuzuki2006}.
Nevertheless, the physical conditions in this early galaxy could be drastically different from the lower redshift quasars given its low-mass and actively accreting black hole as well as its peculiar abundance pattern.
Therefore, we speculate its \feii\ emission could also have different shapes compared to those typically seen lower redshift AGNs.
We note that besides the Balmer continuum and the \feii\ complex, the spectral window between 3000 \AA\ and 3550 \AA\  usually 
%a ``nebular desert'' 
lacks strong nebular emission in both observations and simulations.

%The above points form the whole puzzle we attempted to address.
    
\end{comment}

In the next section, we outline our method to fit different continuum components in the Prism spectrum of GN-z11 and evaluate the goodness of fits based on different assumptions.
\redtxt{We examined different explanations for the presence of the continuum excess, including a Balmer continuum and an \feii\ pesudo-continuum.}
%We then exclude more quantitatively that the bump is associated with a Balmer continuum, hence deriving constraints for the BLR temperature, and then propose the interpretation in terms of Fe complex.

\section{Methodology}
\label{sec:method}

%\subsection{Broken power laws}

%\redtxt{[Subsection newly added]}

%\subsection{Physical models}

\begin{table*}
        \centering
        \caption{Input parameters for \textsc{Cloudy} \feii\ models}
        \label{tab:models}
        \begin{tabular}{l c}
            \hline
                \hline
                Parameter & Values \\
                \hline
                %\multicolumn{2}{|c|}{JY20 model \citep[fiducial model in this work,][]{ji2020b}} \\
                %\hline
                SED & $T_{\rm BB} = 10^6$ K, \redtxt{$\alpha _{\rm ox} = -1.8$}, $\alpha _{\rm uv} = -0.5$, $\alpha _{\rm x} = -1.0$\\
                log(U) & $-3.5$, $-3.0$, $-2.5$, $-2.0$, $-1.5$, $-1.0$\\
                %$\rm [O/H] \equiv \log (O/H)-\log(O/H)_{\odot}$ & $-1$\\
                %$\rm [Fe/O] \equiv \log (Fe/O)-\log(Fe/O)_{\odot}$ & $-1$, $-0.5$, 0.0, 0.5, 1.0, 1.5\\
                $\log (Z/Z_{\odot})$ & $-2$, $-1.5$, $-1$, $-0.5$, 0.0, 0.5\\
                Solar abundance set & \citet{grevesse2010} solar abundance set\\
                $\log (n{\rm_H} / {\rm cm}^{-3})$ & 8, 9, 10, 11, 12\\
                Geometry & Plane-parallel; $\log (N{\rm_H} / {\rm cm}^{-2}) = 21-25$\\
                Turbulence $v_{\rm turb}/(\rm km/s)$ & 
%                100, 600, 1100, 1600, 2100
                {100, 200, 300, 400, 500, 600}\\
                Dust & No dust and no depletion onto dust\\
                \hline
        \end{tabular}
\end{table*}

%In this section we present our analysis results for GN-z11 and GS-z9.
%We start our analyses by fitting the NIRSpec/Prism spectra of these targets using a multi-component model that includes 
To characterize the observed UV continuum of GN-z11, we considered three major components summarized below.
\redtxt{The fits we performed can be broadly divided into two categories under two different assumptions, which are an AGN-dominated UV continuum and an SF-dominated UV continuum, respectively.}
\begin{enumerate}
    \item 
    %A power law continuum that follows $F_{\lambda} = \alpha \lambda ^{\beta}$ \redtxt{[tbd]}, 
    \redtxt{A smooth continuum}
    corresponding to the thermal emission from the AGN accretion disk (the ``big blue bump'') or the UV continuum of young stellar populations. 
    \redtxt{We considered two physically motivated choices for modeling the continuum.
    One choice is AGN spectral energy distributions (SEDs) parametrized by black hole masses and Eddington ratios computed by \citet{pezzulli_agn_2017}, where we used the SED with a black hole mass of $M_{\rm BH}=10^6~M_\odot$ and an Eddington ratio of $\lambda _{\rm Edd} \equiv L_{\rm bol}/L_{\rm Edd} =1$ matching the parameter space estimated by \citet{maiolino2023} for GN-z11.
    We note that varying the black hole parameters in a range of $M_{\rm BH}=10^{6-7}~M_\odot$ and $\lambda _{\rm Edd} =1-10$ does not significantly change the continuum fitting results (Fabian et al. in prep.).
    The other choice is the stellar SED fitted by \citet{tacchella2023} for the extended component in GN-z11 using photometric data from NIRCam, which is likely associated with the stellar population in GN-z11 and contributes roughly 1/3 of the total fluxes in the rest-frame UV \citep{tacchella2023,maiolino2023}.
    This part of the extended emission is likely outside the MSA (and not accounted for by the slit losses).
    This choice is to test the scenario where the point source component that dominates the MSA observations also has a stellar origin in the UV, although we note that this is in disagreement with the emission-line analyses by \citet{maiolino2023}.
    For the AGN and SF SED models above, we set their normalization as a free parameter during the fits.
    In Appendix~\ref{appendix:Balmer_optthick}, we also discuss fits assuming a power-law continuum that does not distinguish between the AGN and SF scenarios.
    }
    %\redtxt{broken pl; AGN SED; SF SED}
    \item A \redtxt{nebular continuum including a hydrogen free-bound continuum (with a} Balmer continuum truncated at 3646 \AA) \redtxt{and a two-photon continuum. The nebular continuum is computed with \pyneb\ \citep{luridiana2015,morisset_2020} assuming a range of temperatures of $\rm 6\times 10^3~K<T_e<3\times 10^4~K$ and a fixed density of $n_{ e}=100~{\rm cm^{-3}}$.}
    While the UV continua of early galaxies might have strong contributions from two-photon emission \citep[e.g.,][]{cameron_9422_2024}, based on the high electron density inferred from \niii\ and \niv\ \redtxt{lines} \citep[{\color{black} $n_e \gtrsim 10^{9}~{\rm cm^{-3}}$},][]{maiolino2023},  
    \redtxt{the ionized gas producing these UV lines should not have a strong two-photon continuum, which has a critical density of $n_e \sim 10^3~{\rm cm^{-3}}$.}
    Also, \redtxt{the observed} $F_\lambda$ \redtxt{of GN-z11} does not peak at $\sim 1430$ \AA\ 
    %(or at $\sim 1620$ \AA\ for $F_\nu$) 
    as would be the case if a two-photon continuum dominated the UV spectrum \citep{gaskell1980}.
    \redtxt{Regardless, there could still be two-photon emission from low-density regions \citep{byler2017} and we computed it consistently with the HI free-bound continuum under the same nebular conditions.
    In Appendix~\ref{appendix:Balmer_optthick}, we discuss the fits with alternative nebular continuum models.
    }
    %We used \pyneb\ \citep{luridiana2015} to generate two-photon continua by setting the hydrogen density to $n_{\rm H} = 100~{\rm cm^{-3}}$.
    %For consistency, we varied the temperature in accordance with the Balmer-continuum temperature, and we used Balmer continua from \pyneb\ with the same nebular conditions.}
    \redtxt{The dominant part of the HI free-bound continuum is the Balmer continuum, which can be described by the parametric form of \citet{grandi1982}}
    \begin{equation}
        F_{\lambda } = F_{\lambda}^{\rm BE} (\frac{\lambda _{\rm BE}}{\lambda })^2 e^{-\frac{hc}{k_B T_e}(\frac{1}{\lambda} - \frac{1}{\lambda _{\rm BE}})}~{\rm ;}~\lambda \leq \lambda _{\rm BE},
        \label{eq:thin}
    \end{equation}
    where BE means the Balmer continuum edge and $\lambda _{\rm BE} = 3646$ \AA.
    The only free parameter is the electron temperature, $T_e$.
    At fixed $T_e$, the flux at the Balmer edge, $F_{\lambda}^{\rm BE}$, is set by the measured fluxes of Balmer lines.
    In our case, we used the strongest available Balmer line, H$\gamma$,
    \redtxt{to normalize the free-bound continuum at given temperatures.
    We note that in previous studies of local nebulae, the flux of H11 ($\lambda _0 = 3771$ \AA) is often used to normalize the Balmer jump due to their close proximity in the wavelength space hence insensitivity to dust attenuation \citep[e.g.,][]{liu_balte_2001}. From Table~\ref{tab:fluxes}, the flux ratio of $\rm H\gamma /H\delta = 0.53\pm 0.05$ is consistent with the Case B ratio of 0.55 at $T_e\approx 10^4$ K and $n_e\approx 10^3~{\rm cm^{-3}}$ \citep{draine11_book}.
    Therefore, the Balmer emission of GN-z11 indicates little or no dust attenuation, consistent with the previous finding of \citet{bunker2023} based on the observations on February, 2023.
    }
    %and computed a series of \textsc{Cloudy} \citep[c17.03;][]{ferland2017} models to relate the model predicted $\lambda _{\rm BE}F_{\lambda}^{\rm BE}$ to the volume-average $T_e$ over a range of \redtxt{$6\times 10^3~{\rm K} \lesssim T_e \lesssim 3\times 10^4$ K} assuming the optically thin limit for the Balmer continuum.
    %and over a density range from $\rm 10^3~cm^{-3}$ to $\rm 10^{11}~cm^{-3}$. 
    %The parameter space we chose covers the physical conditions from the interstellar medium (ISM) to BLRs.
    %\redtxt{Nebular continuum from \pyneb.}
    We discuss an alternative choice of a partially optically thick Balmer continuum as initially suggested by \citet{grandi1982} for fitting quasar spectra \redtxt{specifically} in Appendix~\ref{appendix:Balmer_optthick}.
    %\redtxt{Second, we used the Balmer continuum computed by \pyneb\ \citep{morisset_2020} at a density of $n_{\rm H} = 100~{\rm cm^{-3}}$ normalized to the flux of H$\gamma$.
    %We performed separate fits with and without the two-photon continuum self-consistently computed by \pyneb.}
    We note that our main conclusions do not depend on the choice of the Balmer continuum model.
    \item An \feii\ complex in the UV.
    We computed \textsc{Cloudy} \redtxt{\citep[v17.03,][]{ferland2017,sarkar2021}} \feii\ models assuming a range of parameters listed in Table~\ref{tab:models}, which are typical parameters used to model nebular emission from BLRs \citep[e.g.,][]{verner1999,baldwin2004,ferland2009,sarkar2021}.
    \redtxt{We describe the model assumptions in detail in Section~\ref{sec:feii}.}
    \redtxt{In brief}, we used a standard AGN SED with a big blue bump temperature of $T_{BB} = 10^{6}$ K, a UV-to-X-ray slope of \redtxt{$\alpha _{\rm ox} = -1.8$}, a UV slope of $\alpha _{\rm uv}=-0.5$, and an X-ray slope of $\alpha _{\rm x} = -1.0$. The full SED is described by the following function
    \begin{equation}
        F_\nu = \nu ^{\alpha _{\rm uv}}{\rm exp}(-h\nu /kT_{BB}){\rm exp}(0.01~{\rm Ryd}/h\nu) + a\nu ^{\alpha _{\rm x}},
        \label{eq:sed}
    \end{equation}
    where $a$ is automatically adjusted by \textsc{Cloudy} to produce the preset $\alpha _{\rm ox}$.
    We varied the ionization parameter, $U$, defined as the dimensionless ratio between the flux of the hydrogen ionizing photons and the hydrogen density, from \redtxt{$10^{-3.5}$ to $10^{-1.0}$}.
    We varied the overall chemical abundances from $1\%$ of solar values to three times solar values, assuming the \cite{grevesse2010} solar abundance set\footnote{Although the abundance pattern in GN-z11 is likely different from the solar abundance pattern \citep{cameron2023}, we found the most important elemental abundance in our models is the relative abundance of Fe.}.
    The hydrogen density covers a range from $\rm 10^{8}~cm^{-3}$ to $\rm 10^{12}~cm^{-3}$, and the hydrogen column density ranges from $\rm 10^{21}~cm^{-2}$ to $\rm 10^{25}~cm^{-2}$.
    These values are typical of the BLR clouds or their vicinity, and are those that can provide significant \feii\ emission.
    According to \cite{baldwin2004}, a microturbulence velocity $> 100$ km/s is usually needed to describe the \feii\ emission in AGNs. 
    The main role of microturbulence is to increase the escape fraction of line photons and increase the strength of continuum pumping, which can increase the equivalent width (EW) of the \feii\ emission and smooth the \feii\ pseudo-continuum.
    We considered a broad range for the miroturbulence velocity from 100 km/s to {600} km/s possible for the BLR in GN-z11 \citep{bottorff2000,maiolino2023}.
    Finally, we assume the \feii\ emitting region to lie within the dust sublimation radius and thus has no dust.
    %Such a high big blue bump temperature is suitable to describe small accretion disks as expected in the low-mass black hole in GN-z11.
    %We discuss the effects of different model parameters in Section~\ref{sec:feii}.
    %Besides theoretical models, we also compared the excess with the small blue bumps observed in the spectra of lower-redshift AGNs.
    %the empirical template comes from \cite{mejia-restrepo2016}, which is a modified version of the \cite{tsuzuki2006} template that includes \feii\ emission between 3500 \AA\ and 3646 \AA.
\end{enumerate}

%For the \textsc{Cloudy} \feii\ models, we used a standard AGN SED with a big blue bump temperature of $10^{6}$ K, a UV-to-X-ray slope $\alpha _{\rm ox}$ of $-1.4$, a UV slope $\alpha _{\rm uv}$ of $-0.5$, and an X-ray slope $\alpha _{\rm x}$ of $-1.0$.
%Such a high big blue bump temperature is suitable to describe small accretion disks as expected in the low-mass black holes in our sample galaxies.

To evaluate the effect of different components, we performed two sets of fits. In the first set, we only considered a \redtxt{smooth} component \redtxt{(i.e., AGN or SF SED models)} and \redtxt{nebular} continuum. In the second set, we added the \feii\ emission.
%When fitting the continua, we exclude spectral regions with major emission lines.
\redtxt{During these fits, we used the emission-line subtracted continuum based on our emission-line models shown in Figure~\ref{fig:gnz11_ppxffit}.
The continuum and pesudo-continuum models were first convolved by the LSF from \citet{degraaff2023} and then resampled to the wavelength grid of the Prism spectrum of GN-z11.
}
We note that the unresolved higher order Balmer lines \redtxt{that are not subtracted ($n>9$)} would reach an intensity similar to the Balmer edge at 3646 \AA, making the Balmer jump hardly visible at 3646 \AA\ \citep{wills1985}.
\redtxt{Therefore, the apparent location of the Balmer jump might be slightly \textit{redshifted} due to the blending of lines with the Balmer edge, but not blueshifted \citep[see e.g.,][]{cameron_9422_2024,tacchella2024}.}

\section{Constraints on the Balmer continuum}
\label{sec:balmer_con}

\begin{comment}
    
\begin{figure}
    \centering\includegraphics[width=0.47\textwidth]{Figs/rebin5_original_fit0.png}
    
    \caption{Best-fit model of the observed continuum of GN-z11 assuming a power-law function combined with an optically thin Balmer continuum (magenta).
    We only used the black part of the spectrum during the fit while excluding the yellow part due to the presence of prominent emission lines.
    The grey diamonds represent the residuals. The grey shaded region shows the $1\sigma$ uncertainty of the spectrum.
    The black diamonds and the black shaded region are the residuals obtained by rebinning every 5 pixels and the similarly rebinned uncertainty, respectively.
    The Balmer jump as shown by the model is offset from the jump in the continuum excess and the model does not reproduce the flattening of the continuum in the continuum excess region.
    }
    \label{fig:gnz11_bcfit}
\end{figure}

\end{comment}

\begin{figure*}
    \centering
    
    %\includegraphics[width=0.33\textwidth]{Figs/lamfree_fit0.png}
    %\centering\includegraphics[width=0.33\textwidth]{Figs/lamfree_fit1.png}
    %\centering\includegraphics[width=0.33\textwidth]{Figs/lamfree_fit2.png}

    \includegraphics[width=\columnwidth]{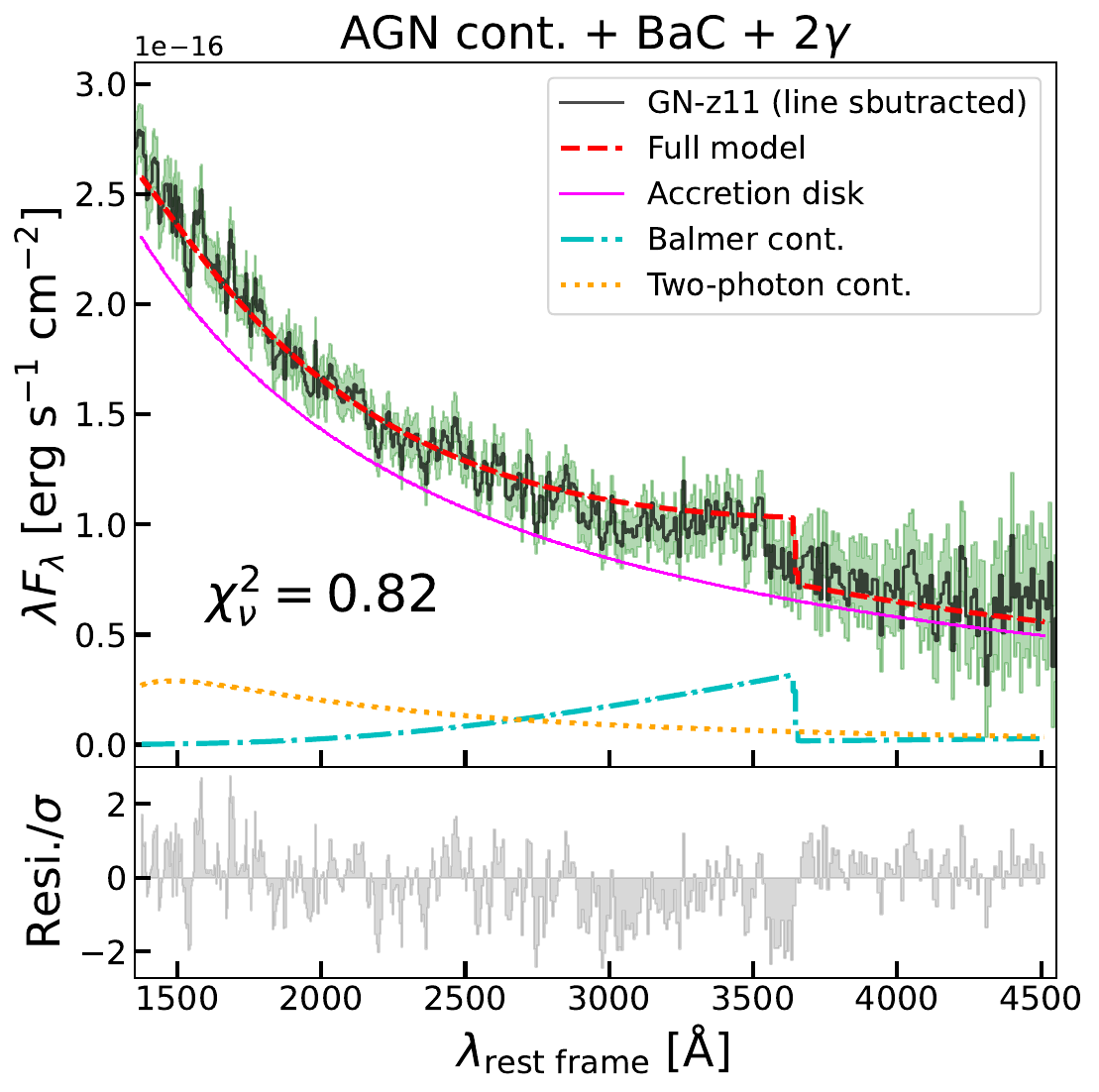}
    \includegraphics[width=\columnwidth]{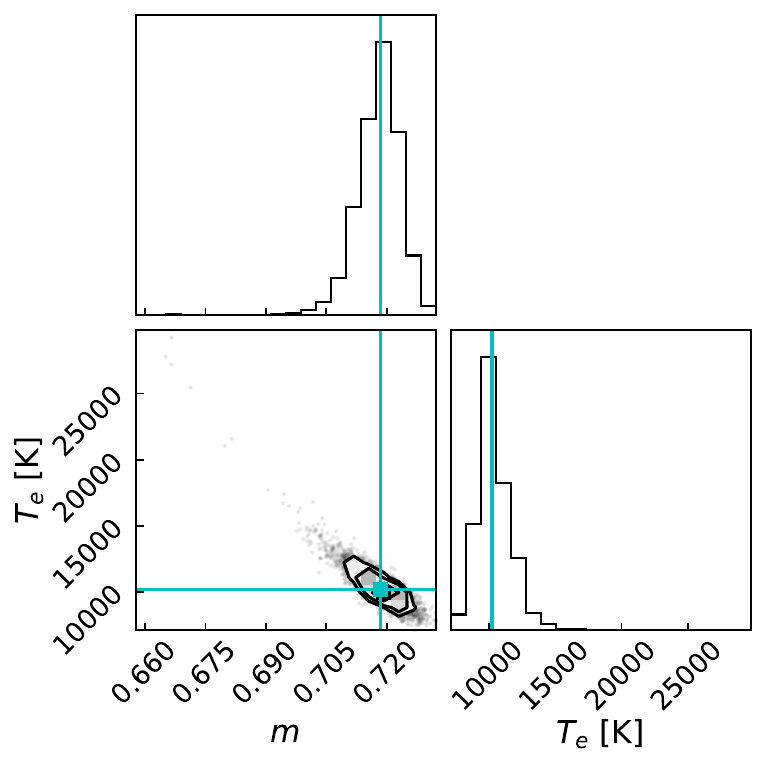}

    \includegraphics[width=\columnwidth]{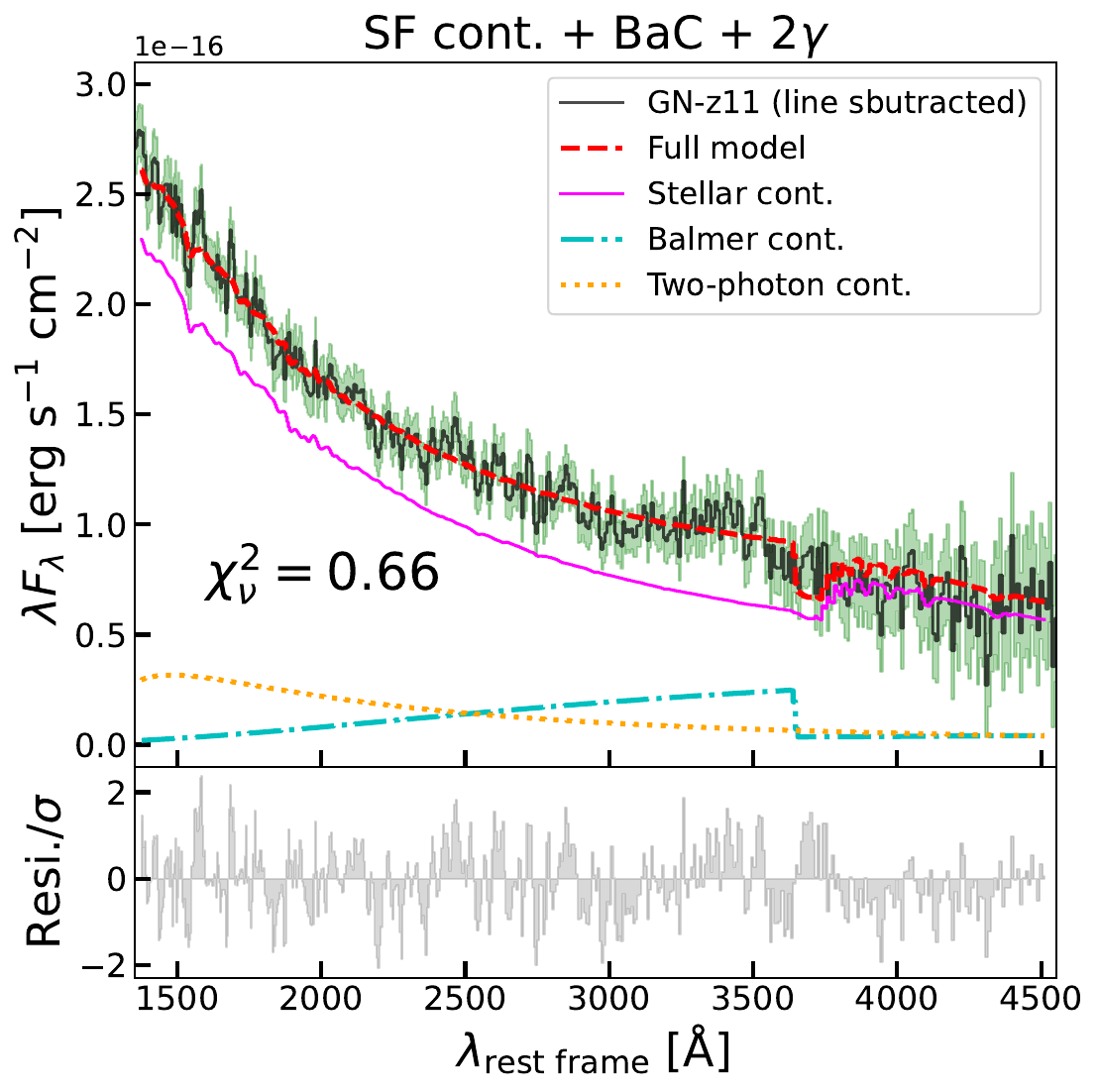}
    \includegraphics[width=\columnwidth]{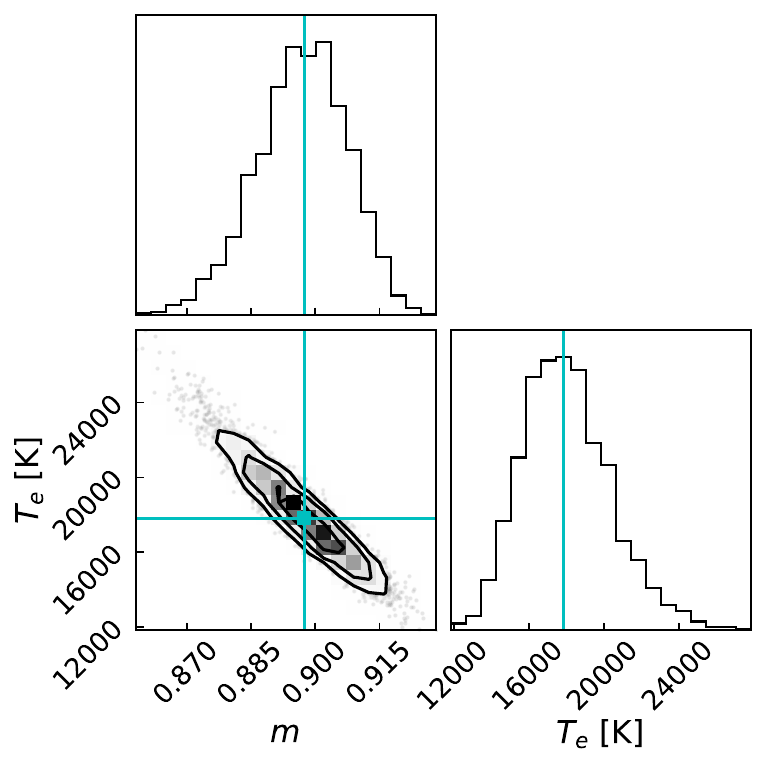}
    
    \caption{\textit{Left columns:} best-fit continuum models for the line-subtracted Prism spectrum (from \textsc{pPXF}) of GN-z11.
    The corresponding reduced $\chi ^2$ are also shown.
    The fitting residuals normalized by the flux uncertainties are shown in the bottom.
    \textit{Right columns:} posterior distributions of model parameters, including the electron temperature of the nebular continuum, $T_e$, and the normalization of the AGN or SF continuum model, $m$.
    The AGN and SF models are first normalized to the Prism flux of GN-z11 at $\lambda = 4260$ \AA\ in the rest frame and then multiplied by $m$. The cyan lines mark the median of the model parameters, which were taken as the best-fit parameters.
    Regardless of the origin of the observed continuum, a significant excess in the near UV (in these cases dominated by the Balmer continuum) is preferred by both fits.
    }
    \label{fig:gnz11_bcfit}
\end{figure*}

\begin{figure*}
    \centering
    
    %\includegraphics[width=0.33\textwidth]{Figs/lamfree_fit0.png}
    %\centering\includegraphics[width=0.33\textwidth]{Figs/lamfree_fit1.png}
    %\centering\includegraphics[width=0.33\textwidth]{Figs/lamfree_fit2.png}

    \includegraphics[width=\columnwidth]{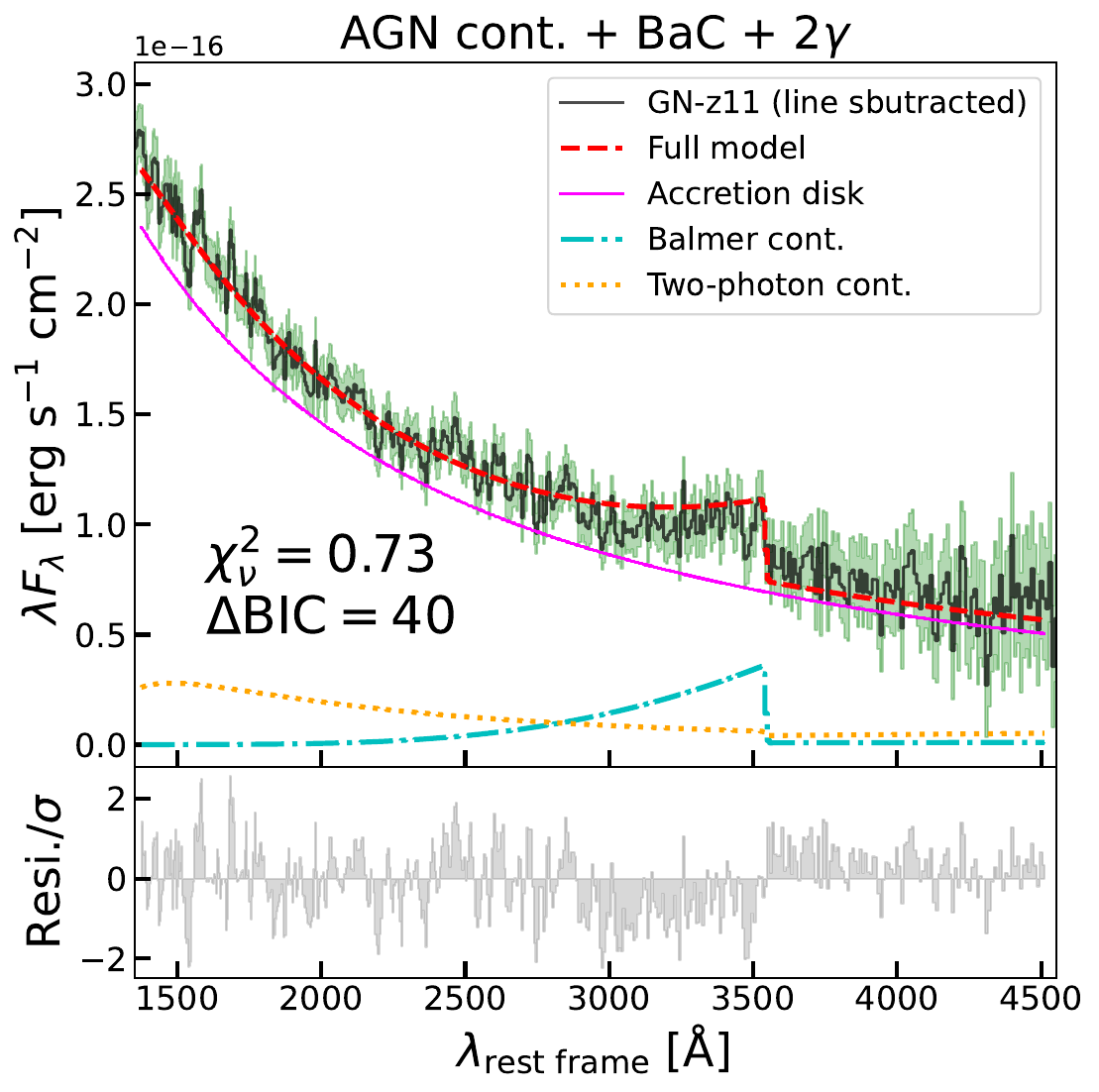}
    \includegraphics[width=\columnwidth]{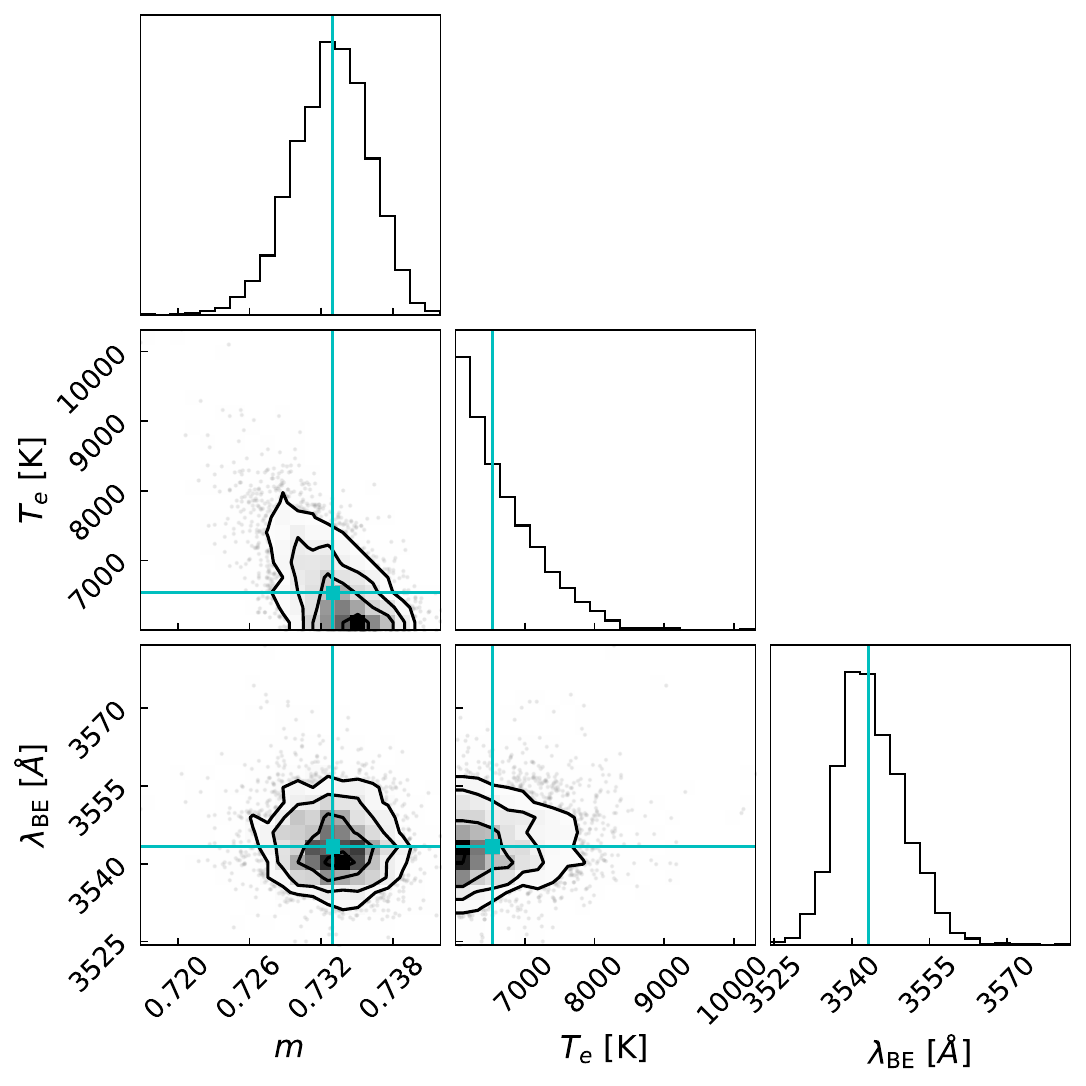}

    \includegraphics[width=\columnwidth]{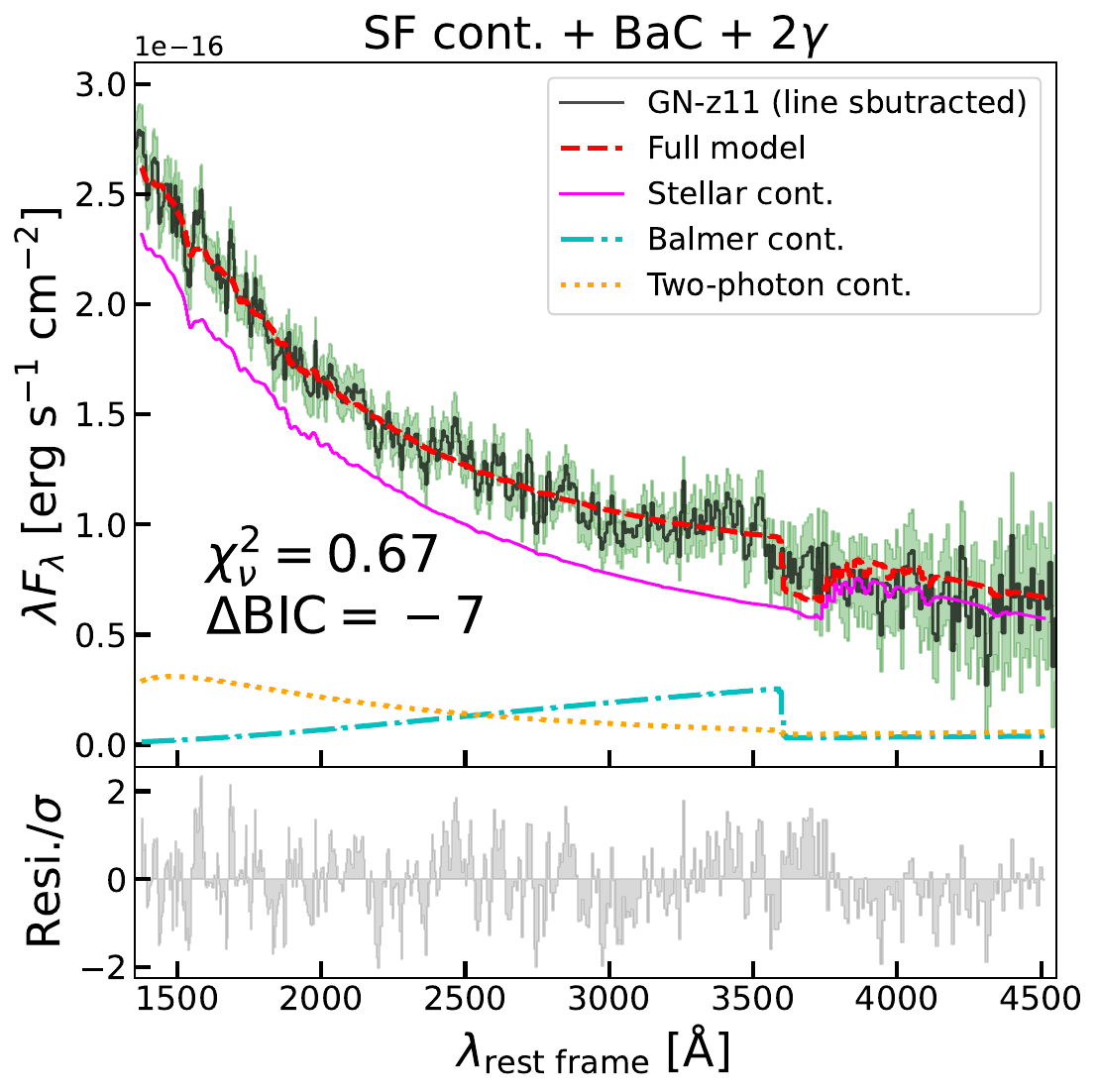}
    \includegraphics[width=\columnwidth]{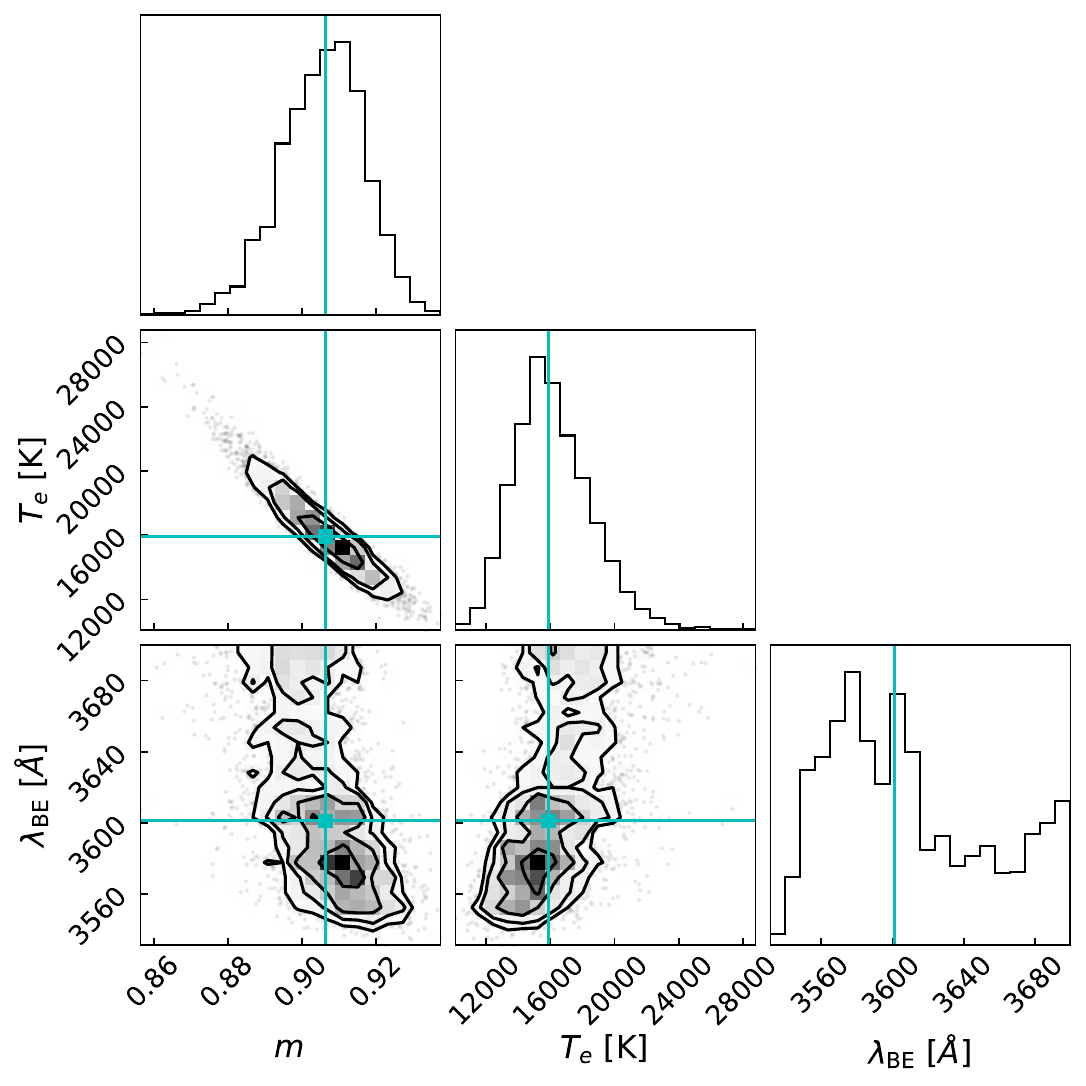}
    
    \caption{Same as Figure~\ref{fig:gnz11_bcfit} but making the location of the Balmer continuum edge, $\lambda _{\rm BE}$, another free parameter.
    In the left columns, besides the best-fit models under the assumptions of an AGN- or SF-dominated continuum, we also show the change in the Bayesian Inference Criterion (BIC) after adding $\lambda _{\rm BE}$ as a free parameter.
    We consider the new fit is significantly better than the old fit only when \rrtxt{$\rm \Delta BIC > 10$}.
    Clearly, in the AGN-continuum scenario, the fit is significantly improved and a blueshifted Balmer edge is preferred.
    In contrast, in the SF-continuum scenario, the shift in the Balmer edge is not significant due to the presence of a small Balmer break.
    %Best-fit model for the observed continuum of GN-z11 assuming a power-law function combined with an optically thin Balmer continuum with a free Balmer continuum edge. The continuum part of the observed spectrum is plotted in black and the part with emission lines is plotted in yellow. The dashed green line is the best-fit continuum model, with the shaded green region indicating the $1\sigma$ uncertainty of the location of the Balmer edge, $\lambda _{\rm BE}$. The dotted orange line marks the location of the Balmer limit at 3646 \AA. The grey diamonds and the grey shaded region represent the residuals of the fit and the $1\sigma$ uncertainty of the spectrum, respectively. From left to right, the contribution of the continuum excess region to the total $\chi^2$  is re-weighted by a factor of 1, 1.5, and 2, respectively. Overall the blueshifted Balmer continuum models produce smaller residuals compared to the one that has a fixed Balmer edge, but the required blueshift is too large to be physical.
    }
    \label{fig:lamfree_bcfit}
\end{figure*}

\begin{table}
        \centering
        \caption{Best-fit model parameters for the Prism spectrum of GN-z11 under different assumptions. Among the parameters listed, $m$ is the normalization of the continuum, $T_e$ is the electron temperature of the nebular continuum, $\lambda _{\rm BE}$ is the wavelength of the Balmer continuum edge.
        For reference, the temperature measured from emission lines is $T_e=1.36\pm 0.13\times 10^4$ K and the theoretical Balmer edge is at $\lambda _{\rm BE}=3646$ \AA.
        }
        \label{tab:best_fit}
        \begin{tabular}{l c c c c c}
                \hline
                \multicolumn{6}{|c|}{AGN ($M_{\rm BH}=10^6~M_\odot$, $\lambda _{\rm Edd}=1$) + nebular continuum; $\chi^2_{\nu} = 0.82$} \\
                \hline
                Parameter & $m$ & & $T_e$ [$10^4$ K] & & \\
                Value & $0.719_{-0.006}^{+0.004}$ & & $1.02^{+0.14}_{-0.09}$ & & \\
                \hline
                \multicolumn{6}{|c|}{AGN ($M_{\rm BH}=10^6~M_\odot$, $\lambda _{\rm Edd}=1$) + nebular continuum} \\
                \multicolumn{6}{|c|}{(free Balmer continuum edge); $\chi^2_{\nu} = 0.73$} \\
                \hline
                Parameter & $m$ & & $T_e$ [$10^4$ K] & & $\lambda _{\rm BE}$ [\AA] \\
                Value & $0.733_{-0.003}^{+0.003}$ & & $0.65^{+0.07}_{-0.04}$ & & $3543^{+7}_{-5}$ \\
                \hline
                \multicolumn{6}{|c|}{SF + nebular continuum; $\chi^2_{\nu} = 0.66$} \\
                \hline
                Parameter & $m$ & & $T_e$ [$10^4$ K] & & \\
                Value & $0.90_{-0.01}^{+0.01}$ & & $1.78^{+0.25}_{-0.21}$ & & \\
                \hline
                \multicolumn{6}{|c|}{SF + nebular continuum} \\
                \multicolumn{6}{|c|}{(free Balmer continuum edge); $\chi^2_{\nu} = 0.67$} \\
                \hline
                Parameter & $m$ & & $T_e$ [$10^4$ K] & & $\lambda _{\rm BE}$ [\AA] \\
                Value & $0.91_{-0.01}^{+0.01}$ & & $1.60^{+0.26}_{-0.20}$ & & $3601^{+67}_{-36}$ \\
                \hline
        \end{tabular}
\end{table}

\begin{comment}

\begin{figure}
    \centering\includegraphics[width=0.48\textwidth]{Figs/masked_bace_fit_zoom.png}
    
    \caption{
    Best-fit model for the observed continuum of GN-z11 assuming a power-law function combined with an optically thin Balmer continuum.
    The continuum part of the spectrum is plotted in black and the part with emission lines is plotted in yellow.
    Different from Figure~\ref{fig:gnz11_bcfit}, the continuum region between 3550 \AA\ and 3646 \AA\ is masked during the fit.
    The solid magenta line is the best-fit continuum model.
    The dotted orange line marks the location of the Balmer limit at 3646 \AA.
    The grey diamonds and the grey shaded region represent the residuals of the fit and the $1\sigma$ uncertainty of the spectrum, respectively.
    The black diamonds and the black shaded region are the residuals obtained by rebinning every 5 pixels and the similarly rebinned uncertainty, respectively.
    }
    \label{fig:lammask_bcfit}
\end{figure}
    
\end{comment}

\begin{figure*}
    \centering\includegraphics[width=0.47\textwidth]{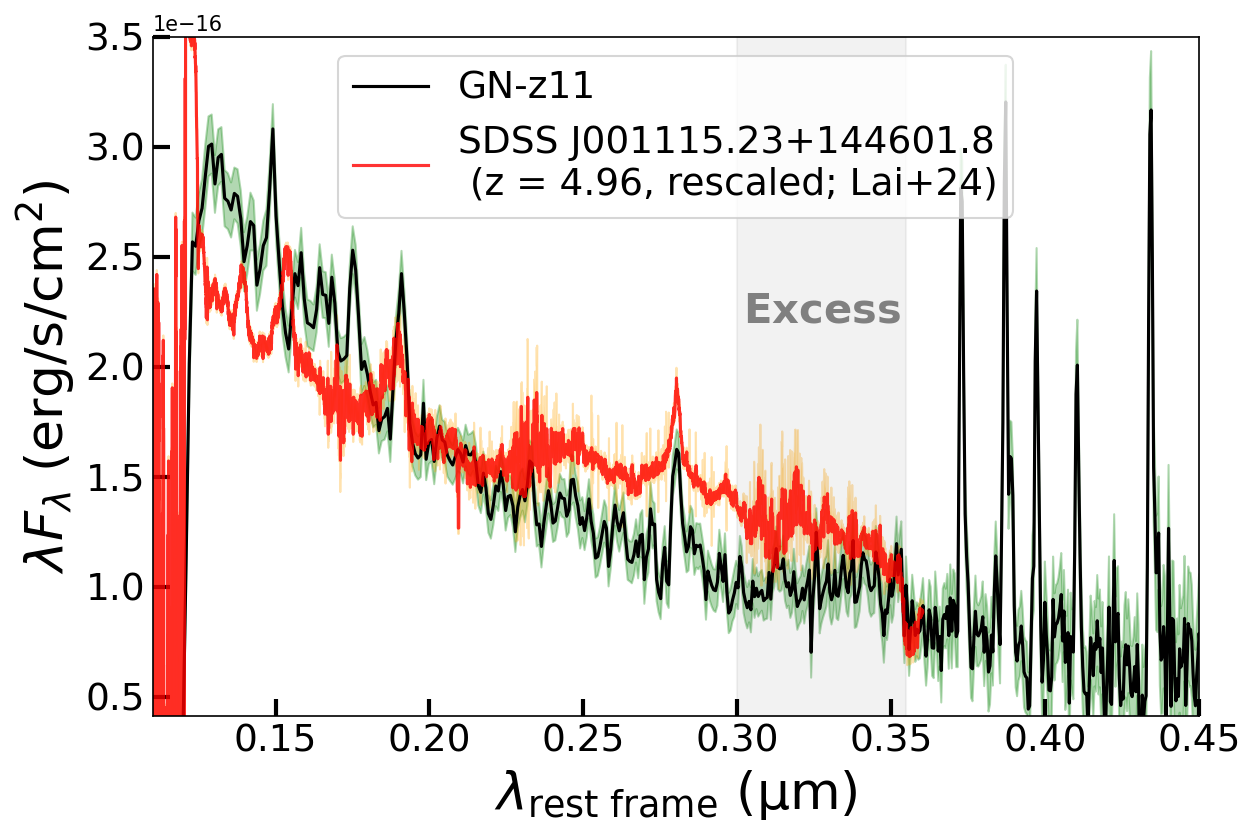}
    \centering\includegraphics[width=0.47\textwidth]{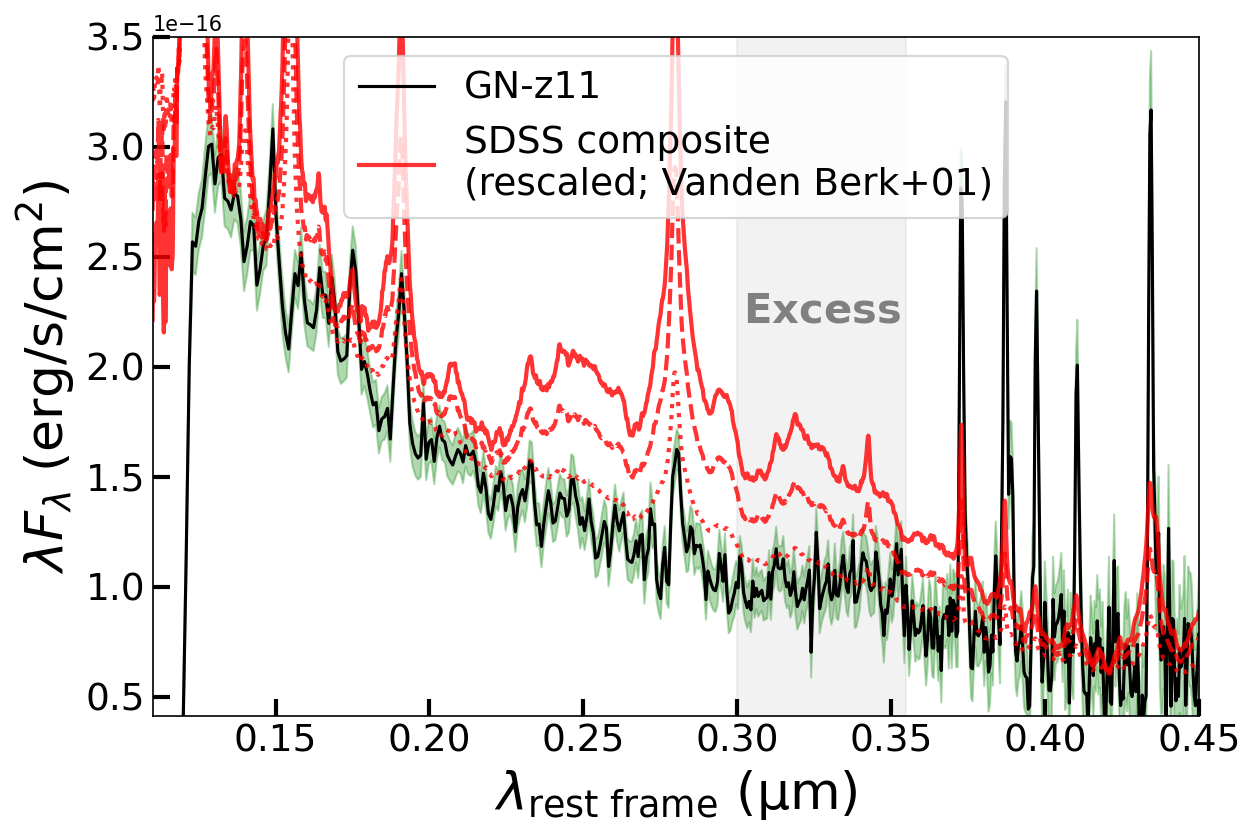}
    \caption{\textit{Left}:
    comparison between the rest-frame spectrum of GN-z11 with that of a quasar at $z = 4.96$, SDSS J001115.23+144601.8, from the XQz5 catalog of \citet{lai2024}. 
    The flux of the quasar spectrum is rescaled to compare with the GN-z11 spectrum.
    The green shaded region and the orange shaded region represent the $1\sigma$ uncertainties for the two spectra, respectively.
    The continuum excess region in GN-z11 is marked by the grey shaded region.
    \textit{Right}:
    comparison between the rest-frame spectrum of GN-z11 with that of a modified composite spectrum constructed from 2200 quasars in SDSS by \citet{vandenberk2001}.
    To better compare the nebular emission, the power-law continuum of the composite spectrum has been replaced with the power-law we fitted for GN-z11.
    The dotted, dashed, and solid red lines correspond to different \redtxt{constant} scaling factors of \redtxt{1, 2, and 3 for} the \redtxt{normalized} nebular emission, \redtxt{respectively}.
    The grey shaded region marks the continuum excess identified in GN-z11.}
    \label{fig:quasar_com}
\end{figure*}

To fit the continuum spectrum, we performed a Markov Chain Monte Carlo (MCMC) calculation using the \textsc{Python emcee} package \citep{emcee}. We used the following likelihood function
\begin{equation}
    \log L = \sum _{\lambda} -0.5 (F_{\lambda}^{\rm cont.} + F_{\lambda }^{\rm neb.} - F_{\lambda})^2/\sigma _{\lambda}^2.
\end{equation}
In the equation above, $F_{\lambda}^{\rm cont.}$ is \redtxt{the continuum of the AGN accretion disk or stellar populations}, $F_{\lambda }^{\rm neb.}$ is \redtxt{the nebular continuum}, $F_{\lambda}$ is the observed flux, and $\sigma _{\lambda}$ is the uncertainty of the observed flux.
\redtxt{We set flat priors for all model parameters and ran a 5000-step chain for each trial.
For each chain, we removed the first 150 steps and resampled the chain with an interval of 50 steps, which is larger than the autocorrelation times of all parameters.
The posterior distributions of the parameters were then extracted from the resampled chains.
}

\subsection{Optically thin Balmer continuum case}

\subsubsection{\color{black} Balmer jump temperature}

In the first fit, we assumed an optically thin Balmer continuum. During the fit, there are \redtxt{two} free parameters: the normalization of the \redtxt{AGN or SF continuum model}, $m$, and the electron temperature \redtxt{for the nebular continuum}, $T_e$.
\redtxt{We note that the AGN and SF models are first normalized to the observed Prism flux of GN-z11 at $\lambda = 4260$ \AA\ avoiding major emission lines in the rest frame, and then multiplied by $m$.}
We assumed flat priors for these parameters.
%and the normalization of the Balmer continuum, $F_{\lambda}^{\rm BE}$.
%\redtxt{Although $F_{\lambda}^{\rm BE}$ is in principle determined by the electron temperature and our \textsc{Cloudy} models, we set both of them as free parameters in an iterative way.}
%We began each iteration by fixing $T_e$ while making $F_{\lambda}^{\rm BE}$ a free parameter.
%After finding the best-fit parameters, we compared $F_{\lambda}^{\rm BE}$ with the value expected from \textsc{Cloudy} models. If the difference is greater than \redtxt{10\%}, we increased $T_e$ by 100 K and started a new iteration.
%We adopted a lower limit of 5000 K for $T_e$ and an upper limit of 30,000 K.
%We found the fit generally did not converge and reached the upper temperature limit.
%Overall, as the temperature increases, the agreement between the fit and the theoretical predictions from \textsc{Cloudy} becomes better.

We plotted our best-fit model in Figure~\ref{fig:gnz11_bcfit}, and list the model parameters as well as the corresponding $1\sigma$ uncertainties estimated from the posterior distributions in Table~\ref{tab:best_fit}.
%The best-fit temperature is $T_e = 1.69 ^{+0.29}_{-0.23}\times 10^4~{\rm K}$.
%These fits include the two-photon continuum as well as the HI free-bound continuum. For fits without the two-photon continuum, we refer the readers to Appendix~TBA, which shows the fits are not significantly improved.
Both fits prefer excess fluxes in the near UV, which, in these cases, are interpreted as mainly the Balmer continuum (dashed-dotted cyan curves).
The AGN continuum produces a larger reduced $\chi ^2$ compared to the SF continuum, but we caution that the hyper parameters that produce these continua are not taken into account during the fits.
\rrtxt{In addition, all reduced $\chi ^2$ calculated are less than 1, implying the flux uncertainties might be overestimated \citep[][]{Bevington_Robinson_2003}.}
The AGN continuum fit also prefers a stronger Balmer jump.
Based on the AGN continuum model, the best-fit temperature is $T_e = 1.02^{+0.14}_{-0.09}\times 10^4$ K.
The stellar continuum model, on the other hand, gives a higher best-fit temperature of $T_e = 1.78^{+0.25}_{-0.21}\times 10^4$ K associated with a weaker jump.
The difference between the two fits is mainly driven by the small Balmer break in the stellar continuum model that allows for a larger fraction of the UV light to be contributed by nebular continuum.

We compared the Balmer continuum temperature, $T_e({\rm H^{+}})$, with the temperature estimated from the auroral line \oiii$\lambda 4363$, $T_e({\rm O^{2+}})$.
\redtxt{We compared three different methods to estimate the temperature.
The first two methods used MSA spectra from JADES.
}
%Since the JADES spectra of GN-z11 do not cover \oiii$\lambda \lambda 4959, 5007$, we used \redtxt{two alternative methods.}
\redtxt{First, we used} the empirical relations between \neiii$\lambda 3869$/\oii$\lambda \lambda 3726,3729$ and \oiii$\lambda 5007$/\oii$\lambda \lambda 3726,3729$ fitted by \cite{witstok2021} to estimate the flux of \oiii$\lambda 5007$\footnote{This approach is motivated by the observations that Ne/O remains roughly constant over a wide range of metallicities \citep{berg2020} and $\rm Ne^+$ and $\rm O^+$ have similar ionization potentials.}. The empirical relations were calibrated using star-forming (SF) galaxies and Seyferts from the Sloan Digital Sky Survey \citep[SDSS,][]{york2000}.
We thereby predicted the strengths of \oiii$\lambda 5007$ assuming an SF origin and a Seyfert origin separately.
Then we estimated \oiii$\lambda 4363$/\oiii$\lambda 5007$, \rrtxt{and} we used \textsc{PyNeb} to compute the temperature at the low density limit as the lines we used mainly come from forbidden transitions of $\rm O^{2+}$.
The estimated temperatures are $T_e({\rm O^{2+}}) = 1.28^{+0.10}_{-0.12}\times 10^4$ K for an SF origin and $T_e({\rm O^{2+}}) = 1.33^{+0.11}_{-0.13}\times 10^4$ K for an AGN origin, which are consistent with the estimation of \cite{cameron2023}.
\redtxt{Second, we used the flux ratio of \oiiip]$\lambda \lambda 1661,1666$/\oiii$\lambda 4363$ to infer the electron temperature.
In the Prism spectrum, the blend \heii$\lambda 1640$+\oiiip]$\lambda \lambda 1661,1666$ is marginally resolved at a spectral resolution of roughly $18$ \AA\ at the central wavelength of the blend.
If we assume \heii\ and \oiiip] have the same kinematics, our fit gives $F_{\rm OIII]}=3.35\pm 0.81\times10^{-19}$ \ergscm.
Combining this value with the flux of \oiii$\lambda 4363$ (Table~\ref{tab:fluxes}), at the low-density limit ($n_e<10^4~{\rm cm^{-3}}$), we obtained $T_e({\rm O^{2+}}) = 1.25^{+0.30}_{-0.16}\times 10^4$ K with \textsc{PyNeb}, consistent with the values derived from the first method.
We note that \citet{bunker2023} reported detection of resolved \heii\ and \oiiip] in the (central 3-pixel extraction) medium resolution ($R\sim 1000$) spectrum with $F_{\rm HeII}=4.2\pm 1.3\times10^{-19}$ \ergscm and $F_{\rm OIII]}=4.3\pm 1.5\times10^{-19}$ \ergscm.
If we took the flux ratio of \heii\ and \oiiip] given by \citet{bunker2023} to deblend \heii+\oiiip] and combined it with the \oiii$\lambda 4363$ flux we measured, we obtained $T_e({\rm O^{2+}}) = 1.29^{+0.26}_{-0.23}\times 10^4$ K, again consistent with the values using other methods.
\rrtxt{We also} note that there is a detection of a line at $\lambda \approx 2321$ \AA in the rest frame, which might be matched with the auroral line \oiii$\lambda 2320$.
Both \oiii$\lambda 4363$ and \oiii$\lambda 2320$ are produced from the same upper level and they should have a fixed flux ratio of roughly 4:1.
However, in Table~\ref{tab:fluxes} one can see the flux ratio is roughly 1:1 and is inconsistent with the expectation, meaning the identification of \oiii$\lambda 2320$ might not be exact.
Therefore, we did not use \oiii$\lambda 2320$ to infer the electron temperature.
Finally, we note that \citet{alvarez-marquez_gnz11_2024} recently reported \oiii$\lambda 5007$ measured in the JWST/MIRI MRS spectrum for GN-z11.
They derived a temperature of $T_{e}=1.40\pm 0.21\times 10^4$ K by combining their measured flux for \oiii$\lambda 5007$ and the flux of \oiii$\lambda 4363$ reported by \citet{bunker2023}.
For consistency, we rederived the temperature with \pyneb\ using our measured \oiii$\lambda 4363$ flux and obtained $T_{e}=1.36\pm 0.13\times 10^4$ K, where the small difference comes from the slightly lower \oiii$\lambda 4363$ flux we measured.
Overall the above temperatures are consistent within $1\sigma$ and we took $T_{e}=1.36\pm 0.13\times 10^4$ K as our fiducial value.
}

%Overall, $T_e({\rm H^{+}})$ is higher than $T_e({\rm O^{2+}})$ by $\sim 1.5\sigma$, in contrast to the usual expectation that $T_e({\rm O^{2+}})>T_e({\rm H^{+}})$ due to the higher ionization potential of $\rm O^{+}$ compared to H \citep{peimbert1967}.
\redtxt{In the case of an AGN-dominated continuum, $T_e({\rm H^{+}})$ is lower than $T_e({\rm O^{2+}})$ by $\sim 1.8\sigma$.
While the difference is insignificant, it has been observed in local \hii\ regions that the recombination line temperatures are systematically lower than the collisionally excited line temperatures, possibly due to temperature inhomogeneities in the ionized gas \citep{peimbert1967,peimbert2017}.
In the case of an SF-dominated continuum, $T_e({\rm H^{+}})$ is higher than $T_e({\rm O^{2+}})$ by $\sim 1.7\sigma$.
A higher recombination line temperature is usually not expected as the recombination lines have higher emissivities at lower temperature regions.
Still, the difference is less than $3\sigma$ and there might be systematic uncertainties associated with the choice of the stellar continuum \citep{tacchella2023}.
Either case, the existence of a nebular continuum indicates the intrinsic UV continuum of GN-z11 is slightly steeper compared to the observed one, as can be seen in Figure~\ref{fig:gnz11_bcfit}.
Also, it appears the actual jump location is slightly blueshifted in the observed spectrum.
}
Indeed, the residuals clearly show a wiggling pattern near the jump.
\redtxt{Next, we investigate this shift in the wavelength space.}
%However, it is not much the temperature difference that is worrying, but the fact that the continuum excess region is not well fitted by the model. Indeed, the residuals clearly show a wiggling pattern near the jump. To make this issue more clear and to highlight its significance,
%{in Figure~\ref{fig:gnz11_bcfit}, the residuals of the fit are rebinned and show clear deviations above and below the rebinned noise level at several locations within the continuum excess region.}

\subsubsection{\color{black} Is the Balmer jump shifted from the theoretical limit?}

%{
%A connected issue, is the fact that the continuum excess does not have a jump at the expected Balmer edge wavelength, but at shorter wavelengths.
To check the significance of the shift, we set the location of the Balmer edge as another free parameter during the fit.
The fitting result is shown in Figure~\ref{fig:lamfree_bcfit}.
\redtxt{Again, we investigated two different scenarios where the UV continuum is dominated by an AGN accretion disk and stellar populations, respectively.}
\redtxt{For the AGN model, the best-fit Balmer edge is at $\lambda _{\rm BE} = 3543^{+7}_{-5}$ \AA, which is significantly blueshifted from the Balmer limit at $3646$ \AA.
However, the best-fit model appears to systematically overpredict the fluxes within the continuum excess region.
The posterior distribution of $T_e$ also hits the lower boundary of 6000 K.
The mismatch might be due to the systematics associated with the synthetic AGN SED, or due to the assumption that the nebular continuum is responsible for the continuum excess.
Compared to the AGN fitting result in Figure~\ref{fig:gnz11_bcfit}, the $\chi^2_{\nu}$ from a free-edge Balmer continuum fit is improved.
In addition, we calculated the Bayesian Information Criterion (BIC) of \citet{liddle_2007} defined as
\begin{equation}
    \rm BIC\equiv \chi^2 + k\,ln\,n,
\end{equation}
where $k$ is the number of free parameters and $n$ is the number of data.
We considered \rrtxt{$\rm \Delta BIC>10$ (i.e., p-value of $\sim 0.01$; \citealp{kass1995bayes})} as a requirement for an improved fit with more free model parameters.
As a result, adding the Balmer continuum edge as a free parameter gives $\rm \Delta BIC\approx 40$ \rrtxt{for the AGN model}, meaning the data do prefer an edge blueshifted from the Balmer limit.
}
\redtxt{The SF fit, on the other hand, does not prefer a significantly blueshifted Balmer edge. As shown in the bottom panels of Figure~\ref{fig:lamfree_bcfit}, the best-fit SF model has $\lambda _{\rm BE} = 3601^{+67}_{-31}$ \AA. Also, compared to the fit with a fixed Balmer edge, we obtained $\rm \Delta BIC\approx -7$, which \rrtxt{again implies that statistically the non-blueshifted Balmer continuum is preferred by the stellar model}.
We summarize the values of the best-fit model parameters from Figures~\ref{fig:gnz11_bcfit} and \ref{fig:lamfree_bcfit} in Table~\ref{tab:best_fit}.
}

\redtxt{As a result, under the assumption of a ``Balmer continuum-like'' continuum excess, whether the edge of the Balmer continuum is shifted from the theoretical location depends on the prior on the nature of the continuum.
If the continuum of GN-z11 is dominated by AGN emission, the continuum excess cannot be reproduced by a typical Balmer continuum as the continuum edge is significantly blueshifted.
If the continuum of GN-z11 is dominated by a stellar continuum with a similar shape inferred from the extended component in its image, the continuum excess does not prefer a blueshifted Balmer edge.
}

\subsection{Other possible scenarios involving the Balmer continuum}

We also considered the possibility of a partially optically thick Balmer continuum. This model includes an additional free parameter, the optical depth at the Balmer edge, $\tau _{\rm BC}$. We discuss this case in Appendix \ref{appendix:Balmer_optthick}. Here we only summarize that the resulting fits (Figure \ref{fig:gnz11_bcfit_optthick}) have a shape very similar to those with the optically thin case, and plagued by the same issues, with $\tau _{\rm BC}$ being degenerate with the normalization of the continuum and temperature.

From the fits above, one can see the key feature that prevented a good fit with a Balmer continuum model \redtxt{for an AGN dominated spectrum} is around the edge of the small blue bump.
We also note that a potential trough centering at $\sim 3480$ \AA\ (or an emission feature at  $\sim 3522$ \AA) and a potential emission feature at $\sim 3133$ \AA\ cannot be described by the simple continuum model.
If we considered a scenario where a Balmer continuum did exist, the question becomes whether to trust a model with a strong jump but somehow blueshifted to around 3550 \AA,
or a model with a noise-limited weak jump where other emission features dominate the continuum excess.

For the strong-jump scenario, one needs a relatively low temperature and a small optical depth at fixed $F_{\rm H\gamma}$.
One might wonder whether a warm cloud in the foreground can absorb part of the Balmer continuum and create the blueshifted jump.
Self-absorption of the Balmer emission is unlikely to produce a blueshifted Balmer jump.
A potential source of absorption for the Balmer continuum is $\rm Fe^+$ \citep{wills1985}. 
However, there is a lack of $\rm Fe^+$ lines between 3550 \AA\ and 3646 \AA. 
Also, one expects emission from $\rm Fe^+$ rather than absorption in a typical BLR.
Alternatively, one might speculate a blueshifted Balmer continuum in a warm outflowing cloud along the line of sight.
There are two difficulties with this explanation.
First, an ionized cloud with an outflow velocity of $\sim 8000$ km/s is required, but there are no blueshifted Balmer lines observed.
Second, one expects high order Balmer lines to be blended with the Balmer jump especially given the spectral resolution of Prism\footnote{The spectral resolution near the bump is $R \sim 280$ after taking into account the fact that the source is more compact with respect to the width of the shutter \citep{degraaff2023}.}. Thus, there should be a smoothly declining edge rather than a sharp truncation even if the Balmer continuum is blueshifted.

\subsection{The weak Balmer continuum scenario}

For the weak-jump scenario, the Balmer continuum becomes a secondary component and we need an alternative explanation for the continuum excess, which will be discussed in the next sections.
If we simply exclude the spectral region between 3000 \AA\ and 3550 \AA\ (which we will associated with other emission features but keep the region between 3550 \AA\ and 3646 \AA\ to constrain the strength of the Balmer jump), 
\redtxt{in the AGN case, the fit (not shown) is significantly worse when the nebular continuum is included ($\chi^2_{\nu}=0.80$) compared to when the nebular continuum is not included ($\chi^2_{\nu}=0.66$), implying that a suppressed nebular continuum is preferred.
}
\redtxt{In the SF case, on the other hand, the fit (not shown) is significantly better when there is a nebular continuum ($\chi^2_{\nu}=0.66$ with the nebular continuum versus $\chi^2_{\nu}=1.0$ without the nebular continuum).
However, the temperature of the nebular continuum is driven towards $T_e \sim 3\times 10^4$ K, corresponding to a strongly suppressed Balmer jump.
}
%then the solution is driven towards $T_e \sim 3\times 10^4$ K, corresponding indeed to a strongly suppressed Balmer continuum.
%This implies the temperature of the Balmer-line emitting cloud is $\gtrsim 30,000$ K for both the optically thin case and the partially optically thick case.
%Such a high temperature averaged over $\rm H^+$ can be achieved in a metal-poor cloud ionized by an AGN with an overall abundances being $\sim 1/10$ the solar values and a high ionization parameter, $\log U \sim -1$, which is consistent with a BLR or a NLR in the early Universe.
%Both the BLR and the NLR can have these physical conditions.
%On the other hand, it is difficult for stellar radiation fields to produce such high temperature gas even at low metallicities.
%We further discuss the implications of high-temperature ionized gas in Section~\ref{sec:discussion}.
%\redtxt{Figure TBA shows ranges of temperatures produced by AGN ionization and stellar ionization, respectively, in \textsc{Cloudy} simulations. [TBD: discuss here or in the Discussion instead to not make this section too long.]}
Such a high temperature averaged over $\rm H^+$ is achievable if the ionized gas is either extremely metal-poor and/or the ionizing radiation field is hard enough.
As a result, the temperature constraint under the weak-jump scenario provides another piece of evidence on the properties of the ionizing source in GN-z11.
As we discuss in Section~\ref{sec:discussion}, compared to stellar radiation, ionization from an AGN is easier to achieve this temperature especially given the potentially enriched abundances in GN-z11. We will also discuss that such high temperature is completely inconsistent with the temperature inferred from the forbidden lines (tracing low density gas), in particular from the \oiii$\lambda 4363$ auroral line (which gives \redtxt{$T_e= 1.36\pm 0.13\times 10^4$ K} as discussed above). The only solution in this case is that the bulk of the Balmer emission comes from the high density gas in the BLR, while the forbidden lines come from the narrow-line region (NLR) or the ISM of the host galaxy.

Alternatively, the Balmer continuum can be suppressed at a lower temperature but with a large optical depth.
Both a high ionization parameter and a high density are needed to strongly obscure the Balmer edge.
In fact, our \textsc{Cloudy} models show to produce a similar suppression of the Balmer edge at $T_e \lesssim 20,000$ K to the optically thin case at $T_e \approx 30,000$ K, it is critical to have $n_{\rm} \gtrsim 10^{6}~{\rm cm^{-3}}$, which is atypical for an ISM origin.
Therefore, to produce a suppressed partially optically thick Balmer continuum, one still needs a BLR with AGN ionization.
%\redtxt{[A further thought: non-Case B ratios might become obvious at high densities under certain conditions --> but multiple components of Balmer emission might also exist.]}

\section{Emission line scenarios}
\label{sec:ems_continuumexcess}

\subsection{\oiiip\ Bowen fluorescence}
%, and a high hydrogen density, $n_{\rm H} \sim 10^{11}~{\rm cm^{-3}}$.
With a weak jump, the next question is the identity of the emission that dominates the small blue bump.
It is noteworthy that this continuum excess region is usually lack of strong emission lines.
In this wavelength regime, AGNs can show the coronal line [Ne\,{\sc v}]$\lambda 3426$ and the Bowen florescence line O\,{\sc iii}$\lambda 3133$ \citep{bowen1934}.
There is, however, no detection of [Ne\,{\sc v}]$\lambda 3426$. 
In contrast, there is potential detection of O\,{\sc iii}$\lambda 3133$ at a significance of \redtxt{$3.1\sigma$}.
The resulting flux for this emission feature is \redtxt{$8.5\pm 2.7\times 10^{-20}~{\rm erg~s^{-1}~cm^{-2}}$ (Table~\ref{tab:fluxes})}.

{We note that detection of O\,{\sc iii}$\lambda 3133$ was also reported recently for a JWST observed target at $z = 12.34$, GHZ2/GLASS-z12 \citep{castellano2024}. As we show in Section~\ref{sec:feii}, a redshifted \feii\ bump can also produce the observed emission at 3133 \AA. We further discuss the implication of the emission at 3133 \AA\ in Section~\ref{sec:discussion}.}

However, we note that the potential \oiiip\ fluorescence can only account for one of the features seen in the hump.

%and the potential O\,{\sc iii}$\lambda 3133$ can also be explained by \feii\ emission as we show in Section~\ref{sec:feii}.
% To produce a detectable blend of emission lines in this regime, the most frequently explored source is $\rm Fe^+$.
% %The physical conditions required to produce \feii\ emission are also compatible with a BLR scenario.
% Next, we explore whether the inclusion of \feii\ emission can produce a spectral jump at $\sim 3550$ \AA.
% %As another frequently explored source for the small blue bumps in AGNs, \feii\ emission is able to produce a noticeable pseudo-continuum in the UV.
% %We show how the \feii\ complex might match the continuum excess we observed in the next section.

\subsection{Inconsistency with the Wolf-Rayet scenario}

Previously, it was also suspected that Wolf-Rayet (WR) stars could be present in GN-z11 and contribute to the chemical enrichment through their powerful winds \rrtxt{\citep{cameron2023, senchyna2023,Gunawardhana_2025}}.
\rrtxt{As shown by the recent results from the UV Legacy Library of Young Stars as Essential Standards program \citep[ULLYSES,][]{Bernini-Peron_2024,Roman-Duval_ullyses_2025}, high-ionization P-Cygni profiles (e.g., \civ) caused by stellar winds of metal-poor and massive stars can be strong (i.e., similar to the \civ\ absorption strength seen in Figure~\ref{fig:gnz11_prism}) at $\sim 20\%$ solar metallicities.}
One might wonder whether the emission typically associated with winds from WR stars can produce the continuum excess we see in GN-z11.

It is known that nitrogen sequence WR (WN) stars show strong nitrogen emission lines in the rest-frame UV \citep{abbott1987}.
In the wavelength range of $3000~\AA<\lambda <3550~\AA$, the strongest emission feature in WN stars is usually N\,{\sc iv}$\lambda \lambda 3478,3483$ \citep{garmany1984,conti1989}.
However, at the expected location of this emission line ($\sim 3480 \AA$), there is actually a trough in the continuum excess region in GN-z11.
\redtxt{As shown in Figure~\ref{fig:gnz11_ppxffit},
there is no individual Gaussian-like component within the continuum excess region comparable to the tentative O\,{\sc iii}$\lambda 3133$.
We tried fitting these strongest WN lines with \textsc{pPXF} and found no detection at $>3\sigma$.}
\redtxt{Besides nitrogen lines, broad emission humps around He\,{\sc ii} lines are expected from the winds of WR stars.
Within the spectral range of interest, WR stars can produce emission of \heii$\lambda 3204$ from the winds \citep{stanway2018}.
Again, there is no detection of \heii$\lambda 3204$ from our fit in Figure~\ref{fig:gnz11_ppxffit}.}
Thus, it is unlikely that the small blue bump is caused by WN emission.

In addition, as discussed by \citet{maiolino2023}, the absence of N\,{\sc iv}$\lambda 1718$ in GN-z11, which would be strong in WN stars showing strong \niv$\lambda 1486$, further argues against the presence of WR stars.
\redtxt{Finally, the
medium-resolution rest-frame UV spectrum of GN-z11 combining NIRSpec/MSA and NIRSpec/IFU observations show no detection of any broad \heii$\lambda 1640$ hump (Maiolino et al. in prep.),
which is a typical feature associated with WR winds.
This further rules out the possibility that the UV emission of GN-z11 is dominated by WR stars.
}
\redtxt{Next, we discuss in the context of AGN, whether there is other continuum model that can provide a better fit for the spectrum of GN-z11.}

%{\color{red} Discuss that the some WR galaxies/stars show some hump there, but that this is ruled out by the absence of the hump at $\sim$4686}

\subsection{\feii\ emission}
\label{sec:feii}

\begin{figure}
    \centering\includegraphics[width=\columnwidth]{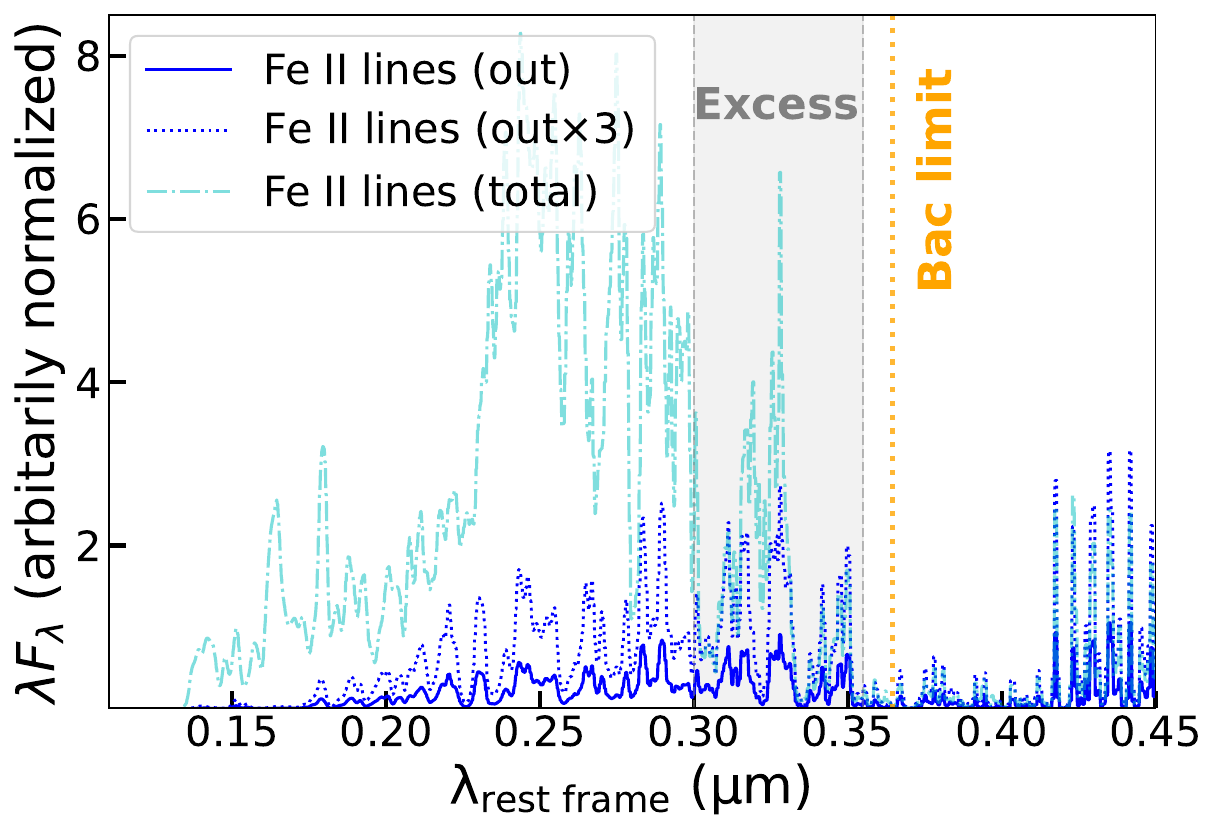}
    
    \caption{Example of an \feii\ pseudo-continuum computed with \textsc{Cloudy}.
    The total \feii\ emission produced by the simulation is plotted as the dashed-dotted cyan line, and the \feii\ emission propagated outwards is plotted as the solid blue line.
    The difference between the two sets of \feii\ emission originates from the wavelength-dependent line optical depth, which leads to a more significant reduction of \feii\ emission at $\lambda _{\rm rest~frame} < 3000~\AA$.
    The dotted blue line was obtained by simply scaling up the flux of the outwards \feii\ emission by a factor of 3.
    The grey shaded region indicates the continuum excess region identified in GN-z11.
    The dotted orange line marks the location of the Balmer limit.
    }
    \label{fig:feii_eg}
\end{figure}

\begin{figure*}
    \centering\includegraphics[width=0.95\textwidth]{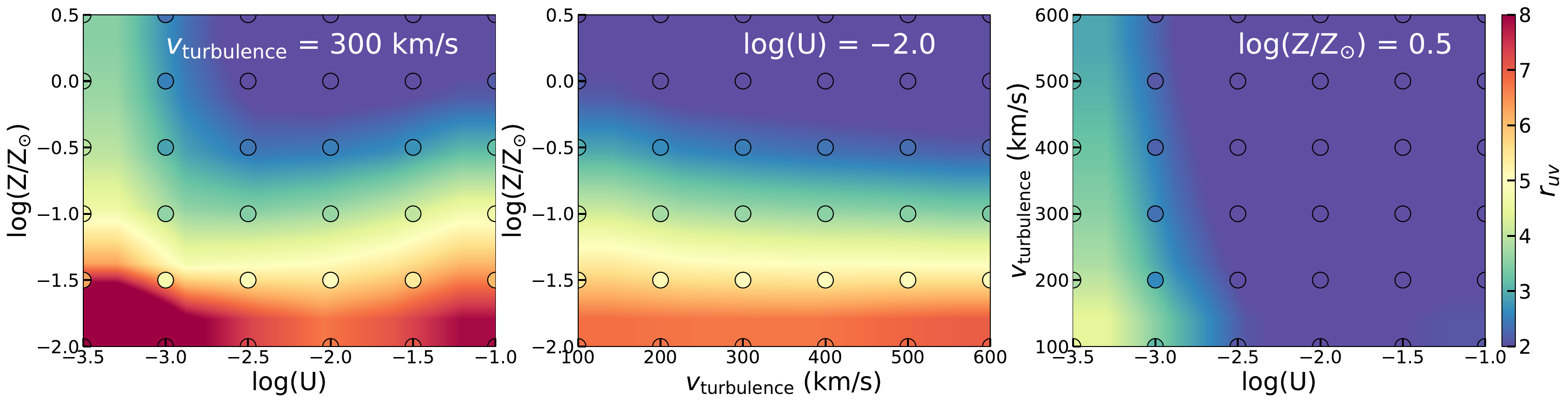}
    \centering\includegraphics[width=0.95\textwidth]{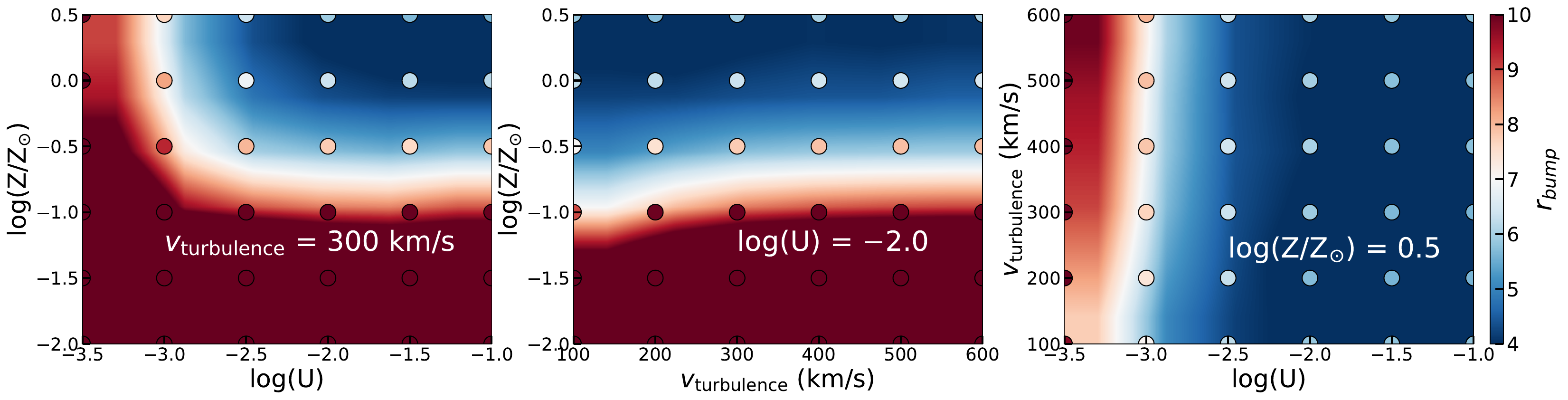}
    \caption{\textit{Top}: the relative strength of the continuum excess, $r_{\rm uv}$, as functions of the metallicity, ionization parameter, and turbulent velocity.
    In each panel, we varied two of the three parameters while keeping the third parameter at the value producing the smallest $r_{\rm uv}$ within the range of the parameter space (as shown by the white text in each panel).
    \redtxt{Individual \cloudy\ models are plotted as colored circles and the background color maps correspond to 2D linear interpolations of the models.
    }
    \textit{Bottom}: the relative strength of the \feii\ emission near the jump, $r_{\rm bump}$, as functions of the metallicity, ionization parameter, and turbulent velocity.
    Definitions of $r_{\rm uv}$ and $r_{\rm bump}$ are given by Equations~\ref{eq:ruv} and \ref{eq:rbump}.
    }
    \label{fig:feii_params}
\end{figure*}

\begin{figure*}
    \centering\includegraphics[width=0.75\textwidth]{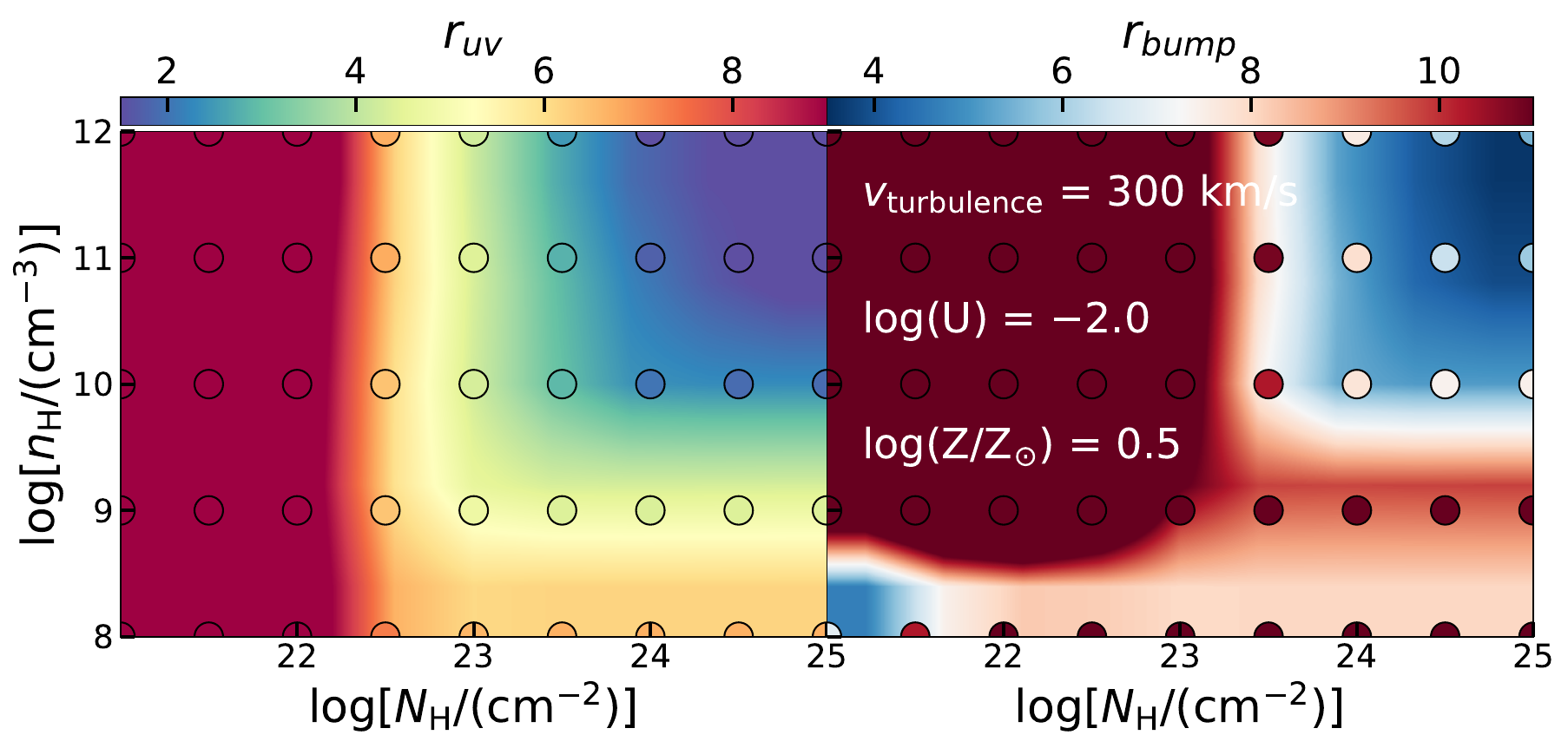}

    \caption{\textit{Left}: the relative strength of the continuum excess, $r_{\rm uv}$, as a function of the hydrogen volume density and the hydrogen column density.
    \redtxt{Individual \cloudy\ models are plotted as colored circles and the background color maps correspond to 2D linear interpolations of the models.
    }
    \textit{Right}: the relative strength of the \feii\ emission near the jump, $r_{\rm bump}$, as a function of the hydrogen volume density and the hydrogen column density.
    The metallicity, ionization parameter, and turbulent velocity were fixed at the values producing the smallest $r_{\rm uv}$ within the range of the parameter space.
    Definitions of $r_{\rm uv}$ and $r_{\rm bump}$ are given by Equations~\ref{eq:ruv} and \ref{eq:rbump}.
    }
    \label{fig:feii_params1}
\end{figure*}

\begin{figure*}
    \centering\includegraphics[width=\columnwidth]{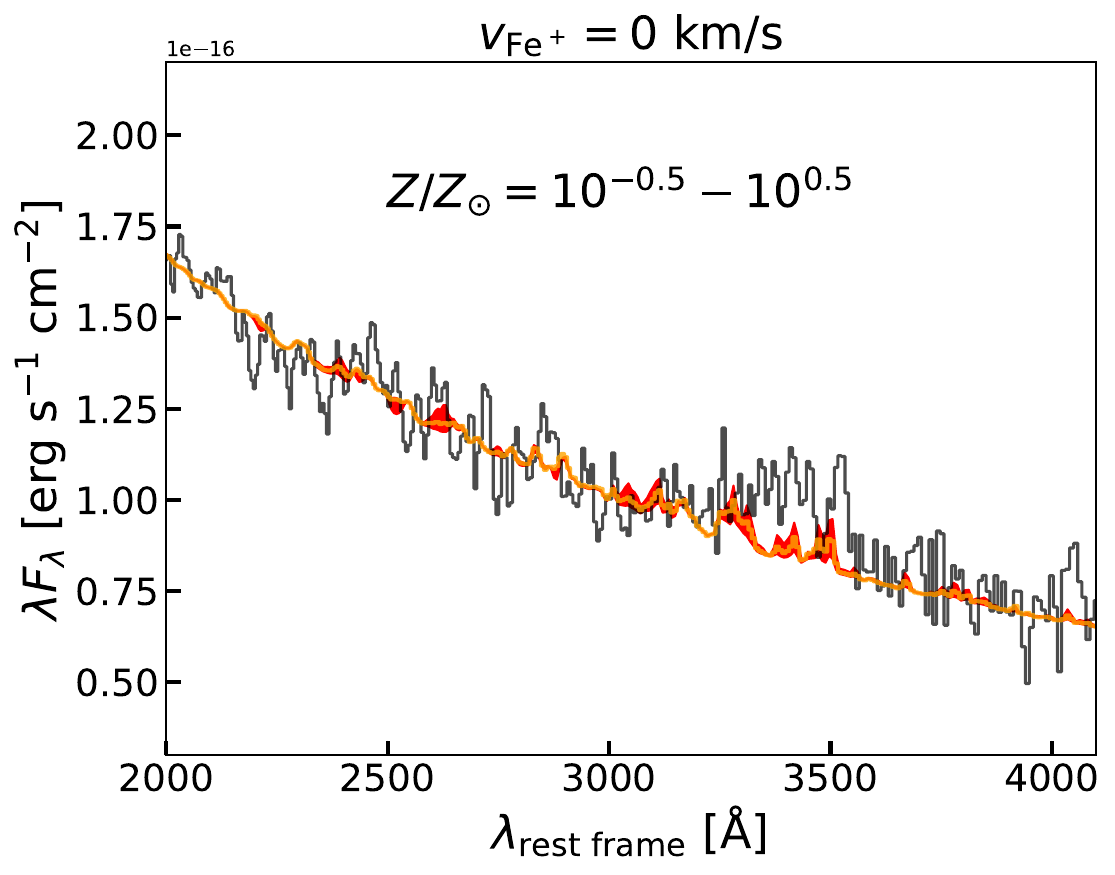}
    \centering\includegraphics[width=\columnwidth]{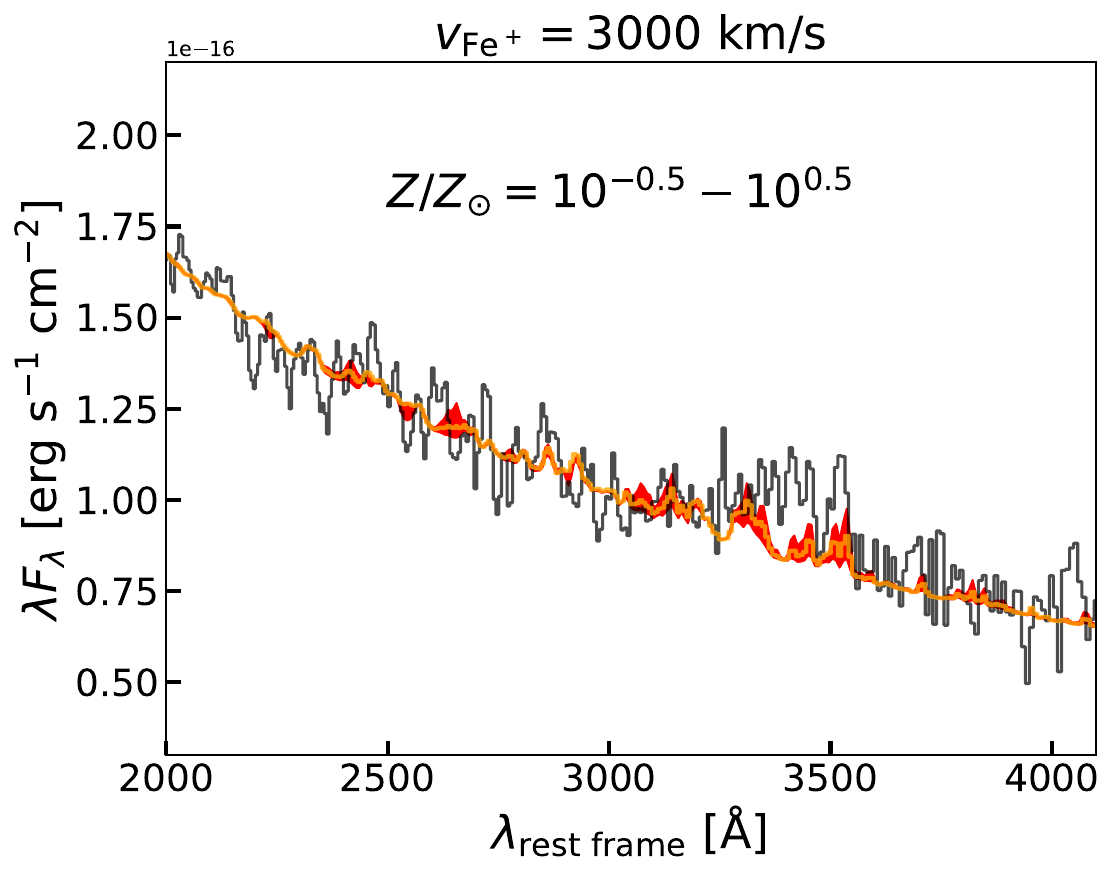}

    \caption{Comparisons between theoretical \feii\ models with the observed bump in GN-z11.
    %\textit{Top}: three \feii\ models with different metallicities (together with a power-law model) are plotted.
    In each panel, the orange solid line represents an \feii\ model with a solar metallicity combined with an \redtxt{AGN} continuum model.
    The red shaded region shows the variation of the \feii\ model within a range of metallicities from roughly 1/3 the solar value to 3 times the solar value (while keeping the normalization of the \feii\ model the same).
    %The solid black and yellow line is the Prism spectrum of GN-z11 in the rest frame.
    \redtxt{The solid black line is the line-subtracted Prism spectrum of GN-z11 in the rest frame.}
    %We only compared the black part of the spectrum with models.
    In the left panel, the relative velocity between the \feii\ emitting clouds and the forbidden-line emitting clouds, $v_{\rm Fe^+}$, is set to 0.
    In the right panel, however, $v_{\rm Fe^+}$ is set to 3000 km/s so that the \feii\ emission is associated with a fast inflow within the BLR.
    %\textit{Bottom}: the same \feii\ models are plotted, but the relative velocity between the \feii\ emitting clouds and the forbidden-line emitting clouds is set to 3000 km/s.
    %The green, cyan, and blue diamonds represent the residuals. The grey shaded region shows the $1\sigma$ uncertainty of the spectrum.
    }
    \label{fig:feii_datacom}
\end{figure*}

\begin{figure*}
    %\centering\includegraphics[width=0.48\textwidth]{Figs/feii_600vturb_fitzd5.png}
    %\centering\includegraphics[width=0.48\textwidth]{Figs/feii_600vturb_fitzd5_zoom.png}

    \centering\includegraphics[width=0.9\textwidth]{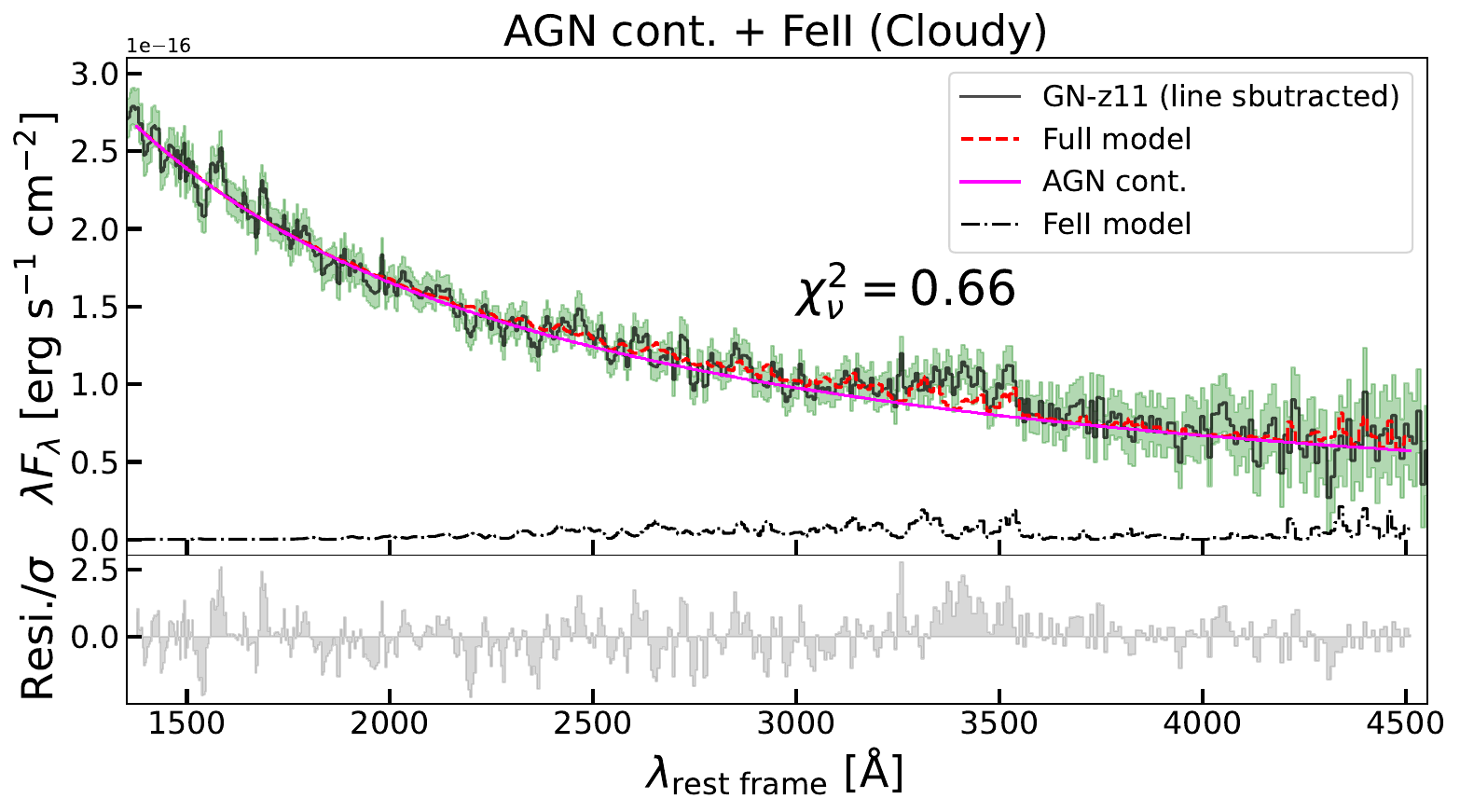}

    \caption{%Comparisons between the theoretical \feii\ model that has the strongest UV jump with the observed bump in GN-z11.
    %\textit{Top}: three \feii\ models with different metallicities (together with a power-law model) are plotted.
    \redtxt{Best-fit model of an AGN continuum combined with theoretical \feii\ emission computed with \cloudy.}
    The solid black line represents the \redtxt{line-subtracted} Prism spectrum of GN-z11.
    %, where the black part is compared to the \feii\ model.
    %The solid red and cyan line is the \feii\ model combined with a power-law model (the solid blue line).
    %The part of the model that poorly matches the observed bump is colored in cyan.
    \redtxt{The dashed red line is the full model composed of an AGN continuum (solid magenta) and \feii\ emission (dashed black).}
    The relative velocity between the \feii\ emitting clouds and the forbidden-line emitting clouds, $v_{\rm Fe^+}$, is set to 3000 km/s.
    The \redtxt{green} shaded region indicates the $1\sigma$ uncertainty of the observed flux.
    \redtxt{The $\chi$ residual of the fit is shown in the bottom panel.}
    %The red and cyan diamonds are the differences between the observed spectrum and the \feii\ model.
    %The left panel shows the comparison over the full spectral range.
    %The right panel is a zoom-in view around the bump.
    %\textit{Bottom}: the same \feii\ models are plotted, but the relative velocity between the \feii\ emitting clouds and the forbidden-line emitting clouds is set to 3000 km/s.
    %We only compared the black part of the spectrum with models.
    %The green, cyan, and blue diamonds represent the residuals. The grey shaded region shows the $1\sigma$ uncertainty of the spectrum.
    }
    \label{fig:feii_datacom1}
\end{figure*}

%{\color{red} [Check with Bartolomeo if he can fit the hump with the empirical mix-and-match method that he is adopting.]}

%{\color{red} [Try to lower vturb to be more consistent with NIV and HeII]}

%{\color{red} [Comment even more that the 3000 km/s shift is probably required also for the lwoer redshift quasars in order to match models and observed/template spectra - maybe add a few additional quasars as examples]}

In a typical BLR, due to the high gas density and the destruction of dust grains, $\rm Fe^+$ can be abundant and produce strong permitted emission lines in the UV.
The resulting \feii\ emission has been considered as the major contributor to the small blue bumps in quasar spectra.
In Figure~\ref{fig:quasar_com}, we make a direct comparison between the spectrum of GN-z11 and quasar spectra at lower redshifts.
In the left panel, we show the spectrum from a luminous quasar at $z = 4.96$ from \cite{lai2024}, where the small blue bump is apparent between 2000 \AA\ and 3550 \AA.
Intriguingly, the small blue bump in the quasar shows a truncation just at 3550 \AA, similar to the truncation of the continuum excess in GN-z11.
This truncation of the small blue bump traces the red end of the UV \feii\ complex and is included in empirical UV \feii\ templates for quasars \citep{mejia-restrepo2016}.
The strength of the \feii\ emission, however, varies strongly among quasars and the underlying physics remains poorly understood due to the complex production mechanisms and radiative transfer of \feii\ lines \citep[see e.g.,][and references therein]{baldwin2004,ferland2009,gaskell2022}.

In the right panel of Figure~\ref{fig:quasar_com}, we show another comparison between an average quasar spectrum constructed by \citet{vandenberk2001} and the spectrum of GN-z11.
The composite quasar spectrum of \citet{vandenberk2001} has a relatively strong and flat continuum.
To better compare the jump feature, we replaced the continuum of the composite spectrum with that of GN-z11 and rescaled the strength of the nebular emission.
\redtxt{To do this, we first normalized the composite quasar spectrum to the flux density of the GN-z11 spectrum at $\lambda = 2300$ \AA. After that, we subtracted the best-fit power law\footnote{The best-fit power law of \citet{vandenberk2001} was obtained by dividing the quasar spectrum into segments with longest possible wavelengths whose ends can be connected by power laws without intersecting the spectrum. We refer the readers to Section 4.1 of \citet{vandenberk2001} for more details.
} of \citet{vandenberk2001} for the spectral range of 1300 \AA\ $<\lambda <$ 5000 \AA, which has a slope of $-1.56$.
We considered that the residual of the quasar spectrum is dominated by nebular emission and, for illustration purposes, multiplied the flux densities by factors of 1, 2, and 3.
Finally, we added the rescaled nebular components to the best-fit power law of the GN-z11 spectrum, which has a slope of $-2.36$ and plotted them in Figure~\ref{fig:quasar_com}.
}
The bump between 3000 \AA\ and 3550 \AA\ is clearly visible in the composite quasar spectrum,
although the continuum excess blueward of the small blue bump is stronger.
The \feii\ jump at $\sim 3550$ \AA\ is also present, and there is an additional jump likely caused by a Balmer continuum around 3800 \AA.
%although not as strong as the specific case in the left panel due to averaging over 2200 SDSS quasars with different shapes of \feii\ emission.
In Appendix~\ref{appendix:feii_obs}, we provide a few other examples of \feii\ jumps in observations of individual quasar spectra.
A group of \feii\ fluorescent lines around 3533 \AA is likely responsible for the \feii\ jump \citep{netzer1983}, although the observed jumps in quasars all appear slightly redshifted to 3550 \AA.

The apparent difficulty associated with the \feii\ explanation for the continuum excess is the atypical shape observed in GN-z11.
In Figure~\ref{fig:quasar_com}, it is clear that \feii\ emissions in both quasar spectra have higher fluxes at shorter wavelengths within the continuum excess region.
In GN-z11, the ``\feii\ complex'' is stronger towards the red end instead.
One might wonder whether significant internal dust attenuation beyond the dust sublimation radius could produce this shape.
While the Balmer decrement is consistent with Case B values, it is possible that the \feii\ emission originates from a different ionized region that has dust in the foreground \citep{shields2010}.
{However, without an independent way to constrain the dust geometry, this approach gives too much freedom to \feii\ models and the resulting optical \feii\ emission might become too strong compared to UV \feii\ emission.
%We further discuss the effect of dust in Section~\ref{sec:discussion}.
}
%However, a typical \feii\ complex in the UV would show strong emission around 3000 \AA\ and it is difficult for any given dust attenuation curve to preferentially attenuated the flux right below 3000 \AA\ while keeping the flux beyond 3000 \AA\ visible.
On the other hand, given that the physical conditions in GN-z11 are very different from lower-redshift quasars, we considered the possibility of producing a strong \feii\ jump by exploring a large grid of dust-less \textsc{Cloudy} photoionization models.
%We caution that although the atomic data for \feii\ is relatively complete, the shape of the \feii\ complex strongly depends on the assumed gas properties \citep{sarkar2021}.

Table~\ref{tab:models} summarizes the range of parameters we adopted for the \feii\ models.
We adopted a standard AGN SED for the ionizing source and a range of ionization parameters typically found in AGN ionized gas.
\redtxt{The SED is assumed to have the functional form of Equation~\ref{eq:sed},
where the accretion disk temperature is set to $T_{\rm BB}=10^6$ K, the X-ray slope is set to $\alpha _{\rm x} = -1$, and the UV slope is set to $\alpha _{\rm uv} = -0.5$.}
\redtxt{Parameter $a$ sets the optical-to-X-ray slope, which is typically $\alpha _{\rm ox} = -1.4$ for local AGN \citep{zamorani1981}.}
\redtxt{However, as recently suggested by \citet{lambrides2024}, the intermediate-to-low luminosity AGN population discovered by JWST have $\alpha _{\rm ox} \sim -1.8$ or even lower as they are much weaker in hard X-ray bands compared to the local AGN with similar bolometric luminosities. We checked SEDs with $\alpha _{\rm ox}=-1.2$, $-1.4$, and $-1.8$, respectively. The shape of the \feii\ emission has a slight dependence on $\alpha _{\rm ox}$.}
\redtxt{When the optical-to-X-ray slope is steeper, the \feii\ emission at $\lambda > 3000$ \AA\ is slightly enhanced with respect to that from the shorter wavelength. We also checked the effect of varying accretion disk temperature.}
\redtxt{When $T_{\rm BB}$ is decreased to $10^{5}$ K, the change in the relative shape of the \feii\ emission is not obvious despite some slight suppression at $\lambda < 2200$ \AA\ and slight enhancement at $\lambda > 3000$ \AA.
We therefore set $\alpha _{\rm ox}=-1.8$ as the fiducial value, but keep $T_{\rm BB}=10^6$ K.
}

For the metallicity, although the estimated oxygen abundance is around 1/10 the solar value based on forbidden lines \citep{bunker2023}, we considered metallicities up to super-solar values to take into account potential early enrichment of metals in the central region of GN-z11 \citep{cameron2023,maiolino2023b}, where the \feii\ emission originates.
To allow strong permitted transitions of $\rm Fe^+$, it is necessary to have high densities, and thus we set $n_{\rm H} \ge 10^8~{\rm cm^{-3}}$.
In addition, we assumed the gas to be density-bounded as expected for typical BLRs and set the stopping column density, $N_{\rm H}$, to range from $10^{21}~{\rm cm^{-2}}$ to $10^{25}~{\rm cm^{-2}}$.
Although a hydrogen column density as high as $10^{25}~{\rm cm^{-2}}$ is atypical for low-$z$ AGNs, such a value might not be uncommon for high-$z$ AGNs that show BLR components in emission lines but lack X-ray emission \citep{maiolino2023b}.
A high column density combined with a high ionization parameter generally boosts \feii\ emission due to the extended warm ionized zone with $\rm Fe^+$.
Finally, the microturbulence velocity is an important parameter that changes the strength and shape of the \feii\ emission \citep[e.g.,][]{netzer1983,verner1999}.
{We allowed the microturbulence velocity\footnote{\redtxt{We note that in \cloudy\ the turbulence velocity is related to the velocity dispersion by $v_{\rm turb}=\sqrt{2}\sigma$.}} to vary from 100 km/s to 600 km/s, encompassing the potential range of gas velocity dispersion in the BLR of GN-z11 probed by \niv$\lambda 1486$ ($\sigma = 200\pm 21$ km/s; \redtxt{\citealp{maiolino2023}}) and \heii$\lambda 1640$ ($\sigma = 510\pm 160$ km/s; \redtxt{\citealp{maiolino2023}}).}
%which is higher than the maximum values adopted by previous modeling attempts (e.g., $1000$ km/s for \citealp{baldwin2004} and \citealp{sarkar2021}) but not unexpected for BLRs \citep{bottorff2000}.
%The microturbulence mainly affects the escape fraction of \feii\ photons and the continuum pumping of the \feii\ emission.

A key mechanism that can produce a strong \feii\ jump at $\sim 3550$ \AA\ relative to the \feii\ emission blueward of 3000 \AA\ is the strong dependence of the \feii\ optical depth on the wavelength.
This generally leads to a greater reduction of the \feii\ emission in the shorter wavelength in the emergent spectrum \citep{ferland2009}.
Figure~\ref{fig:feii_eg} shows an example of an \feii\ model with log(U) = $-2.5$, Z = Z$_{\odot}$, $n_{\rm H} = 10^{11}~{\rm cm^{-3}}$, $N_{\rm H} = 10^{25}~{\rm cm^{-2}}$, and \redtxt{$v_{\rm turb} = 300$ km/s.}
\redtxt{As pointed out by \citet{ferland2009}, it is important to distinguish the total (intrinsic) \feii\ emission from the \feii\ emission that is observed.
For a typical geometry proposed by \citet{ferland2009} where the \feii\ emission is seen from the back of the BLR clouds (i.e., opposite to the illuminated face), one should consider the portion of the emission that is beamed away from the ionizing source, which we plotted as the emergent or ``out'' component.
}
While the total \feii\ emission peaks around 2500 \AA\ in the UV, the emergent \feii\ emission is stronger in the continuum excess region of GN-z11.
Between 3000 \AA\ and 3550 \AA, there are three groups of \feii\ emission separated by two gaps in the emergent spectrum, with the rightmost one producing a jump close to the edge of the continuum excess region.
We discuss below how flexible the shape of the UV \feii\ emission is and whether the range of physical parameters preferred by the observed data is physically plausible.

\subsubsection{Parametrization of the shape of the \feii\ emission}

{To characterize the shape of the \feii\ complex, we defined two parameters. The first parameter, $r_{\rm uv}$, is defined as
%$r_{\rm uv} \equiv F_{2000~\AA< \lambda < 3000~\AA}/F_{3000~\AA< \lambda < 3550~\AA}$, 
\begin{equation}
    r_{\rm uv} \equiv F_{2000~\AA< \lambda < 3000~\AA}/F_{3000~\AA< \lambda < 3550~\AA},
    \label{eq:ruv}
\end{equation}
which describes the relative strength of the \feii\ pseudo-continuum in the continuum excess region with respect to the part at shorter wavelengths. The second parameter, $r_{\rm bump}$, is defined as
%$r_{\rm bump} \equiv F_{3000~\AA< \lambda < 3550~\AA}/F_{3400~\AA< \lambda < 3550~\AA}$, 
\begin{equation}
    r_{\rm bump} \equiv F_{3000~\AA< \lambda < 3550~\AA}/F_{3400~\AA< \lambda < 3550~\AA},
    \label{eq:rbump}
\end{equation}
which describes the relative strength of the \feii\ emission making up the jump compared to the whole \feii\ emission in the continuum excess region. As an example, for the model in Figure~\ref{fig:feii_eg}, 
%$r_{\rm uv} = 1.5$ and $r_{\rm bump} = 4.0$ 
$r_{\rm uv} = 1.7$ and $r_{\rm bump} = 4.2$
for the outwards \feii\ emission, 
%and $r_{\rm uv} = 6.8$ and $r_{\rm bump} = 7.9$
and $r_{\rm uv} = 6.2$ and $r_{\rm bump} = 8.6$ 
for the total \feii\ emission. Given the shape of the observed continuum excess in GN-z11, we looked for models with both \textit{small} $r_{\rm uv}$ and \textit{small} $r_{\rm bump}$.}

To investigate the effect of different parameters, we started with a fiducial set of parameters typical for a Compton-thick BLR with $n_{\rm H} = 10^{11}~{\rm cm}^{-3}$ and $N_{\rm H} = 10^{25}~{\rm cm}^{-2}$, and varied the other three parameters, which are the metallicity, ionization parameter, and turbulent velocity.
%While the total hydrogen column density we adopted is much higher than what is usually adopted for lower redshift AGNs (with $N_{\rm H} \sim 10^{23}~{\rm cm}^{-2}$), it is not unexpected for high-redshift AGNs that show BLR components but are undetected in X-ray \citep{Maiolino2024_Xrays,maiolino2023b}.
%log(U) = $-2.5$, log(Z/Z$_{\odot}$) = $-1$, $n_{\rm H} = 10^{11}~{\rm cm}^{-3}$, $N_{\rm H} = 10^{25}~{\rm cm}^{-2}$, and $v_{\rm turb} = 100~{\rm km/s}$, and varied two parameters at a time.
%Once we determined the major parameters and varying directions that reduce $f_{\rm uv}$ and $f_{\rm bump}$, we computed models with finer grids, thereby saving computational time.
Within the range of the parameter space we considered, the minimum $r_{\rm uv}$ was found at \redtxt{log(U) = $-2.0$}, log(Z/Z$_{\odot}$) = $0.5$, and 
\redtxt{$v_{\rm turb} = 300~{\rm km/s}$,}
and the minimum $r_{\rm bump}$ was found at log(U) = $-1.0$, log(Z/Z$_{\odot}$) = $0.5$, and 
\redtxt{$v_{\rm turb} = 200~{\rm km/s}$.}
While the values of log(U) and $v_{\rm turb}$ are possible for a highly ionized and turbulent BLR (or one having shears/velocity gradients on small scales; \citealp{bottorff2000}),
a super-solar metallicity seems too high for a galaxy $\sim 430$ Myr after the Big Bang and is in conflict with the sub-solar oxygen abundance inferred from forbidden lines \citep{bunker2023}.
Still, based on different model assumptions, the potential range of the metallicity for GN-z11 could reach near-solar or even super-solar values \citep{cameron2023,isobe2023}, especially in the very small region of the BLR \citep{maiolino2023}.
We also note that the iron abundance is more important than the overall chemical abundances in these models, meaning a super-solar Fe/H combined with a sub-solar metallicity can reproduce these results as well.

Figure~\ref{fig:feii_params} shows the impact of different input parameters on the predicted $r_{\rm uv}$ and $r_{\rm bump}$.
We fixed one parameter at the value that produces the minimum $r_{\rm uv}$ or $r_{\rm bump}$, and varied two remaining parameters at a time.
The result shows that a metallicity of log(Z/Z$_{\odot}$)$~\gtrsim -0.5$ is critical to suppress the UV \feii\ emission blueward of the jump while enhancing the apparent jump strength.
Meanwhile, an intermediate value of log(U) and a turbulent velocity of 
%$v_{\rm turb} \gtrsim 1000~{\rm km/s}$ 
$v_{\rm turb} \gtrsim 200~{\rm km/s}$ 
help to achieve the minimum $r_{\rm uv}$ and $r_{\rm bump}$.
%The required turbulent velocity is compatible with the BLR velocity dispersion constrained by the width of \niv$\lambda 1486$ and \heii$\lambda 1640$ \citep{maiolino2023}.
To check whether our choice of $n_{\rm H} = 10^{11}~{\rm cm}^{-3}$ and $N_{\rm H} = 10^{25}~{\rm cm}^{-2}$ significantly affect the determinations of minimum $r_{\rm uv}$ and $r_{\rm bump}$, we further varied $n_{\rm H}$ and $N_{\rm H}$ over a larger range while keeping log(U), log(Z/Z$_{\odot}$), and $v_{\rm turb}$ at values producing the minimum $r_{\rm uv}$.
Figure~\ref{fig:feii_params1} shows the results of varying $n_{\rm H}$ and $N_{\rm H}$.
To minimize $r_{\rm uv}$, one needs $n_{\rm H} \gtrsim 10^{10}~{\rm cm}^{-3}$ and $N_{\rm H} \gtrsim 10^{24}~{\rm cm}^{-2}$.
For $r_{\rm bump}$, while the global minimum is at $n_{\rm H} = 10^{8}~{\rm cm}^{-3}$ and $N_{\rm H} = 10^{21}~{\rm cm}^{-2}$, combined with the constraints from $r_{\rm uv}$, one still needs $n_{\rm H} > 10^{10}~{\rm cm}^{-3}$ and $N_{\rm H} > 10^{24}~{\rm cm}^{-2}$.
The latter implies Compton-thickness, typical of other high-redshift AGNs \citep{brightman2012,lanzuisi2018}.
The large $n_{\rm H}$ is typical of the BLR clouds.
Finally, the high metallicity is also typical of the BLR of AGN, although at such early epochs the very high Fe enrichment might be puzzling, but as much as the high nitrogen enrichment has been puzzling for this and other high-$z$ galaxies \citep{cameron2023,isobe2023,senchyna2023,topping2024}.
% In summary, to minimize the shape parameters we defined for the \feii\ emission, the only parameter that seems incompatible with a high-redshift AGN scenario is the high metallicity.
We discuss this point in detail in Section~\ref{sec:discussion}.

Issues however remain even if we adopt high-metallicity models. The minimum values of parameters predicted by the models are 
%$r_{\rm uv} = 1.2$ and $r_{\rm bump} = 3.4$.
$r_{\rm uv} = 1.5$ and $r_{\rm bump} = 3.9$.
If we fit a power-law to the observed spectrum while excluding the bump region, and took the residual as the potential \feii\ emission, the corresponding $r_{\rm uv}$ is basically noise-limited and $r_{\rm bump} = 1.1 \pm 0.2$, which are significantly smaller than the model predictions.
The differences are in part due to the gaps between groups of \feii\ emission in the wavelength space, which are not observed in typical quasar spectra as well as in GN-z11.
The absence of the gaps could be a result of broadening by the line spread function (LSF) of the instrument, broadening by the intrinsic velocity dispersion of the BLR, and pumping of unaccounted transitions by nearby lines and continua \citep{netzer1983,verner1999,baldwin2004,ferland2009,sarkar2021}.
Next, we directly compared the models with the bump in GN-z11 taking into account the above effects.

\subsubsection{Comparison with the observed bump}

In Figure~\ref{fig:feii_datacom}, we compare three sets of \feii\ models with log(Z/Z$_{\odot}$) = $-0.5$, 0, and $0.5$ combined with \redtxt{the fiducial AGN continuum SED we adopted for GN-z11 from \citet{pezzulli_agn_2017}.}
%a power-law continuum model to the observed spectrum of GN-z11.
Based on the analysis in the previous subsection, we set \redtxt{$\rm log(U) = -2.0$, $v_{\rm turb} = 300$ km/s,} $n_{\rm H} = 10^{11}~{\rm cm^{-3}}$, and $N_{\rm H} = 10^{25}~{\rm cm^{-2}}$ for these models.
We set the normalization of the \feii\ emission as a free parameter to take into the account the possibility that it arises from a region different from other strong lines in the BLR \citep[e.g.,][]{ferland2009} and adopted the normalization that best reproduces the small blue bump.
Before we made the comparisons, we broadened the \feii\ models by an LSF constructed from combining the broadening by the microturbulence in the models and the broadening by the instrumental resolution ($\sim 1.1 \times 10^3$ km/s).
%Before we made the comparisons, we broadened the \feii\ models by an LSF which generate an FWHM of $1.2\times 10^3$ km/s near the location of the bump.
%The broadening corresponds to an intrinsic broadening of $4.7\times 10^2$ km/s, as inferred from the width of N\,{\sc iv}] \citep{maiolino2023}, combined with an instrumental broadening of $1.1 \times 10^3$ km/s.

{Finally, judging from Figures~\ref{fig:quasar_com} and \ref{fig:feii_eg}, the theoretical \feii\ jump generated by \textsc{Cloudy} is slightly blueshifted compared to the observed jump, meaning the \feii\ emission in GN-z11, if it exists, should be redshifted with respect to other strong emission lines.
In fact, \feii\ emission was often found to be redshifted in quasar spectra at lower redshifts and redshift velocities up to 3000 km/s were reported previously \citep{hu2008a,hu2008b}.
As an example, in Figures~\ref{fig:quasar_com} and \ref{fig:more_feii}, the \feii\ jumps in these quasars are not significantly blueshifted with respect to the jump seen in GN-z11.
We therefore performed a separate comparison using redshifted \feii\ models in the right panel of Figure~\ref{fig:feii_datacom}, and further discuss its physical explanation in Section~\ref{sec:discussion}.}

While compared to the Balmer continuum models in Figure~\ref{fig:gnz11_bcfit}, the \feii\ models fail to reproduce the middle part of the small blue bump due to the aforementioned gaps between different groups of \feii\ emission, the \feii\ models are able to recover certain observed features near the boundaries of the bump.
One can see clear \feii\ jumps in the models, which match the observed jump increasingly better with increasing metallicity, especially for those models having a velocity shift of $\sim 3000$ km/s.
On the other hand, the potential emission line at 3133 \AA\ can be explained by a group of redshifted \feii\ emission, as shown in the right panel of Figure~\ref{fig:feii_datacom}.
%The required velocity shift among lines is not uncommon in AGNs given the potentially different origins of lines.
%The velocity shift of the \feii\ models is smaller than that required for Balmer continuum models (about $8\times 10^3$ km/s), but it would still indicate a fast inflow in a BLR.
%We suspect the seemingly shift in the velocity might also come from unaccounted fluorescent (\feii) emission in \textsc{Cloudy} models.
%As an example, in Figures~\ref{fig:quasar_com} and \ref{fig:more_feii}, the \feii\ jumps in these quasars are not blueshifted with respect to the jump seen in GN-z11.
%at $z = 4.96$ from \cite{lai2024} is clearly 
%redshifted with respect to Mg\,{\sc ii}$\lambda \lambda 2796,2804$.
In Figure~\ref{fig:feii_datacom1}, we compare the highest metallicity \feii\ model that produces the strongest UV jump with the observed bump in GN-z11.
While the \feii\ jump in the model well resembles the observed jump at 3550 \AA, the flux of the observed bump between 3300 \AA\ to 3480 \AA\ is clearly underestimated by the model.
\redtxt{Still, the best-fit \feii+AGN model outperforms other AGN fits by giving $\chi^2_{\nu}=0.66$, which is similar to the $\chi^2_{\nu}$ from the best-fit SF model with a nebular continuum.
}

The question now becomes whether it is possible to fill the gaps between the \feii\ emission.
In Figure~\ref{fig:quasar_com}, the \feii\ emission in quasars does not exhibit noticeable gaps in the continuum excess region.
If the absence of gaps is purely due to the broadening of \feii\ line profiles, one needs an intrinsic FWHM $\gtrsim 8\times 10^3$ km/s.
%This would also make the \feii\ emission at shorter wavelengths to be significantly blended and results in a worse fit in the UV.
Although with a larger microturbulence velocity, the \feii\ emission in the continuum excess region would become even stronger with respect to the \feii\ emission at shorter wavelengths (and thus produce a stronger bump), the required broadening is significantly larger than the measured widths of semi-forbidden lines and permitted lines in GN-z11, which presumably come from the same BLR that produces the \feii\ emission.

Alternatively, one might wonder whether any unaccounted line/continuum fluorescence of the \feii\ emission between the gaps could explain the model mismatch.
Indeed, from the observed high-resolution quasar spectra as well as the empirical \feii\ templates constructed from them, the gaps we see in the models are also filled with lines \citep[e.g.,][]{tsuzuki2006,mejia-restrepo2016}.
Thus, it is possible that our current models still do not fully capture the fluorescence of the \feii\ emission between 3000 \AA\ and 3550 \AA.
A full scrutiny of the physical processes in \feii\ modeling is beyond the scope of this work and we leave the theoretical calculation of the \feii\ emission under extreme physical conditions for future investigations.
%will further investigate the theoretical calculation of the \feii\  emission under extreme physical conditions in a future paper.
In what follows, we simply consider the \feii\ emission as a candidate to produce the jump feature while discussing the physical interpretations.
%Indeed, the gaps in the models are not void of \feii\ emission but are simply weak compared to other parts.
%\redtxt{[TBA]}

\section{Discussion}
\label{sec:discussion}

In the previous sections we have attempted fitting the observed continuum in the Prism spectrum of GN-z11 using combinations of an \redtxt{AGN or SF continuum} model, a \redtxt{nebular} continuum model, and \feii\ emission models.
%Although none of the models is able to fully reproduce the small blue bump in GN-z11, the preferred ranges of the model parameters and the failure of a good fit itself indicate an uncommon origin of the small blue bump.
\redtxt{In the SF case, the continuum excess originates from a Balmer continuum that has a temperature marginally higher than that derived from \oiii\ lines.
In the AGN case, the continuum excess is better described by \feii\ emission compared to a Balmer continuum.
Purely based on the $\chi^2_{\nu}$, the best-fit SF model and AGN model do not have significant differences.
}
In this section, we discuss the implications of our results.

\subsection{Presence of different ionized regions}

\subsubsection{Hot gas with hydrogen Balmer emission}
\begin{comment}
\begin{figure*}
    \centering\includegraphics[width=0.98\textwidth]{Figs/cloudy_temperature.png}

    \caption{Predictions on hydrogen Balmer temperatures for clouds ionized by different sources as functions of the metallicity and ionization parameter.
    \textit{Left}: contours of electron temperatures of SF models ionized by a simple stellar population (SSP) from a 1 Myr-old starburst generated by BPASS assuming the default initial mass function (IMF) with an upper mass limit of 300 $M_{\odot}$ described in \citet{stanway2018}.
    Both the dashed lines and dotted lines are plotted with steps of 3000 K. Adjacent dashed lines and dotted lines are separated by 1500 K.
    \textit{Middle}: contours of electron temperatures of NLR models ionized by an AGN SED the same as the one in Table~\ref{tab:models}.
    The hydrogen density of the models is set to $10^3~{\rm cm^{-3}}$.
    \textit{Right}: contours of electron temperatures of BLR models ionized by an AGN SED the same as the one in Table~\ref{tab:models}.
    The hydrogen density of the models is set to $10^{11}~{\rm cm^{-3}}$ and the hydrogen column density is set to $10^{22}~{\rm cm^{-2}}$ as appropriate for the $\rm H^+$ region within the BLR.
    While the SF models and NLR models include dust depletion, the BLR models do not. \purpletxt{TBD.}
    }
    \label{fig:cloudy_tem}
\end{figure*}

\end{comment}

\begin{figure}
    \centering
    \includegraphics[width=0.48\textwidth]{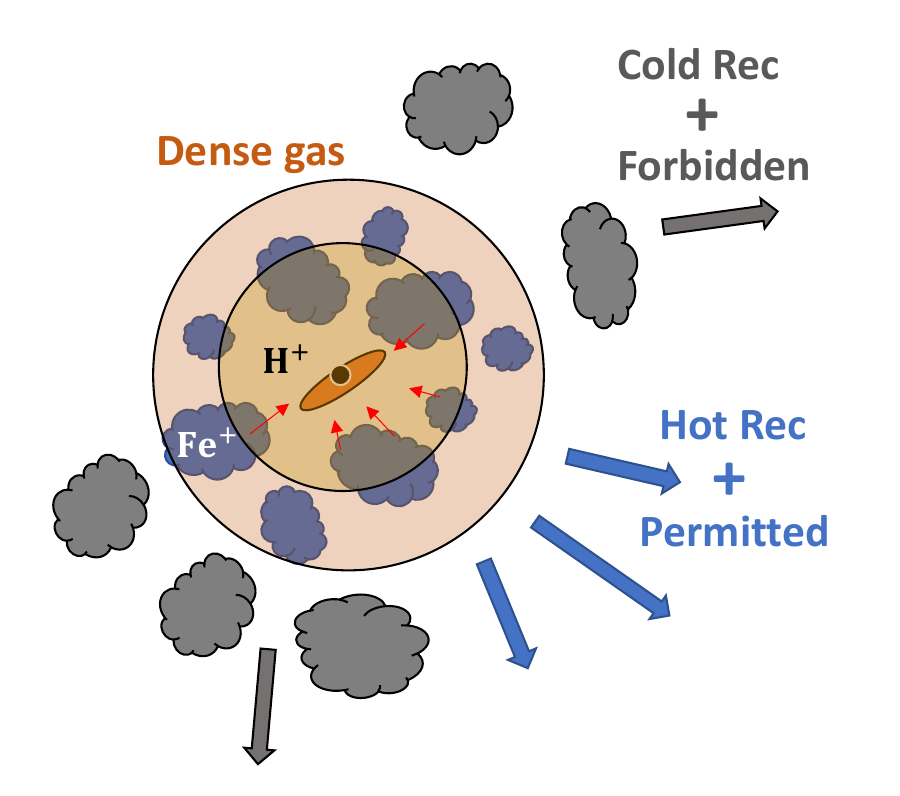}
    \caption{Schematic plot illustrating the inferred ionization structure within the nuclear region of GN-z11.
    In the central BLR that hosts dense gas clouds, Balmer recombination emission including continua and lines comes from the inner and hot region.
    In contrast, permitted emission of \feii\ mainly comes from partially ionized region at large column densities, where the clouds might be infalling.
    Finally, outside the BLR and in the rest of the galaxy, there is also Balmer recombination emission from low density gas with lower temperatures, which is co-spatial with forbidden lines of metals.
    The plot does not reflect the actual physical scales of different regions.}
    %\purpletxt{[TBA]}}
    \label{fig:cartoon}
\end{figure}

As shown by \cite{maiolino2023}, gas clouds with different densities are likely present in GN-z11. On the one hand, constraints from the semi-forbiddden \niv$\lambda 1486$ relative to [\niv$\lambda 1483$, and the relative strengths of the semi-forbidden \niii\ multiplet indicate these UV emission lines to originate from high density clouds with $n_{\rm H} \gtrsim 10^{9}~{\rm cm^{-3}}$, consistent with BLR clouds
On the other hand, the presence of strong, forbidden \oiii$\lambda 4363$ requires densities much lower than the UV-line emitting clouds, and consistent with NLR (or simple ISM) clouds.
% The high density of the UV-line emitting gas is consistent with the physical condition in a BLR.
% What remains unclear is the main ionizing source for optical emission lines.

\redtxt{In the SF case, the hydrogen emitting clouds might be marginally hotter than the \oiii-emitting clouds and they could still come from the same spatial region.}
\redtxt{In the AGN case,} we have shown, given the observed strength of H$\gamma$, if the Balmer-line emitting gas has a temperature similar to the \oiii-line emitting gas, \redtxt{$T_{e}({\rm O^{2+}})= 1.36\pm 0.13\times 10^4$ K}, one expects a detectable Balmer jump clearly different from the shape of the continuum excess.
If the continuum excess is dominated by some other mechanisms, which is our preferred solution, then the Balmer jump is weak enough that the best-fit temperature would be $T_{e}({\rm H^{+}})\approx 3\times 10^4$ K, implying the observed Balmer emission to originate from gas much hotter than the forbidden-line emitting gas.

We note that low-redshift observations of individual \hii\ regions or Planetary nebulae frequently show $T_{e}({\rm H^{+}}) < T_{e}({\rm O^{2+}})$ \citep[e.g.,][]{peimbert1967}, clearly the opposite to what we found in GN-z11.
In the low-redshift cases, the temperature difference potentially rises from micro-fluctuations of temperature or chemical abundance inhomogeneities within individual nebulae \citep{peimbert2017}.
In the case of GN-z11, however, we argue the temperature difference is created by the spatial distributions of gas clouds responsible for different emission lines.

\redtxt{In the AGN case, the} observed Balmer emission can have two components, one from hot and dense gas close to the accretion disk (the BLR), and the other from less-dense and colder clouds at a larger distance from the accretion disk (the NLR or SF-ionized ISM).
These two components are not separable kinematically due to the limited spectral resolution.
The hot-dense component has a negligible Balmer jump due to its high temperature, and the Balmer jump in the cold and low-density component is weak simply because the flux of the cold component is subdominant.
%We can estimate the contributions from the two Balmer components in the following way.
%Assuming the cold component has the same temperature as the that of forbidden lines
%Basically, Balmer lines are mainly produced by dense and hot gas close to the accretion disk and forbidden lines including \oiii\ come from less-dense and colder clouds at a larger distance from the accretion disk.
Forbidden lines such as \oiii$\lambda 4363$ are only present in the cold and low-density gas, together with the cold Balmer component, due to their low critical densities.
One might wonder whether this explanation leads to an unphysically high \oiii$\lambda 4363$/H$\gamma$ in the cold gas.
Interestingly, high \oiii$\lambda 4363$/H$\gamma$ values have been observed in a number high-$z$ galaxies by JWST, especially AGNs \citep{ubler2023b,mazzolari2024}, with a few galaxies even showing \oiii$\lambda 4363$/H$\gamma$ $\sim$ 1.
A hard ionizing radiation from AGN combined with metal-poor gas can give rise to a high \oiii$\lambda 4363$/H$\gamma$ in an NLR.
For a rough estimation, if we simply use \oiii$\lambda 4363$/H$\gamma$ $<$ 1 to set a prior for the flux of the cold H$\gamma$, and fit the continuum assuming both a hot component and a cold component (whose temperature is fixed to $1.3\times 10^4$ K), we obtain ${\rm H}\gamma _{\rm cold} = 0.43^{+0.24}_{-0.14}{\rm H}\gamma _{\rm total}$.
%$0.23~{\rm H}\gamma _{\rm total} < {\rm H}\gamma _{\rm cold} < {\rm H}\gamma _{\rm total}$.
This calculation also slightly lowers the temperature estimation for the hot component to $T_e = 2.6^{+0.4}_{-0.6}\times 10^4$ K.
%The reduced intrinsic flux of the Balmer emission from the cold component also makes its corresponding Balmer continuum weaker and explains the undetection of a cold Balmer jump.

%Even with the lowered temperature limit, the hot component is still more consistent with an origin of AGN ionization rather than SF ionization.
We emphasize that for the hot component, a high density is necessary to explain the missing Balmer jump while avoiding overpredicting the strength of \oiii$\lambda 4363$.
This requirement rules out the possibility of the hot component being the ionized gas within the low-density NLR or the ISM ionized by young stars.
In addition, for the ISM ionized by young stars, it is difficult to reach a temperature of $T_e \sim 3\times 10^4$ K.
One might wonder whether the BLR can reach such a high temperature.
Indeed, due to the high density of the gas and the absence of dust depletion, the cooling through metals within the BLR could be efficient.
However, cooling in the BLR would be limited to semi-forbidden and permitted transitions due to the high density.
With \textsc{Cloudy}, we calculated volume-average electron temperatures in clouds ionized by young stellar populations and AGN, as shown in Figure~\ref{fig:cloudy_tem}.
While SF models cannot produce high enough temperatures even if log(Z/Z$_{\odot}$) $\sim -3$, both NLR and BLR models are able to reach $T_e ({\rm H^+}) \sim 3\times 10^4$ K at log(Z/Z$_{\odot}$) $\sim -1$ and log(U) $\sim -1$.
The problem with the NLR model, however, is that the corresponding \oiii$\lambda 4363$ is too strong compared to what has been observed in GN-z11.

Another complicating factor for BLRs is the physical conditions for different emission line specifies can well be different as they are optimally emitted by different clouds \citep{baldwin1995}.
%For this model, $T_e ({\rm H^+})$ can reach $3\times 10^4$ K at log(Z/Z$_{\odot}$) $\sim -1$ and log(U) $\sim -1$.
Again, we caution that when the Balmer edge become partially optically thick in the BLR, it can be suppressed at the same level but at a lower temperature \citep{grandi1982}.
Regardless, the requirement for the gas to be dense remains the same, and the presence of an BLR is preferred given the little contribution of the Balmer continuum to the small blue bump.
%We note that high \oiii/H$\gamma$ in individual clouds could simply be a result of the ionized gas being density-bounded rather than ionization-bounded
%\redtxt{[Oh, temperature!]}

{Figure~\ref{fig:cartoon} shows our inferred ionization structure in the nuclear region of GN-z11. The BLR produces the continuum excess we see in the near-UV as well as permitted and semi-forbidden emission lines. The Balmer continuum originates in a hot and dense region within the BLR. In contrast, \feii\ emission likely originates in a partially ionized zone with high column densities. Finally, forbidden lines and part of the Balmer lines in the optical come from a less dense and colder region in the NLR or the ISM.}
%\purpletxt{
Although we propose the Balmer lines to have origins from different regions, they are kinematically indistinguishable, especially with the limited signal-to-noise of current observations. Indeed, the broadening of lines in the BLR of GN-z11 by virial motions is only on the order of a few hundred kilometers per second due to its small black hole mass \citep{maiolino2023}.
%}
However, it may be possible to spatially resolve some of the NLR/ISM components with integral field spectroscopy, as illustrated in \cite{Maiolino2023_HeII}.

The presence of two Balmer components would also indicate biases in abundance estimations based on the assumption that the entire Balmer emission comes from the forbidden-line emitting clouds.
Our estimated ${\rm H}\gamma _{\rm cold}$ corresponds to \oiii$\lambda 4363$/H$\gamma _{\rm cold} \approx 0.53^{+0.33}_{-0.09}$,
resulting in an abundance of {$\rm 12+\log (O/H) = 8.18^{+0.48}_{-0.27}$}
assuming higher ionization states of oxygen are negligible.
The uncertainties in the above value include uncertainties in the flux of ${\rm H}\gamma _{\rm cold}$ as well as uncertainties in the $\rm O^{2+}$ temperature.
Although there is another uncertainty associated with converting the $\rm O^{2+}$ temperature to the $\rm O^{+}$ temperature in the NLR, it is at the level of $\sim 0.01$ dex in our case assuming the temperature relations for AGNs in \citet{dors2020}.
While our lower abundance estimation is broadly consistent with the value obtained by \cite{cameron2023}, the fiducial value we obtained is about 30\% of the solar value and is about two times the previously estimated value. High chemical abundances in high-$z$ AGNs are commonly seen \citep{nagao2006a,nagao2006b,juarez2009,matsuoka2009,matsuoka2011,matsuoka2018,guo2020}.
We caution, however, that the range of the abundance we derived depends on the prior on the range of \oiii$\lambda 4363$/H$\gamma _{\rm cold}$.
A lowered upper limit for \oiii$\lambda 4363$/H$\gamma _{\rm cold}$ would produce an overall lower metallicity estimation.
To settle this problem, it is vital to have high spectral resolution observations of Balmer lines, which are unfortunately missed due to the truncation of the high-resolution NIRSpec grating spectrum in the previous observation \citep{maiolino2023}.

\subsubsection{Infalling gas with a high column density}

{As we show in the previous section, \redtxt{under the AGN-dominated assumption}, if \feii\ emission is responsible for the observed bump in GN-z11, the gas with \feii\ emission should be redshifted at a velocity of $\sim 3000$ km/s with respect to other line-emitting regions.}
Such a high velocity would indicate a fast inflow.
As shown by \cite{hu2008a,hu2008b}, it is not uncommon for \feii\ emission to be redshifted with respect to narrow emission lines and broad emission lines in quasars.
Within a sample of 4037 quasars from SDSS \citep{york2000} at $z<0.8$ inspected by \citet{hu2008b}, the typical velocity shift of \feii\ is $\sim 400$ km/s, and the largest velocity shift of \feii\ clearly exceeds $2000$ km/s and even reaches $3000$ km/s.

To explain the ubiquitous shift in \feii\ velocity, \citet{ferland2009} proposed a scenario where \feii\ emission comes from infalling clouds with large column densities viewed from their shielded faces, to which the gravitational pull overcomes radiation pressure from accreting supermassive black holes (SMBHs).
In fact, for the SMBH in GN-z11 that is likely accreting at a super Eddington rate of $L/L_{\rm Edd} \sim 5$ \citep{maiolino2023}, the required column density would be $7.5\times 10^{24}~{\rm cm^{-2}}$ according to the equation given by \citet{ferland2009}.
The column density needed is in good agreement with the best-fit \feii\ model we adopted.

Finally, the infalling \feii\ emitting clouds should have a high Fe abundance as well to enhance the \feii\ emission at $3000~\AA < \lambda < 3550~\AA$.
An interesting question is whether the potentially enriched Fe is related to the ``abundance anomaly'' in GN-z11, where nitrogen appears particularly enriched compared to oxygen, which we discuss in the next subsection.
%the potential presence of \feii\ emission based on the observed jump indicates a relatively high Fe abundance or high overall chemical abundances for the gas much closer to the accretion disk.
%We discuss the physical origin of the abundance difference together with the previously reported ``abundance anomaly'' in GN-z11 in the next subsection. 
%One might question the validity of our $T_{e}({\rm O^{2+}})$ estimation based on \oiii$\lambda 4363$, \neiii$\lambda 3869$, and \oii$\lambda \lambda 3726,3729$.

\subsection{Revisiting the ``abundance anomaly''}

One of the most puzzling features in the spectrum of GN-z11 is the strong \niv\ and \niii\ emission in the UV.
Based on the high \niii/O\,{\sc iii}] in the UV, \cite{bunker2023} and \cite{cameron2023} derived a super-solar N/O, indicating a fast and/or exotic chemical enrichment in GN-z11.
With the new evidence from the potential \feii\ emission, it is possible that iron is also enriched in GN-z11.
{It is noteworthy that high Fe abundances were also found in nitrogen-loud quasars at lower redshifts and were predicted by chemical enrichment models of quasars \citep[e.g.,][]{hamann1993}.
In addition, recent JWST observations have revealed potential early enrichment of Fe as traced by forbidden transitions of Fe in a galaxy showing a prominent Balmer jump at $z = 5.94$ \citep{cameron_9422_2024,tacchella2024}, and in an AGN host galaxy at $z = 5.55$ \citep{ubler2023a,ji2024}.
}
To explain the high N/O, \cite{maiolino2023} suggest a possibility that the UV emission mainly originates in the BLR of the AGN in GN-z11, making the fast chemical enrichment within such a small region ($\sim 10^{-2}$ pc) and small gas mass (a few solar masses) possible without invoking any exotic enrichment channels.
Similarly, since the \feii\ emission mainly comes from clouds in the BLR, a fast enrichment of Fe towards the solar value is also compatible with the scenario above.
%Combining a high Fe abundance, a large column density expected for high-redshift AGNs, and a possibly highly turbulent environment, we naturally obtained a \feii\ complex showing a strong \feii\ jump in the UV, as shown in the previous section.

In addition to the fast enrichment of the overall chemical abundances within a small region, it is possible to achieve high Fe/O at low metallicities, which would give rise to the a strong \feii\ complex without invoking a high oxygen abundance.
According to the models of \cite{vanni2023}, it is possible for high-mass Pair Instability Supernovae (PISNe) to produce super solar Fe/O even when Fe/H is as low as 1 -- 10\% the solar value.
{An early enrichment of Fe is also possible through ejecta from hypernovae (HNe), especially given a potentially dense star formation environment of GN-z11 \citep{isobe2023,watanabe2024}.
It is noteworthy that solar abundance ratios of Fe/O have also been observed in several extremely metal-poor galaxies in the local Universe and it is suspected that the enrichment of Fe is produced by PISNe, HNe, or SNe of supermassive stars with stellar masses of $M_* \gtrsim 300~M_\odot$ \citep[e.g.,][]{kojima2021,isobe2022}.
The production of supermassive stars, on the other hand, might come from the stellar encounters in the dense nuclear environment of GN-z11 as suggested previously for dense globular clusters \citep[CGs;][]{gieles2018}.
In the local Universe, Fe is mainly enriched by Type-Ia SNe that has a typical delay time from 100 Myr to 1 Gyr.
However, we note that the first Type-Ia SNe can occur only $\sim 30$ Myr after the initial star formation \citep[see e.g.,][and references therein]{maiolino2019}.
According to the SED fitting of the JWST/NIRCam photometry of GN-z11 by \cite{tacchella2023}, the extended part of the source has a SFH with a rising SFR starting from a look-back time of $\sim 50$ Myr, which then shows a peak of SFR at a look-back time of $\sim 20$ Myr.
Therefore, it is still possible that the initial enrichment of Fe by Type-Ia SNe from the progenitors created in the rising phase of the past star formation starts to take effect at the time of observation.
However, such a scenario would also require suppression of oxygen enrichment by previous core-collapse supernovae (CCSNe), or a dilution of the oxygen abundance, which might be achieved through inflow of pristine gas.
}
Still, considering the mismatch between our photoionization models and data due to gaps in the models, it is important for future work to have more detailed modeling of the emission lines in the BLR in the early Universe.

Finally, a caveat of interpreting the strong high-ionization UV lines in GN-z11 as a sign of a high abundance is the optical-depth effect in BLRs.
As the density and ionization parameter in an BLR become too high, the emission lines start to be dominated by a Black-body component near the surface of the cloud and their ratios no longer reflect the actual chemical abundances \citep{temple2021}.
To check this effect, we computed a series of photoionization models for BLRs having a range of hydrogen densities and ionization parameters while adopting a low metallicity and N/O.
We found at $\log(Z/Z_{\odot}) = -1$ and log(N/O) = $-1.5$, both \niii/O\,{\sc iii}] and \niv/O\,{\sc iii}] can be greater than one when log(U) reaches 0 and the surface emission starts to dominate.
However, at such a high ionization parameter, \niv\ would be significantly stronger than \niii, which contradicts the similar fluxes of \niv\ and \niii\ observed in GN-z11.
Therefore, with a BLR origin for the UV nitrogen lines, we still need chemically enriched gas in order to match models with the observation.
%We will further explore abundance calibrations using high ionization lines excited by AGNs in a forthcoming paper.

\subsection{Potential implications of the Bowen emission}

%{\color{red} some discussion here about the interpretation of the Bowen emission if present?}

While in the previous section, we fitted the emission feature detected at $\sim 3\sigma$ at 3133 \AA\ with a redshifted peak within the UV \feii\ complex, it is also possible that the emission feature is the Bowen fluoresence line, \oiiip$\lambda 3133$.
This line is produced via fluorescence of \oiiip$\lambda 303.8$ with \heii\ Ly$\alpha$ \citep[i.e., Bowen fluorescence;][]{bowen1934}, and it
is usually the strongest Bowen line of \oiiip\ in the near UV.
Strong \oiiip$\lambda 3133$ emission has been observed in AGNs, planetary nebulae (PNs), symbiotic stars \citep{kastner1996}, and more recently, in a galaxy at $z=12.34$ \citep{castellano2024}.

The strength of \oiiip$\lambda 3133$ strongly depends on
the fraction of \heii\ Ly$\alpha$ photons that are resonantly absorbed by \oiiip$\lambda 303.8$ and converted to Bowen emission, or the ``Bowen efficiency'' \citep{kallman1980}.
Thus, the flux of the Bowen line is sensitive to the shape of the ionizing SED as well as the metallicity, which greatly impacts the resulting abundance ratio of $\rm He^{++}/O^{++}$ in the ionized gas.
The physical conditions of the BLR in GN-z11 could provide the environment that boosts the strength of the Bowen emission.
According to the photoionization calculations by \citet{netzer1985}, in dense PNs or BLRs of AGNs, the flux of \oiiip$\lambda 3133$ can be similar or even several times higher than that of \heii$\lambda 4686$.
While the JWST spectra of GN-z11 do not cover \heii$\lambda 4686$, we can use the flux of \heii$\lambda 1640$ to make a rough estimation.
This is motivated by the fact that the flux ratio of \heii$\lambda 1640$/\heii$\lambda 4686$ is typically $7-8$ in all kinds of nebulae.
According to the measurements by \citet{maiolino2023}, $F({\rm HeII}\lambda 1640) = 6.0^{+1.4}_{-1.3}\times 10^{-19}~{\rm erg/s/cm^{-2}}$.
Our measured flux of the emission line at 3133 \AA is \redtxt{$8.5\pm 2.7\times 10^{-20}~{\rm erg~s^{-1}~cm^{-2}}$}, which is roughly 1/7 of the flux of \heii$\lambda 1640$.
Therefore, the strength of the Bowen emission line is likely similar to that of \heii$\lambda 4686$ and is indeed what one would expect for a BLR origin.

To better understand the origin of this line, it is important to reexamine JWST-confirmed high-$z$ Type-1 AGNs with strong \heii$\lambda 4686$ emission.
Also, it would be useful to compute photoionization models with physical conditions appropriate for AGNs in early galaxies.
However, this is beyond the scope of the current paper, and we leave it for future investigations.

\subsection{Is the small blue bump unique?}

\begin{figure}
    \centering\includegraphics[width=0.47\textwidth]{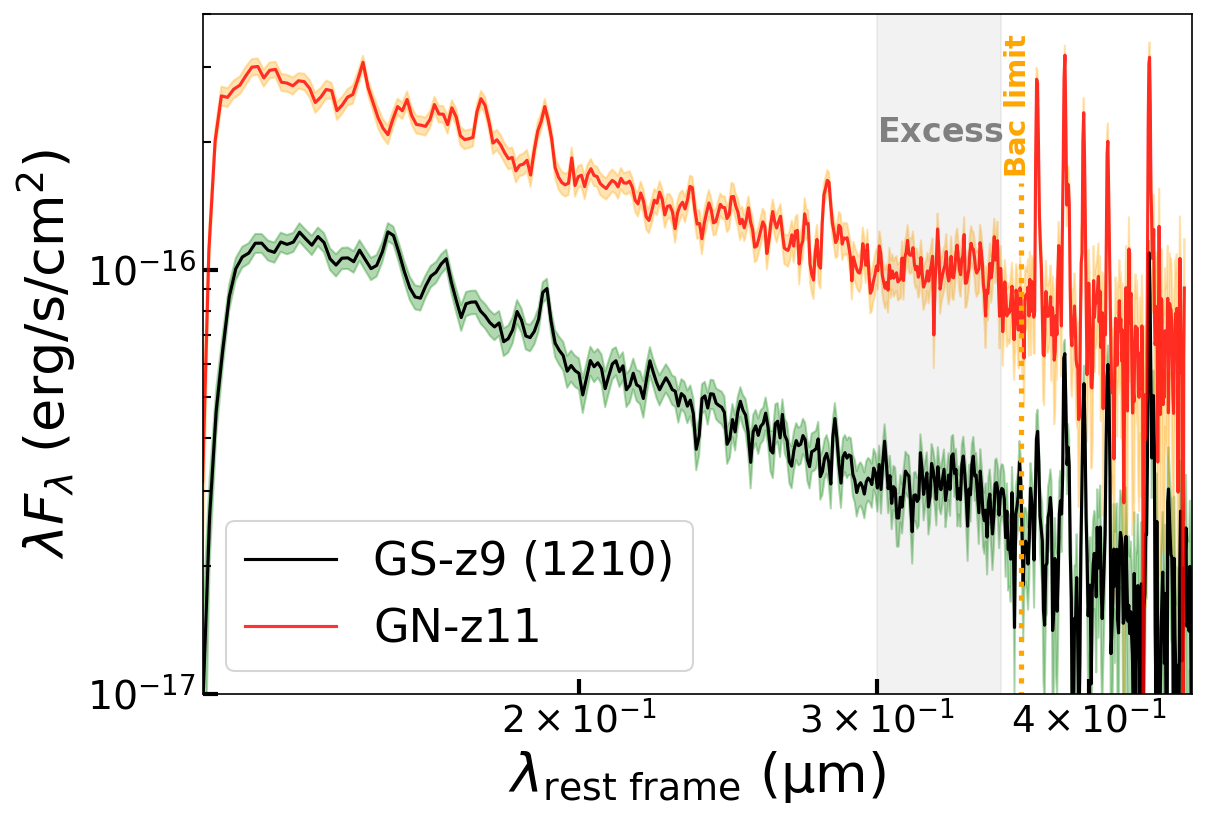}
    
    \caption{Comparison between the Prism spectrum of JADES-GS-z9-0 and that of GN-z11 on a log scale. The shaded regions correspond to the $1\sigma$ uncertainties of the spectra. The grey band marks the continuum excess region and the orange dotted line markes the location of the Balmer limit.
    }
    \label{fig:gsz9_gnz11_com.png}
\end{figure}

\begin{figure*}
    \centering\includegraphics[width=0.95\textwidth]{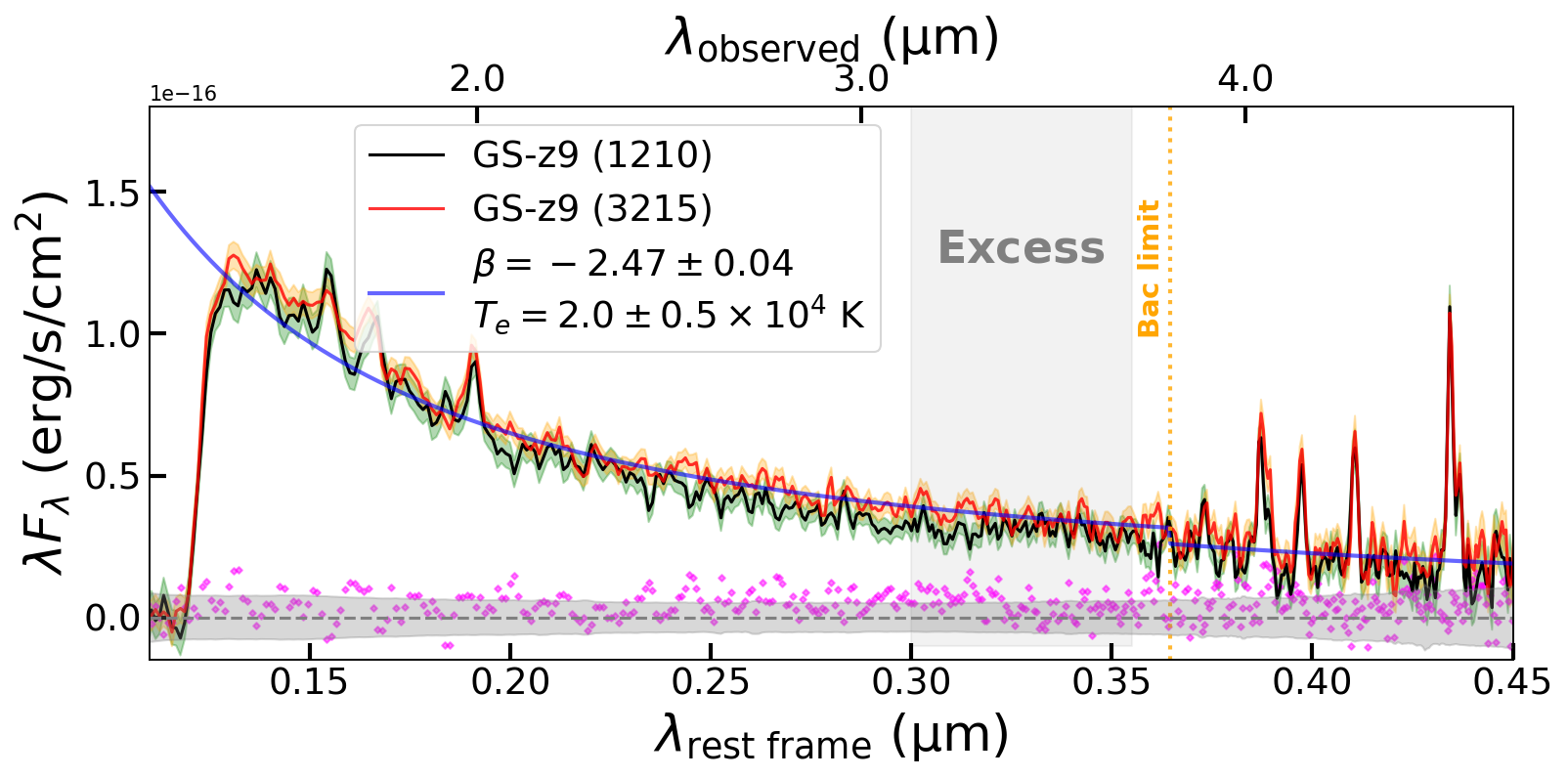}
    
    \caption{NIRSpec/Prism spectra of JADES-GS-z9-0 from the Cycle 1 Program 1210 and the Cycle 2 Program 3215.
    The two observations are separated by $\sim 1$ month in the rest frame of JADES-GS-z9-0.
    The shaded regions represent $1\sigma$ uncertainties of the spectra.
    The magenta diamonds are the differences between the fluxes the two spectra.
    The horizontal grey shaded band indicates the $1\sigma$ uncertainty of the spectral difference.
    The blue line is a fit to the continuum of the JADES-GS-z9-0 in 3215 using a power-law function and an optically thin Balmer continuum model.
    The vertical grey shaded band indicates the continuum excess region in 1210.
    The orange dotted line indicates the location of the Balmer limit at 3646 \AA.
    }
    \label{fig:gsz9_prism}
\end{figure*}

In previous sections, we discussed potential physical origins for the small blue bump in GN-z11.
Specifically, we investigated the scenarios where the bump is \redtxt{part of a Balmer continuum or} a blend of \feii\ lines.
One might wonder whether there is other observational evidence of similar bumps in the spectra of early galaxies.
Given the extreme physical conditions of GN-z11, it is possible that the small blue bump is a rare case.
Interestingly, we did find another galaxy in JADES that show a very similar continuum excess in the UV.
This galaxy, JADES-GS+53.11243-27.77461, has a redshift of 9.433 and was observed in the GOODS South field (hereafter JADES-GS-z9-0) in both Cycle 1 Program 1210 (P.I.: N. Luetzgendorf; \citealp{bunker_dr1_2024}) and Cycle 2 Program
3215 (P.I.: D. Eisenstein; \citealp{eisenstein2023b}).
Its nebular emission was analyzed by \citet{cameron_jwstism_2023},
and it was later classified as a Type-2 AGN by \cite{scholtz2023} based on UV emission-line diagnostics and the presence of the high ionization line [Ne\,{\sc iv}]$\lambda 2422$ (as well as the absence of broad emission lines), and more recently by \citet{curti_gsz9_2024} based on the detection of the coronal line [Ne\,{\sc v}]$\lambda 3426$.

Figure~\ref{fig:gsz9_gnz11_com.png} shows the Prism spectrum of JADES-GS-z9-0 from the Program 1210 along with the Prism spectrum of GN-z11.
One can see a bump within the continuum excess region in JADES-GS-z9-0, whose shape, however, is different from that in GN-z11 and does not exhibit a sharp jump at $\sim 3550$.
While the bump in JADES-GS-z9-0 cannot be well described by a Balmer continuum, \feii\ models cannot provide a good fit as well due to the smoothly declining edge of the bump.

Intriguingly, the small blue bump in JADES-GS-z9-0 disappears in the observation of the Program 3215.
Figure~\ref{fig:gsz9_prism} shows a comparison between the Prism spectrum in the Program 1210 and that in the Program 3215 for JADES-GS-z9-0.
While the overall UV continuum rises in 3215 (as shown by the overall positive residuals), the UV slope slightly flattens.
By fitting power-law functions to the UV continua of the two spectra, we obtained a slope of $\beta = -2.47\pm 0.04$ for the 3215 spectrum when including the Balmer continuum, and $\beta = -2.35\pm 0.03$ when excluding the Balmer continuum.
In comparison, the UV slope of the 1210 spectrum is $\beta = -2.58\pm 0.03$ if the bump region is excluded.
We checked the shutter positions of the two observations and found the positioning is unlikely the cause to the difference of the UV flux.
{Also, the two Prism spectra were obtained with the same data reduction procedure}.

In the rest frame of JADES-GS-z9-0, the two observations are separated by $\sim 1$ month and we speculate this systematic change in the UV is indicative of spectral variability in the AGN.
While JADES-GS-z9-0 was a Type-2 AGN classified from the previous observations, we might be witnessing the rise of of the underlying UV continuum from the AGN accretion disk, a process found in ``changing-look quasars'' in lower redshifts \citep{macleod2019}.
The small blue bump either has responded to the change of the UV continuum in 3215, or is simply overwhelmed by the rising continuum.
According to the standard BLR model with a stratified ionization structure, emission from different ionic species respond to the continuum variability differently \citep{ulrich1997}.
Therefore, the variability of the small blue bump could provide clues on its origin, which requires follow-up observations of this target.
In fact, the UV continuum of JADES-GS-z9-0 in 3215 appears more smooth and can be described by a power-law function and an optically thin Balmer continuum with $T_e = 2.0\pm 0.5\times 10^4$ K, as shown in Figure~\ref{fig:gsz9_prism}.
The best-fit Balmer temperature is consistent with the temperature inferred from \oiii$\lambda 4363$ as well as the temperature inferred from O\,{\sc iii}]$\lambda 1666$ \citep{curti_gsz9_2024}.

In summary, the Prism spectrum JADES-GS-z9-0 shows a small blue bump at a similar spectral location as that in GN-z11, although with a slightly different shape.
The bump in JADES-GS-z9-0 seems to be linked to its potential spectral variability.
Future observations of the UV spectrum of JADES-GS-z9-0 is needed to verify the variability and provide more evidence on the identify of the small blue bump.

\section{Conclusions}
\label{sec:conclude}

In this work we investigated the physical origin of the continuum excess between $3000$ \AA\ and $3550$ \AA in the JWST/NIRSpec Prism spectrum of GN-z11.
The continuum excess is significantly detected in the spectrum at $7.8\sigma$ and shows a sharp jump around $3550$ \AA.
This excess in the UV partly resembles the shape of the small blue bump observed in typical quasar spectra, typically associated with \feii\ complex emission from the Broad Line Region in the vicinity of the accreting supermassive black hole.
\redtxt{We investigated the origin of the continuum excess by fitting the observed spectrum with a combination of AGN/SF continuum models, nebular continuum models, and \feii\ emission models.}
%Since GN-z11 is also hosting an AGN, we investigated whether the continuum excess has the same origin as the small blue bump by fitting the continuum excess using a combination of a Balmer continuum and \feii\ emission. 
We summarize our findings in the following.
\begin{itemize}
    \item \redtxt{If the UV continuum of GN-z11 is dominated by a stellar continuum, the continuum excess can be fitted by a Balmer continuum with a temperature of $T_e=1.78^{+0.25}_{-0.21}\times 10^4$ K, which is marginally higher than the temperature derived from the optical \oiii\ lines of $T_e=1.36\pm 0.13\times 10^4$ K.}
    \item \redtxt{If the UV continuum of GN-z11 is dominated by AGN continuum emission,} a Balmer continuum alone cannot fit the continuum excess due to the significantly blueshifted jump of the continuum excess with respect to the Balmer limit.
    This indicates that the continuum excess in GN-z11 is not dominated by the Balmer continuum and that the Balmer continuum, if exists, is intrinsically weak.
    %Additionally, both an optically thin Balmer continuum and a partially optically thick Balmer continuum cannot recover the shape of the continuum excess.
    %\item {\color{red} Discuss here that the weakness of the Balmer continuum implies high temperatures, inconsistent with the forbidden lines, hence Balmer emission coming from the dense-hot BLR, while forbidden lines coming from less dense-cooler NLR or ISM}
    \item The weakness of the Balmer continuum \redtxt{in the AGN case} implies part of the Balmer emission should be associated with hot gas at a temperature of $T_e = 2.6^{+0.4}_{-0.6}\times 10^4$ K, which is much higher than the temperature traced by forbidden lines.
    Hence the Balmer emission likely originates in the dense and hot part of the BLR, while forbidden lines are from the less dense and cool NLR or ISM. We note that, due to the small black hole mass in GN-z11, it is difficult to kinematically distinguish the two components in integrated spectra, as they have similar widths, but they may be separated in spatially resolved spectroscopy.
    \item While the \feii\ emission is typically stronger at $\lambda < 3000$ \AA, we explored a series of \textsc{Cloudy} models and found the \feii\ emission at $3000 ~\AA < \lambda < 3550 ~\AA$ can be significantly enhanced under certain conditions.
    Specifically, our fiducial models that show a prominent \feii\ complex within the continuum excess region have $n_{\rm H} \sim 10^{11}~{\rm cm^{-3}}$, $N_{\rm H} \sim 10^{25}~{\rm cm^{-2}}$, \redtxt{$v_{\rm turb}\sim 300$ km/s, $\rm log(U)\sim -2.0$}, and log(Z/Z$_{\odot}$) $\gtrsim -0.5$ or [Fe/H] $\gtrsim -0.5$.
    These \feii\ models show a prominent jump around $3500$ \AA, which can match the observed jump if the \feii\ emitting clouds are infalling at a velocity of $\sim 3000$ km/s.
    Still, the \feii\ models fail to reproduce the middle part of the jump due to the gaps in the models.
    Continuum/emission-line florescence not taken into accounted by current models might lead to this mismatch.
    Future modeling efforts focusing on the comparisons between models and observations between $3000$ \AA\ and $3550$ \AA\ are needed to fully solve this problem.
    \item The redshifted \feii\ emission was also observed previously in low-redshift quasars, with the highest velocity shift reaching 3000 km/s.
    A potential explanation is that \feii\ emission in BLRs is usually associated with infalling clouds with large column densities viewed from the shielded faces.
    The minimum column density required to drive the inflow is $\sim 7.5\times 10^{24}~{\rm cm^{-2}}$, consistent with the \feii\ model we adopted and the likely Compton-thick nature of the BLR in GN-z11, and as inferred for several other AGN at high-z discovered by JWST \citep{Maiolino2024_Xrays}.
    %In addition, the potential emission feature at 3133 \AA\ can be explained by the \feii\ emission as well.
    \item The density of the \feii\ models is consistent with the density inferred from high ionization lines, indicating their common origins in the BLR. On the other hand, the seemingly enriched Fe abundance could either be due to the fast enrichment of the overall chemical abundances within the small BLR, or result from a high Fe/O enriched by Type-Ia supernovae or massive HNe/PISNe in the nuclear region.
    %This enriched Fe abundance is also compatible with the enriched N abundance inferred from high ionization UV lines, if their lines come from the same BLR.
    %\item 
    % With the addition of \feii\ models, the contribution from the Balmer continuum is suppressed, corresponding to an optically thin Balmer continuum models with $T_e \sim 3\times 10^4$ K or a partially optically thick Balmer continuum with a lower temperature.
    % In both cases, the Balmer continuum would arise from the BLR, despite not necessarily from the same ionized zone that produces \feii\ emission.
    %Typical ISM surrounding young massive stars can neither produce prominent \feii\ emission nor a significantly suppressed Balmer continuum.
    %Thus, the existence of the continuum excess further supports the UV spectrum of GN-z11 is dominated by AGN ionization.
    \item We found a continuum bump at a similar spectral range in another AGN found in JADES, JADES-GS-z9-0. However, the continuum excess in JADES-GS-z9-0 has a slightly different shape.
    Interestingly, the UV spectrum of JADES-GS-z9-0 shows variability between observations at two different epochs, with the continuum excess disappear under the rising continuum at the later epoch.
    We speculate the varying continuum is driven by a changing-look AGN.
    Whether and how the continuum excess responds to the continuum variability could provide further clues on its origin.
    It is thus important to have follow-up monitoring of the target to verify the variability as well as characterizing its pattern.
\end{itemize}

We conclude by emphasizing that the presence of this small blue bump in GN-z11 and its peculiar properties demonstrates the need to reexamine previous theoretical models over a less frequently explored spectral regime, using typically subdominant spectral features in lower redshifts, and in an entirely new parameter space.
In the meantime, new observational evidence from higher S/N spectra, especially at high spectral resolution and high spatial resolution with the IFU (e.g., $R\sim 2700$ IFU spectroscopy with NIRSpec), are needed to verify the current interpretations on GN-z11, especially the presence of distinct ionized regions and its exotic chemical abundance pattern in the very early Universe.

\section*{Acknowledgements}
\redtxt{We thank the anonymous referee, whose thoughtful comments and suggestions improved the clarity of this work.}
XJ, RM, FDE, JS, and JW acknowledge ERC Advanced Grant 695671 “QUENCH” and support by the Science and Technology Facilities Council (STFC) and by the UKRI Frontier Research grant RISEandFALL.
RM acknowledges funding from a research professorship from the Royal Society.
JC acknowledges funding from the ``FirstGalaxies'' Advanced Grant from the European Research Council (ERC) under the European Union’s Horizon 2020 research and innovation programme (Grant agreement No. 789056).
ECL acknowledges support of an STFC Webb Fellowship (ST/W001438/1).
BER acknowledges support from the NIRCam Science Team contract to the University of Arizona, NAS5-02015, and JWST Program 3215.
B.R.P. acknowledges support from the research project PID2021-127718NB-I00 of the Spanish Ministry of Science and Innovation/State Agency of Research (MICIN/AEI/10.13039/501100011033).
ST acknowledges support by the Royal Society Research Grant G125142.
The research of CCW is supported by NOIRLab, which is managed by the Association of Universities for Research in Astronomy (AURA) under a cooperative agreement with the National Science Foundation.

\section*{Data Availability}

The JWST/NIRSpec data used in this paper are available through the MAST portal.
\rrtxt{The reduced Prism spectrum of GN-z11 we used for our main analyses can be accessed using this \href{https://zenodo.org/records/15682467?token=eyJhbGciOiJIUzUxMiJ9.eyJpZCI6IjkwYzE3MDIyLTI1ODAtNDBhYS1iZTZmLTU3MjViMjg4N2I4MyIsImRhdGEiOnt9LCJyYW5kb20iOiIwYjQ1NGQ5MjE4NWRlYzE1MTlhMDhjYTRiMTdhOTIxNiJ9.VzO50HVFiYmA48OxCsqHnl7uqxc3ilCORgensZeRzKbyCKkmd1x9a9VnDuLd7JsFtKQEUHZ06rUp0JM6qXr5hg}{link}.
}
\rrtxt{The reduced Prism spectra for individual observations of GN-z11 listed in Table~\ref{tab:obser} are available from the JADES} \href{https://jades-survey.github.io/scientists/data.html#jades-data-release-3}{DR3}.
All analysis results of this paper will be shared on reasonable request to the corresponding author.

\begin{appendix}

\end{appendix}

% WARNING
%-------------------------------------------------------------------
% Please note that we have included the references to the file aa.dem in
% order to compile it, but we ask you to:
%
% - use BibTeX with the regular commands:
   \bibliographystyle{mnras} % style aa.bst
   \bibliography{ref} % your references ref.bib
%
% - join the .bib files when you upload your source files
%-------------------------------------------------------------------

\appendix

%\section{Bump detection over different visits of observations}
\section{Detection of the small blue bump in Jackknifed Prism spectra}
\label{appendix:a}

To check whether the small blue bump in the Prism spectrum of GN-z11 is caused by any detector artefact during observations, we Jackknifed the Prism spectra from independent measurements.

The May observation of GN-z11 has 7 unique MSA positions and a total of 9 exposures.
We considered exposures with the same shutter position as an independent trial, which results in 7 trials for the May observation.
Similarly, we identified 12 trials from a total of 24 exposures in February.
In total, we have 19 measurements for Jackknifing.

\begin{figure*}
    \centering\includegraphics[width=0.97\textwidth]{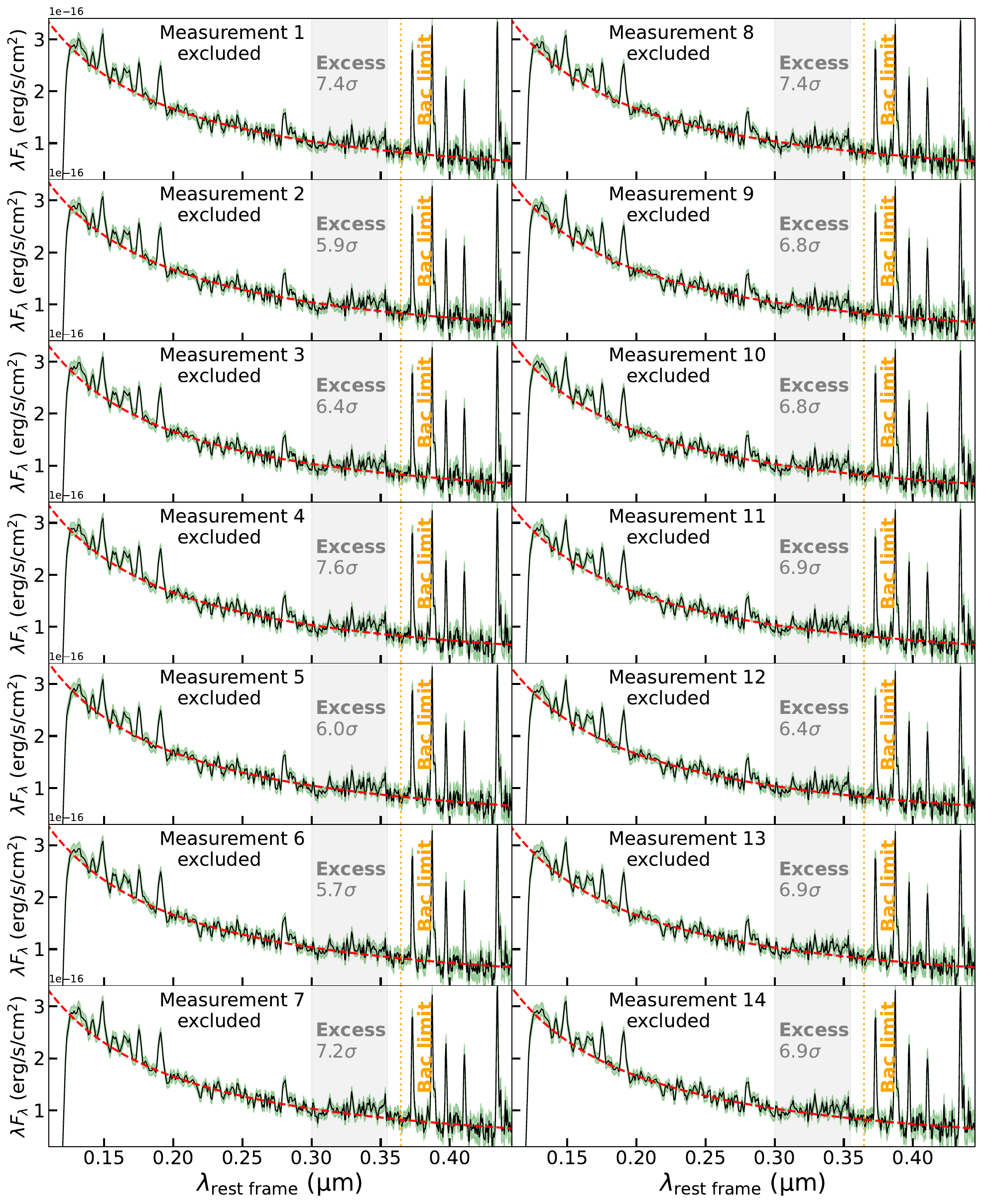}
    
    \caption{Jackknifed Prism spectra of GN-z11 generated from a total of 19 measurements. For each panel, one of the measurements is excluded and the rest of the measurements is combined via an inverse-variance weighted method with 3-$\sigma$ clipping.
    The green shaded regions indicate the combined $1\sigma$ uncertainties.
    \redtxt{In each panel, the dashed red line is the best-fit power law for the continuum excluding the excess region.}
    The grey shaded region marks the continuum excess region identified in the final spectrum combining all measurements.
    \redtxt{The significance of the integrated flux excess in the grey shaded region with respect to the power-law continuum is also shown.}
    The orange dotted line marks the location of the Balmer continuum limit.
    The first 14 Jackknifed spectra are shown.
    }
    \label{fig:jack_prism0}
\end{figure*}

\begin{figure*}
    \centering\includegraphics[width=0.8\textwidth]{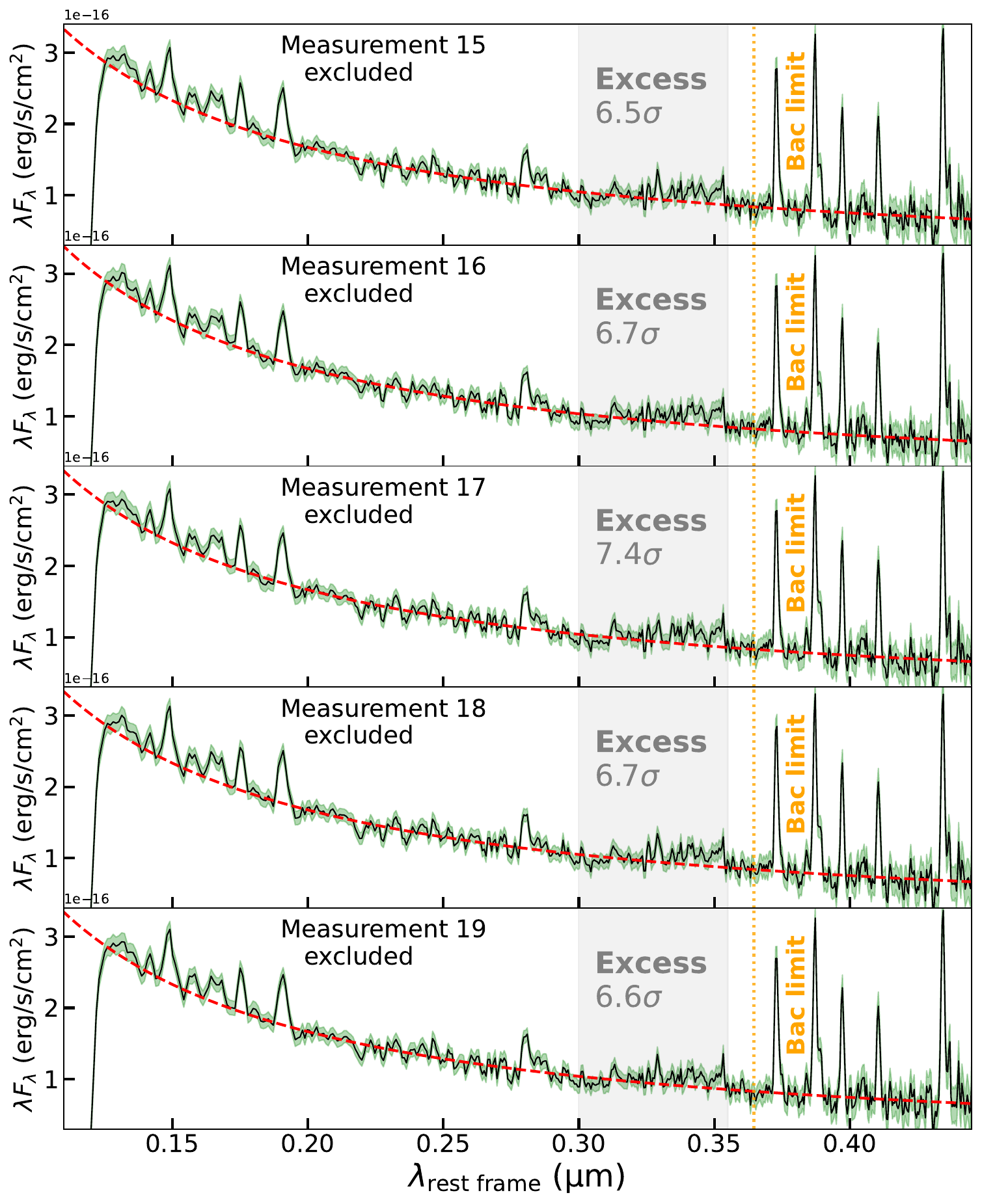}
    
    \caption{Jackknifed Prism spectra of GN-z11 generated from a total of 19 measurements. The remaining 5 Jackknifed spctra are shown.}
    \label{fig:jack_prism1}
\end{figure*}

Figures~\ref{fig:jack_prism0} and \ref{fig:jack_prism1} show the 19 Jackknifed spectra from the measurements.
Each spectrum was obtained by excluding one measurement and combining the rest of the measurements through inverse-variance weighting and 3-$\sigma$ clipping.
It is clear that the continuum excess in the near-UV as well as the jump near 3550 \AA\ persists in all Jackknifed spectra \redtxt{and is significant}, confirming it is not an artefact driven by a single measurement.
Meanwhile, the emission at 3133 \AA\ is also visible in all the spectra.
In contrast, a potential emission line at $\sim 3300$ \AA\ seemed disappeared when measurement 16 was excluded, meaning it is mainly from this measurement.

%\redtxt{[TBD: show Jackkinifed spectra from different visits if there are enough visits (no?); or simply present the two observations?]}

%Figure~\ref{fig:twoobs_gnz11} shows 1D spectra of GN-z11 obtained from two different observations within Program 1181. The continuum excess is clearly present in two spectra.

%\section{Optically thick Balmer jump case}
\section{Fits with alternative continuum models}
\label{appendix:Balmer_optthick}

\subsection{Power-law continuum model}

\begin{figure*}
    \centering\includegraphics[width=\columnwidth]{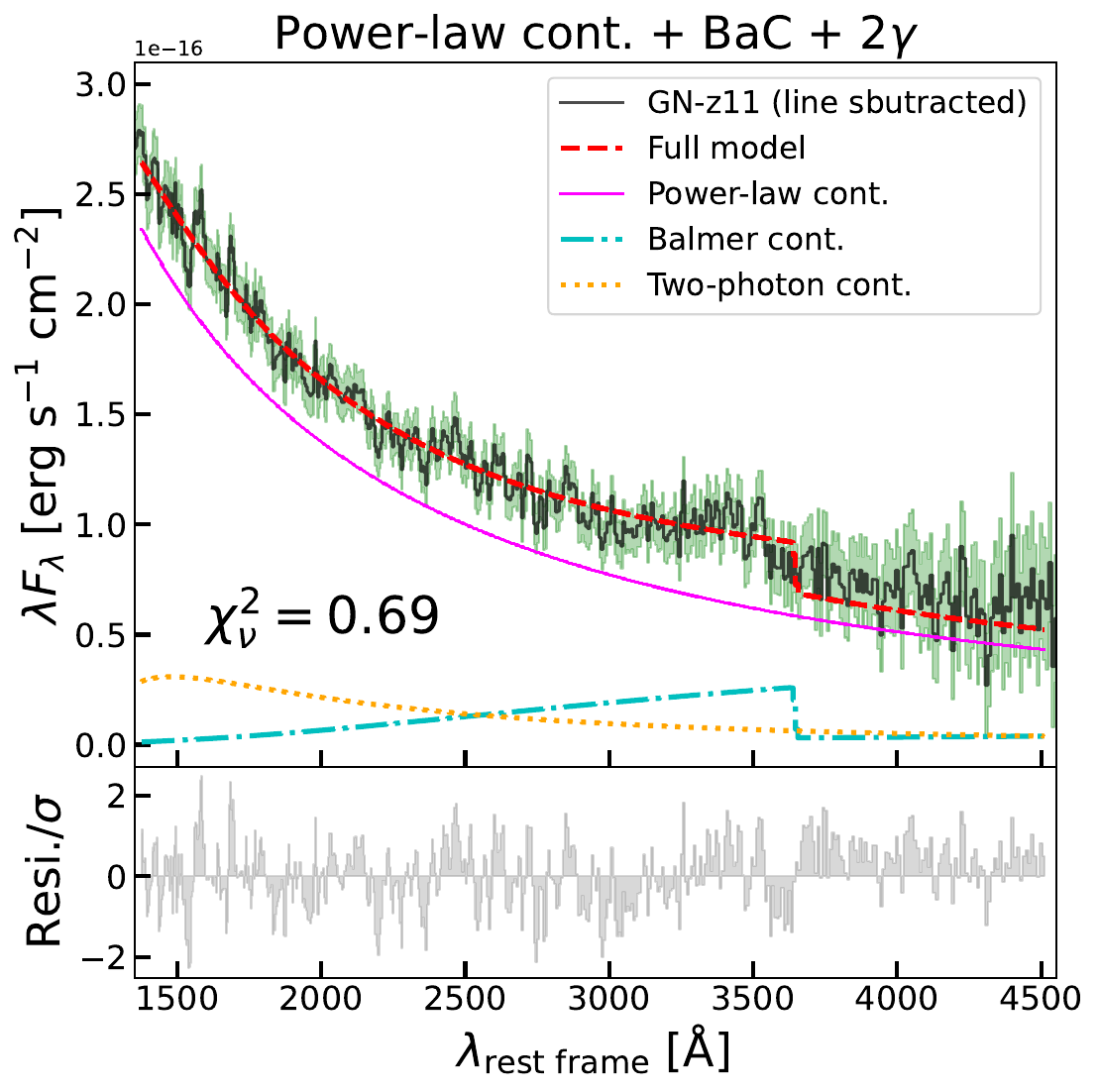}
    \includegraphics[width=\columnwidth]{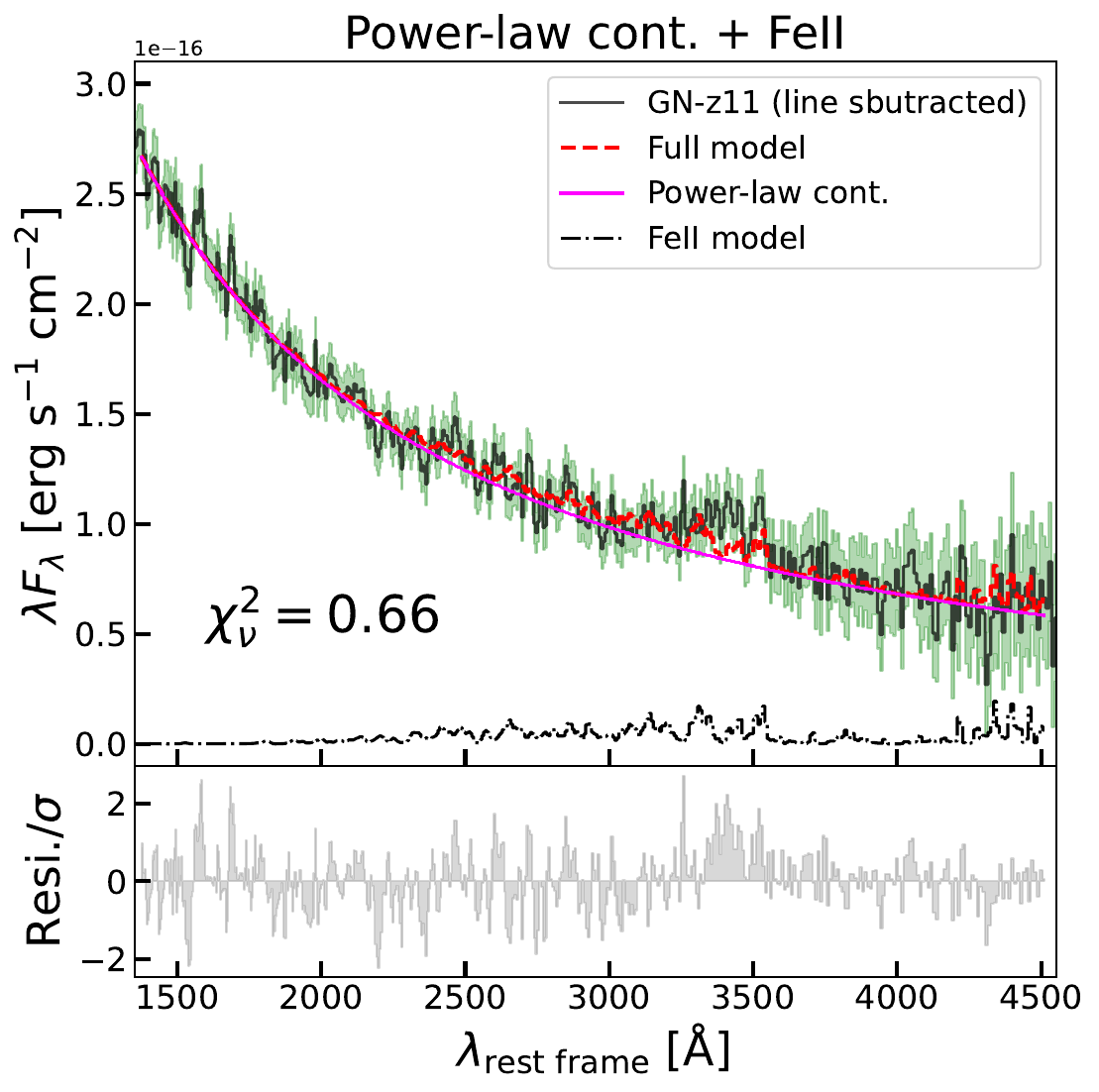}
    
    \caption{Best-fit models for the line-subtracted Prism spectrum of GN-z11, where the continuum is assumed to be a power law. 
    The top panels show the models with the reduced $\chi^2$. The bottom panels show the $\chi$ residual of the fit.
    \textit{Left:} fit combining a power-law continuum and nebular continuum.
    \textit{Right:} fit combining a power-law continuum and \feii\ emission.
    The fit involving \feii\ produces a slightly smaller reduced $\chi^2$.
    }
    \label{fig:pl_eg0}
\end{figure*}

\redtxt{In this Appendix we describe the fitting results by assuming the continuum of GN-z11 is a power law, without assuming any physical origin as we did in the main text.
Figure~\ref{fig:pl_eg0} shows the fitting results from two models, where we used nebular continuum and \feii\ emission to describe the continuum excess, respectively.
The nebular continuum model and the \feii\ model are described in Section~\ref{sec:method}.
The fit using the nebular continuum model yields a best-fit power-law slope of $\beta=-2.43\pm 0.02$ and a best-fit electron temperature of $T_e=1.61^{+0.26}_{-0.20}\times 10^4$ K, close to the temperature estimate of the stellar-continuum fit ($T_e=1.78^{+0.25}_{-0.21}\times 10^4$ K).
The fit using the \feii\ model yields a best-fit power-law slope of $\beta=-2.28\pm 0.02$.
Overall, the reduced $\chi^2$ is slightly smaller in the \feii\ fit.
}

\subsection{Optically thick Balmer continuum model}

In this Appendix we describe an attempt to fit the small blue bump by assuming a partially optically thick Balmer continuum, which includes an additional free parameter, the optical depth at the Balmer edge, $\tau _{\rm BC}$.
This model was initially proposed by \cite{grandi1982} to describe line-emitting clouds that are heated towards large optical depths in quasars.
\redtxt{Therefore, we considered the continuum to be dominated by an AGN SED from \citet{pezzulli_agn_2017} in this case (see Section~\ref{sec:method}).
}

\begin{figure}
    \centering\includegraphics[width=\columnwidth]{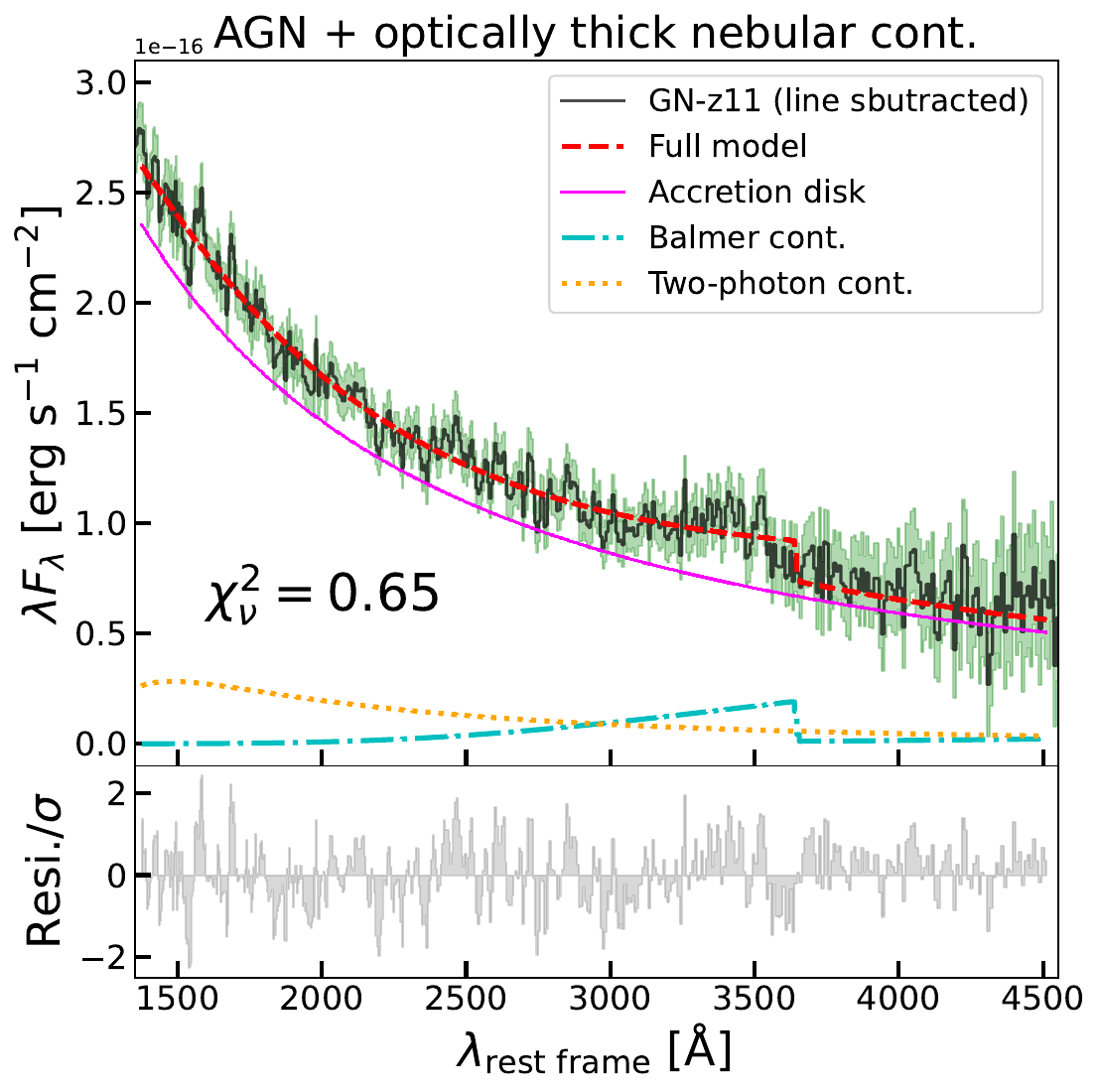}

    \caption{As Figure~\ref{fig:gnz11_bcfit} but where the Balmer continuum has been replaced by the partially optically thick case given by Equation~\ref{eq:thick}.
    The top panel shows the best-fit models.
    The bottom panel shows the $\chi$ residual of the fit.
    %In this case the optical depth and the normalization of the Balmer edge were jointly fitted.
    %The cyan line and the magenta line represent models with $T_e = 5000$ K and $T_e = 30,000$ K, respectively.
    }
    \label{fig:gnz11_bcfit_optthick}
\end{figure}

A partially optically thick Balmer continuum is increasingly suppressed at longer wavelengths due to the cubic dependence of the absorption cross section on the wavelength.
\redtxt{We adopted the functional form proposed by \citet{grandi1982}
 \begin{equation}
        F_{\lambda ; \rm thick} = F_{\lambda}^{\rm BE} \frac{B_{\lambda}(T_e)(1-e^{-\tau_{\lambda}})}{B_{\lambda}^{\rm BE}(T_e)(1-e^{-\tau_{\rm BE}})} ~{\rm ;}~\lambda \leq \lambda _{\rm BE},
        \label{eq:thick}
    \end{equation}
where $F_{\lambda}^{\rm BE}$ is the flux at the Balmer edge, $B_{\lambda}(T_e)$ is the Planck function, and $\tau _{\lambda} = \tau _{\rm BE} (\lambda/\lambda _{\rm BE}) ^3$ is the optical depth of the Balmer continuum at a given wavelength.
The optically thick Balmer continuum is related to the optically thin one by
\begin{equation}
    F_{\lambda ; \rm thick} = F_{\lambda ; \rm thin}\left(\frac{\lambda _{\rm BE}}{\lambda}\right)^3 \left[\frac{e^{-\tau_{\rm BE}}-e^{-(\tau_{\lambda}+\tau_{\rm BE})}}{1-e^{-\tau_{\rm BE}}}\right],
\end{equation}
and it is reduced to the optically thin case when $\tau_{\rm BE}\rightarrow0$.
}
Figure~\ref{fig:gnz11_bcfit_optthick} shows the best-fit model with \redtxt{$T_e = 8.2^{+1.2}_{-1.0}\times 10^3$ K and $\tau _{\rm BE} = 0.65^{+0.10}_{-0.09}$.}
\redtxt{The temperature is significantly lower than the forbidden-line temperature inferred from \oiii.
If we interpret this as the optically thick Balmer continuum arising from dense BLR gas,
the BLR gas needs to be colder than the low-density gas that emits \oiii.
However, from Appendix~\ref{appendix:cloudy_tem}, such a low temperature is implausible at a moderate ionization parameter of $\log(U)\gtrsim -3$ and a metallicity of $Z/Z_\odot \lesssim 1$.
}
%While the model has a similar shape to that of the model in the optically thin limit, the best-fit temperature is lower.
\redtxt{The reduced $\chi^2$ of the fit is significantly smaller compared to the optically thin case shown in Figure~\ref{fig:gnz11_bcfit}.}
%and it is atypical for photoionized gas.
%This is due the normalization of the Balmer edge is set as a free parameter.
%There exists a degeneracy between the normalization of the Balmer edge and the optical depth.
%In practice, the temperature is usually fixed to a typical value when adopting this flexible model \citep{grandi1982}.
%It is clear that even with a partially black-body component and (possibly unphysical) freedom to vary the normalization of the Balmer continuum, the Balmer continuum cannot fully recover the shape of the small blue bump.
\redtxt{Still, one caveat of this model is the treatment of the two-photon continuum. If the two-photon continuum has the same origin as the Balmer continuum (i.e., from the BLR), it should be significantly suppressed.
It is also possible that the two-photon continuum comes from the low-density gas outside the BLR.
In the latter case, however, the normalization of the two-photon continuum becomes uncertain.
In the next Appendix, we discuss fitting results with a suppressed two-photon continuum.
}

\begin{comment}
\begin{table*}
        \centering
        \caption{Best-fit model parameters for the GN-z11 continuum}
        \label{tab:con_params}
        \begin{tabular}{l c c c}
            \hline
                \hline
                Assumption & Power law (PL) & Balmer continuum (BC) & \feii\ emission (FeII) \\
                \hline
                PL + BC (optically thin) & $\beta = -2.37\pm 0.02$ & $T_e = 1.69 ^{+2.9}_{-2.3}\times 10^4~{\rm K}$ & N/A \\
                \hline
                PL + BC & $\beta = -2.27\pm 0.02$ & $T_e = 5.2 ^{+1.3}_{-1.0}\times 10^3~{\rm K}$ & N/A \\
                (partially optically thick) & & $\tau _{\rm BE} = 4.4^{+3.7}_{-2.7}$ & \\
                \hline
                PL + BC (suppressed) + FeII & $\beta = -2.27\pm 0.02$ & $T_e = 2.6^{+0.4}_{-0.6}\times 10^4$ K & log(Z/Z$_{\odot}$) $\gtrsim -0.5$ \\
                & & & $v_{\rm Fe^+} = 3\times 10^3$ km/s \\
                \hline
        \end{tabular}
\end{table*}
\end{comment}

\subsection{High-density nebular continuum models}

\begin{figure*}
    \centering\includegraphics[width=\columnwidth]{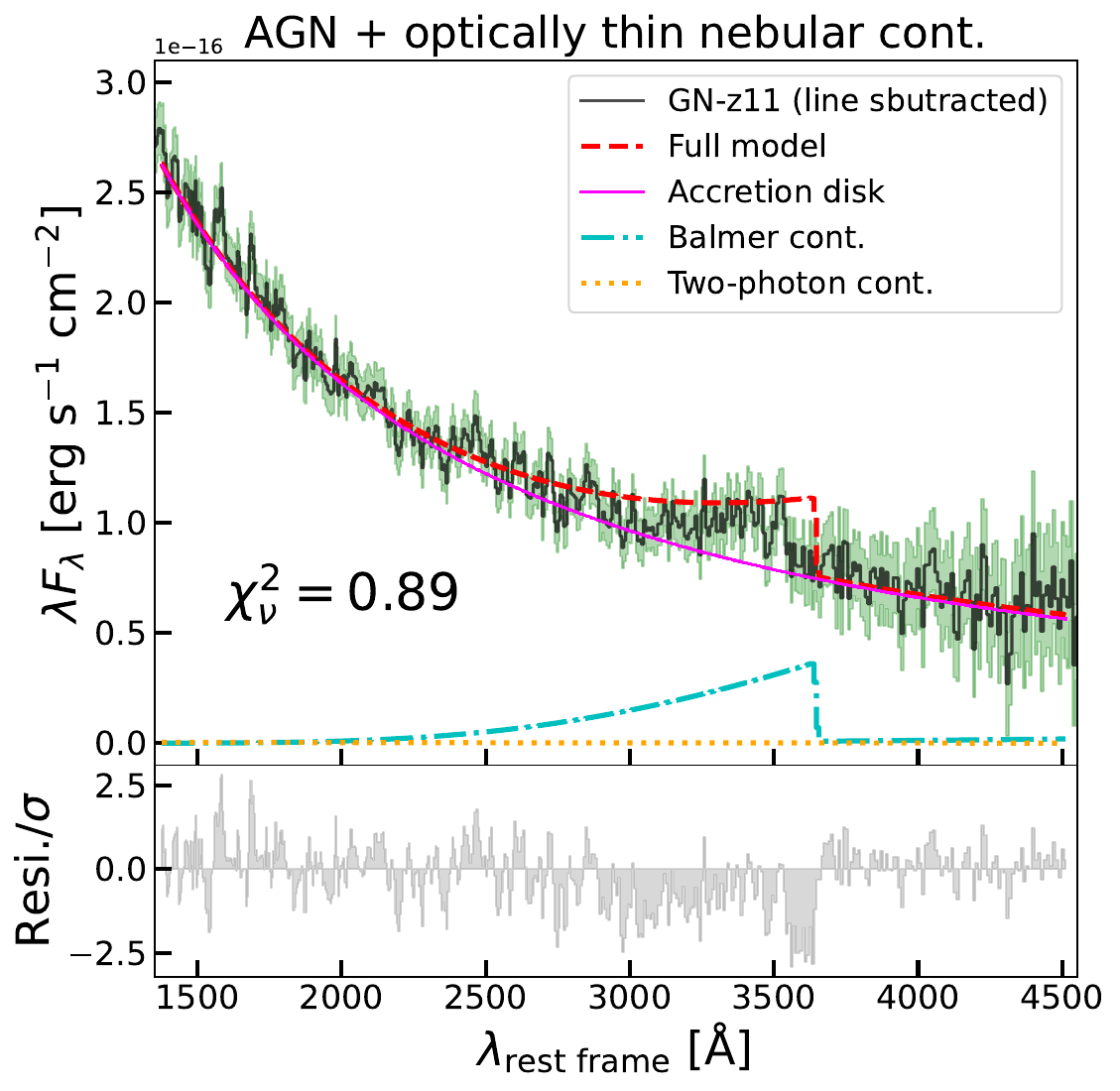}
    \includegraphics[width=\columnwidth]{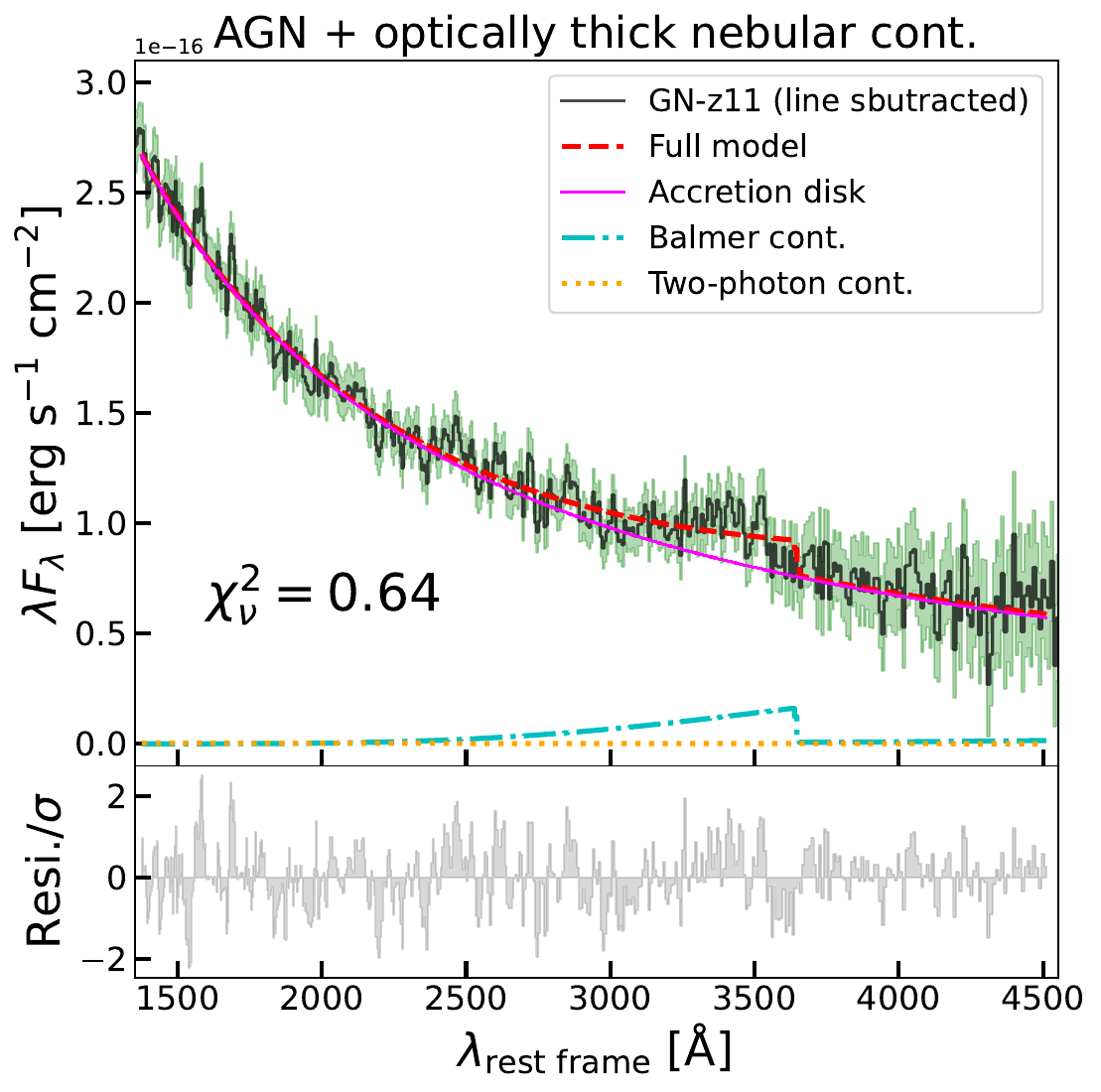}
    
    \caption{Best-fit models for the line-subtracted Prism spectrum of GN-z11, where the two-photon continuum is suppressed due to the high density. 
    The top panels show the models with the reduced $\chi^2$. The bottom panels show the $\chi$ residual of the fit.
    \textit{Left:} fit combining an AGN SED and optically thin nebular continuum.
    \textit{Right:} fit combining an AGN SED and optically thick nebular continuum.
    }
    \label{fig:no2g_eg0}
\end{figure*}

\redtxt{In this Appendix we present fitting results based on nebular continuum models assuming a high gas density of $n_{\rm H}=10^6~{\rm cm^{-3}}$, which strongly suppresses the two-photon continuum.
The condition would be satisfied if the nebular continuum in GN-z11 predominantly comes from high-density gas, which is possible for the AGN scenario where there is a BLR component.
Therefore, we investigate fits assuming an AGN dominated continuum (see Section~\ref{sec:method}) specifically.
Figure~\ref{fig:no2g_eg0} shows the fitting results for the optically thin nebular continuum case and the optically thick nebular continuum case, respectively.
In both models, it is clear that the two-photon continuum is strongly suppressed due to the high density.
The optically thin model produces a worse fit compared to the case with a non-suppressed two-photon continuum as shown in Figure~\ref{fig:gnz11_bcfit} and has a lower best-fit temperature of $\sim 7.5\times 10^3$ K as reflected by the stronger jump.
The optically thick model produces a much smaller reduced $\chi^2$, although the best-fit temperature is even lower, reaching $\sim 6.5\times 10^3$ K, which is unlikely to be found in highly ionized gas as shown in Appendix~\ref{appendix:cloudy_tem}.
}

\section{\feii\ jump in quasar spectra}
\label{appendix:feii_obs}

In this Appendix we show a few more examples of quasar spectra that exhibit UV jumps blueward of the Balmer limit caused by \feii\ emission.
These quasars were selected from a sample of luminous quasars at $4.5<z<5.3$ in the XQz5 catalog compiled by \citet{lai2024}.
The spectra were obtained with X-shooter \citep{vernet2011} at the Very Large Telescope (VLT).
These quasars have prominent UV continuum emission and their Eddington ratios are on average 0.3 dex higher than the lower-redshift quasars in SDSS with similar black hole masses \citep{wu2022,lai2024}.

\begin{figure*}
    \centering\includegraphics[width=0.47\textwidth]{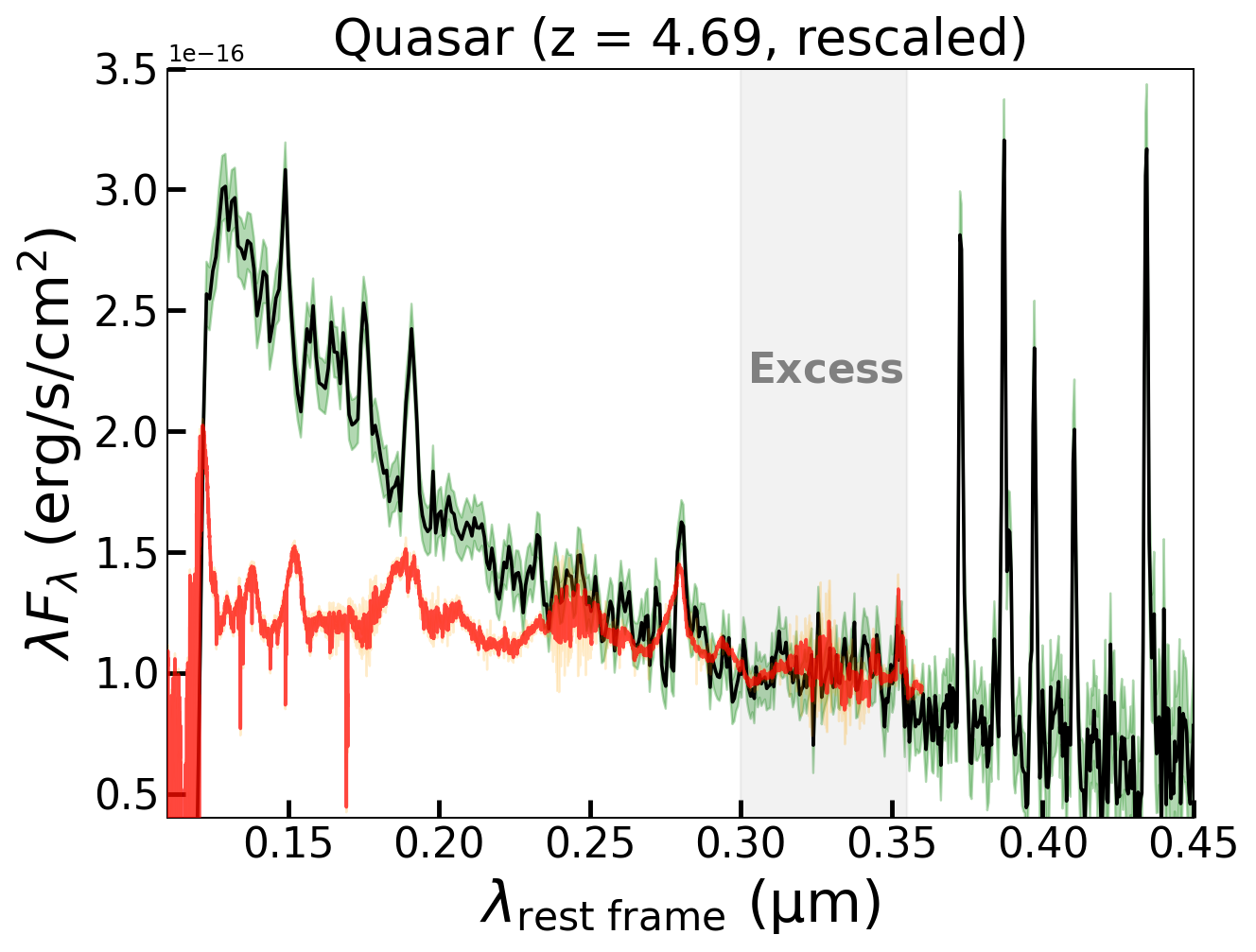}
    \includegraphics[width=0.47\textwidth]{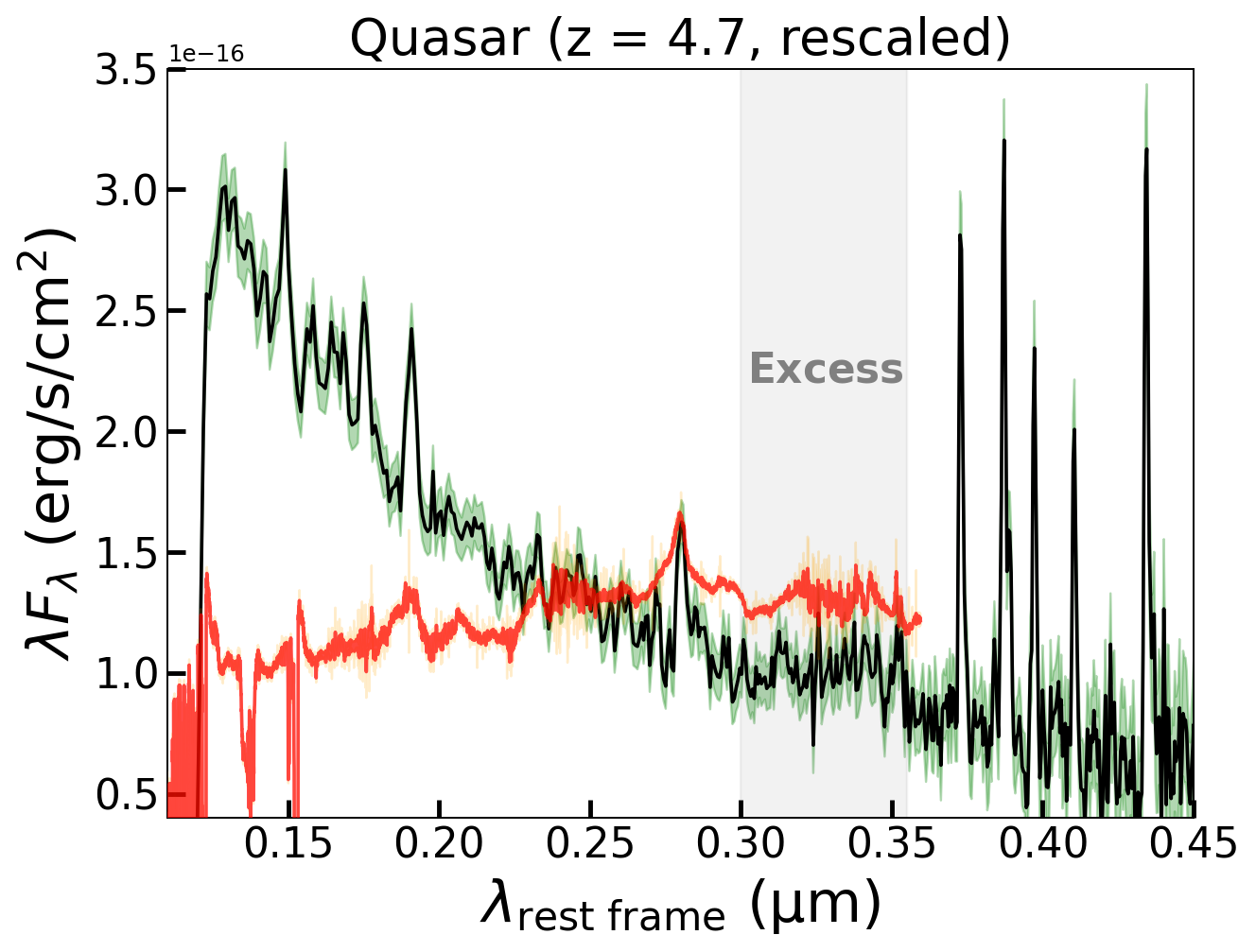}
    \includegraphics[width=0.47\textwidth]{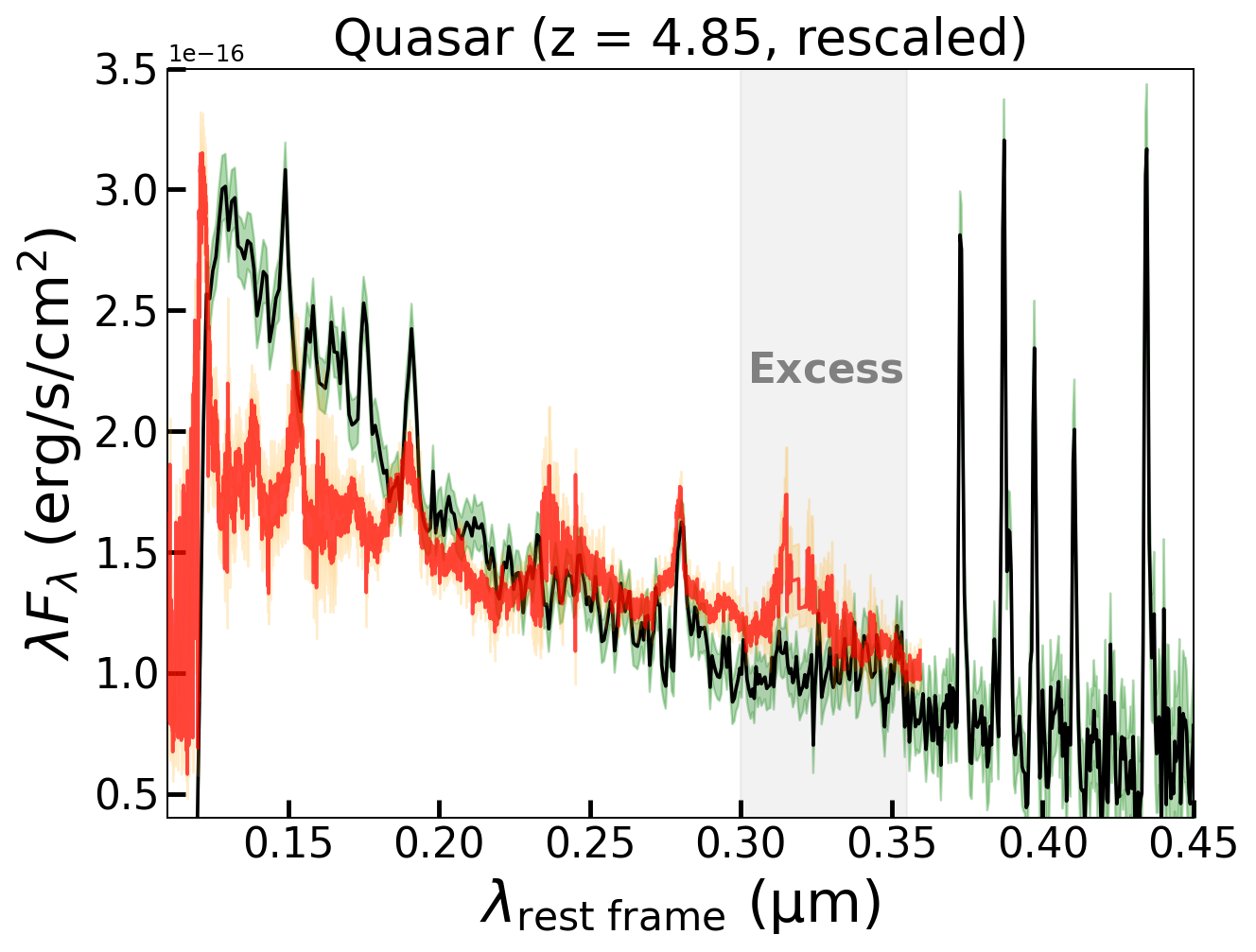}
    \includegraphics[width=0.47\textwidth]{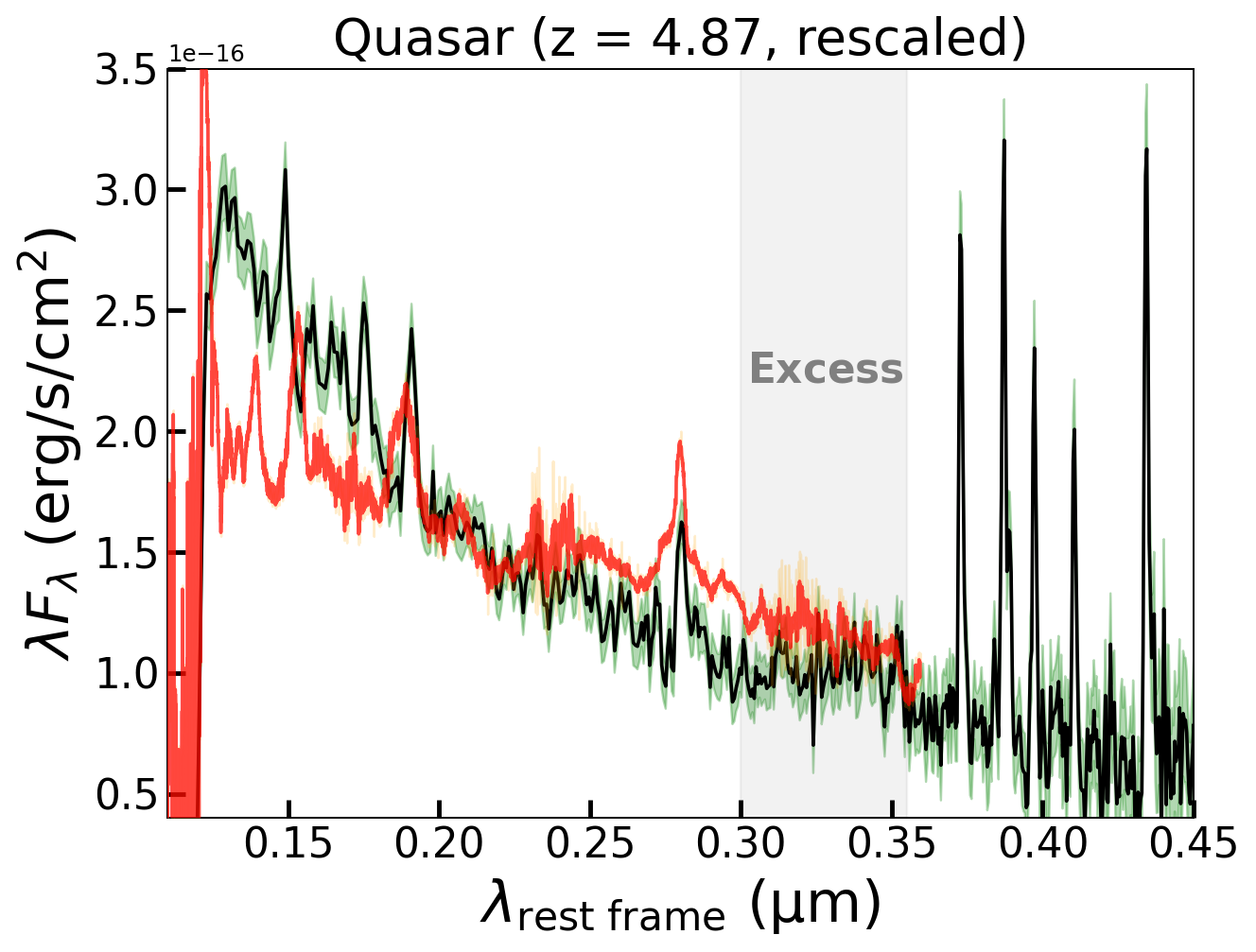}
    
    \caption{Comparison between the rest-frame UV spectra of quasars from \citet{lai2024} and that of GN-z11.
    The quasar spectra are plotted in red with their $1\sigma$ uncertainties indicated by the orange shaded region.
    The fluxes of the quasars are rescaled.
    The spectrum of GN-z11 is shown in black with the $1\sigma$ uncertainty indicated by the green shaded region.
    The continuum excess region we identified in GN-z11 is highlighted with the grey shaded band.
    }
    \label{fig:more_feii}
\end{figure*}

\section{Volume-average electron temperatures in different ionized clouds}
\label{appendix:cloudy_tem}

\begin{figure*}
    \centering\includegraphics[width=0.98\textwidth]{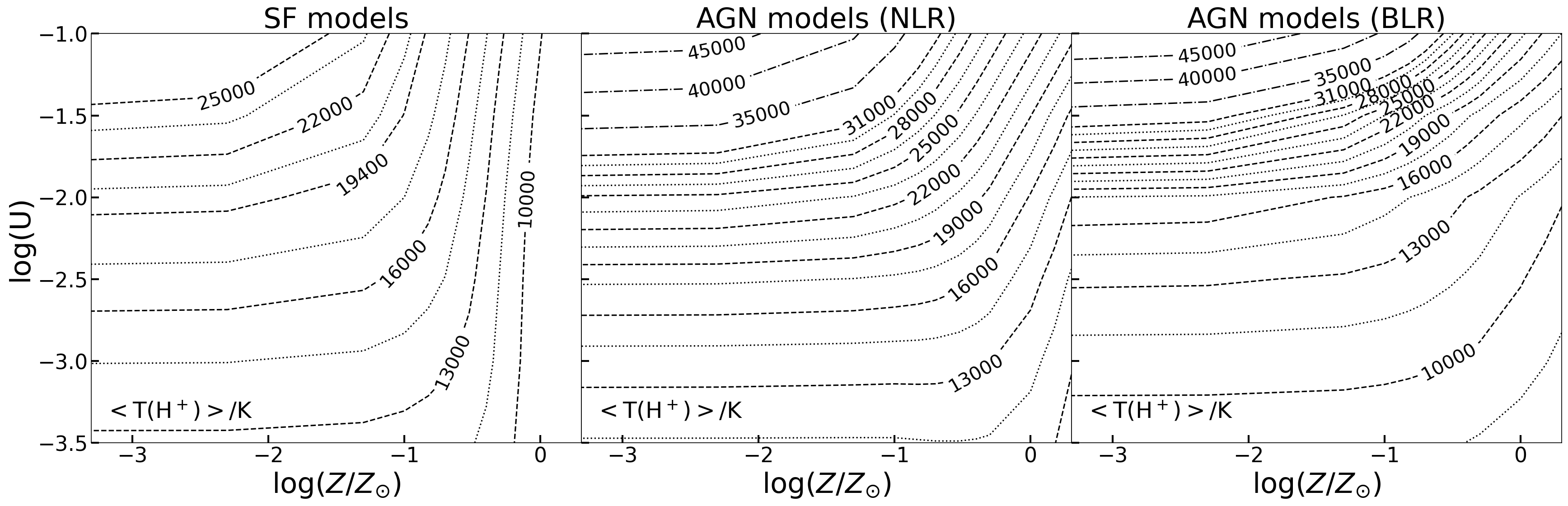}

    \caption{Predictions on hydrogen Balmer temperatures for clouds ionized by different sources as functions of the metallicity and ionization parameter.
    \textit{Left}: contours of electron temperatures of SF models ionized by a simple stellar population (SSP) from a 1 Myr-old starburst generated by BPASS assuming the default initial mass function (IMF) with an upper mass limit of 300 $M_{\odot}$ described in \citet{stanway2018}.
    Both the dashed lines and dotted lines are plotted with steps of 3000 K. Adjacent dashed lines and dotted lines are separated by 1500 K.
    \textit{Middle}: contours of electron temperatures of NLR models ionized by an AGN SED the same as the one in Table~\ref{tab:models}.
    The hydrogen density of the models is set to $10^3~{\rm cm^{-3}}$.
    \textit{Right}: contours of electron temperatures of BLR models ionized by an AGN SED the same as the one in Table~\ref{tab:models}.
    The hydrogen density of the models is set to $10^{11}~{\rm cm^{-3}}$ and the hydrogen column density is set to $10^{22}~{\rm cm^{-2}}$ as appropriate for the $\rm H^+$ region within the BLR.
    While the SF models and NLR models include dust depletion, the BLR models do not. 
    }
    \label{fig:cloudy_tem}
\end{figure*}

Figure~\ref{fig:cloudy_tem} shows the \textsc{Cloudy} predictions on the volume-average $T_e ({\rm H^+})$ for clouds ionized by young stars and AGNs, respectively.
For SF clouds, we adopted a fiducial model which uses the Binary Population and Spectral Synthesis models \citep[BPASS;][]{stanway2018,byrne2022} assuming 1 Myr after a burst of star formation.
Even with log(U) = $-1$ and log(Z/Z$_{\odot}$) = $-3$, the volume-average $T_e ({\rm H^+})$ is still lower than $3\times 10^4$ K.
In comparison, a low-density NLR can reach $T_e ({\rm H^+}) \sim 3\times 10^4$ K at a fiducial metallicity of log(Z/Z$_{\odot}$) = $-0.84$ inferred for GN-z11 \citep{cameron2023} and a log(U) between $-1.5$ and $-1$.
At such a high temperature, however, the electron temperature traced by \oiii$\lambda 4363$ in the NLR would be comparably high, which is inconsistent with what we observed in GN-z11.
%The problem with this NLR model is the strong \oiii$\lambda 4363$ from the high temperature gas inconsistent with the observation.

Finally, for high-density BLR models, we calculate the volume-average temperature, $T_e ({\rm H^+})$, within a column density of $N_{\rm H} = 10^{22}~{\rm cm^{-3}}$ at a fiducial density of $n_{\rm H} = 10^{11}~{\rm cm^{-3}}$, which roughly encloses the $\rm H^+$ zone for the range of ionization parameters we considered \citep{ferland2009}.
%the volume-average $T_e ({\rm H^+})$ is sensitive to the volume density as well as the column density of the gas.
%In Figure~\ref{fig:cloudy_tem}, we show a fiducial model with $n_{\rm H} = 10^{11}~{\rm cm^{-3}}$ and $N_{\rm H} = 10^{22}~{\rm cm^{-3}}$ and has no dust depletion.
%We note that the dependence of $T_e ({\rm H^+})$ on $n_{\rm H}$ and $N_{\rm H}$ is non-monotonic.
While the above parameters can produce $T_e ({\rm H^+}) \sim 3\times 10^4$ K at log(Z/Z$_{\odot}$) $\sim -1$ and log(U) $\sim -1$, it is possible for other combinations of $n_{\rm H}$ and $N_{\rm H}$ to achieve high temperatures.

% Don't change these lines
\bsp	% typesetting comment
\label{lastpage}
\end{document}